\documentclass[10pt,final,twocolumn,a4paper,twoside]{report}

\usepackage{titlesec, blindtext}

\newcommand{\hsp}{\hspace{20pt}}
\titleformat{\chapter}[hang]{\Huge\bfseries}{\thechapter\hsp{|}\hsp}{0pt}{\Huge\bfseries}
\usepackage[Lenny]{fncychap}

\usepackage{palatino}
\usepackage[latin1]{inputenc}
\usepackage{amsfonts}
\usepackage{amsmath}
\usepackage{amssymb}
\usepackage[T1]{fontenc}
\usepackage{textcomp}
\usepackage[breaklinks=true]{hyperref}
\usepackage{pxfonts}

\usepackage{float}
\usepackage{eurosym}

\usepackage[toc,page]{appendix}

\usepackage{lettrine}

\usepackage{graphicx}
\usepackage{float}

 \usepackage[table]{xcolor}

\newcommand{\mail}[1]{{\href{mailto:#1}{#1}}}
\newcommand{\httplink}[1]{{\href{#1}{#1}}}

\def\today{\number\day\space\ifcase\month\or
 January\or February \or March\or April\or May\or June\or
 July\or August\or September\or October\or November\or December\fi
 \space\number\year}
 
 \usepackage{tikz}
 \usetikzlibrary{shapes,arrows}
 
\usepackage{fancyhdr}

\usepackage[a4paper,top=3.2cm,bottom=2.8cm,inner=2cm,twoside]{geometry} 

\fancypagestyle{plain}{%
  \fancyhf{}

\fancyhead[LE,RO]{\slshape \textsc{Kuntzer}, T.} 
\fancyhead[LO,RE]{\slshape Simulation of Stray Light Contamination} 
\fancyfoot[LE,RO]{page | \thepage}

}

\usepackage{rotating} 
\usepackage[nottoc,notlof,notlot]{tocbibind} 
\usepackage{booktabs} 
 \usepackage[font={rm,md}]{caption, subfig}
\newcommand{\HRule}{\rule{\linewidth}{0.5mm}}

\usepackage{natbib}
\usepackage{verbatim}

\begin{document}
\pagenumbering{roman}
\begin{titlepage}
\thispagestyle{empty}
\pagestyle{empty}
\begin{center}
\begin{minipage}{1\textwidth}
 \includegraphics[height=20mm]{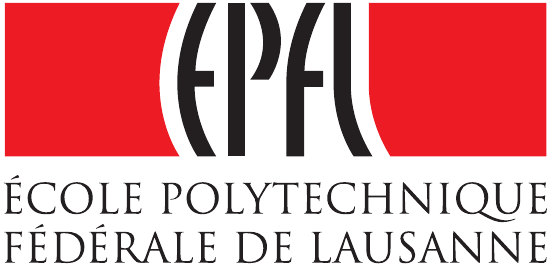}
\hfill
\includegraphics[height=25mm]{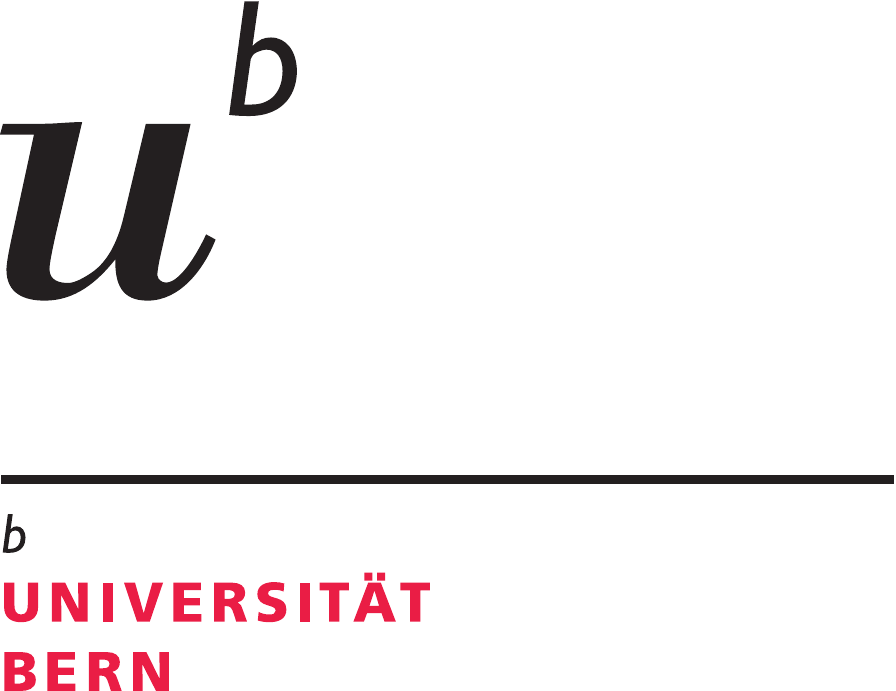}
\\[3.5cm]
\end{minipage}
\textsc{\large Master Thesis}\\[0.2cm]

\HRule \\[0.4cm]
{ \huge \bfseries {\textsc{CHaracterising ExOPlanets Satellite}\\[0.2cm]}}
{\Large Simulation of Stray Light Contamination on CHEOPS Detector}
\\[0.4cm]
\HRule \\[2.5cm]

{\large Thibault \textsc{Kuntzer}\\{\small \mail{thibault.kuntzer@epfl.ch}}}\\[2cm]

{\large\bf\textsc{Master Thesis Carried out at the University of Bern\\in the Theoretical Astrophysics and Planetary Science Group}}\vspace{1em}
{\large
\begin{minipage}[t]{0.4\textwidth}
  \begin{flushleft}
  {\bf\textsc{Supervised by}}\\
     Dr. Andrea \textsc{Fortier}\\{\small \mail{andrea.fortier@space.unibe.ch}}
  \end{flushleft}
\end{minipage}
\begin{minipage}[t]{0.4\textwidth}
\begin{flushright}
{\bf\textsc{Advised by}}\\
  Prof. Willy \textsc{Benz}\\{\small \mail{willy.benz@space.unibe.ch}}
\end{flushright}
\end{minipage}
\\[1.7cm]
{\bf\textsc{Followed at EPFL by}}\\
Prof. Georges \textsc{Meylan}\\{\small \mail{georges.meylan@epfl.ch}}\\[1cm]

\vspace{0.5cm}

 \vfill

{Spring Semester 2013}}
\end{center}
\end{titlepage}

\newpage 
\pagestyle{plain}
\mbox{}
 
 \onecolumn
 
 \chapter*{Abstract} \addcontentsline{toc}{chapter}{Abstract}
 \lettrine[lraise=0., nindent=0em, slope=-.5em]{T}{he} aim of this work is to quantify the amount of Earth stray light that reaches the CHEOPS (CHaracterising ExOPlanets Satellite) detector. This mission is the first small-class satellite selected by the European Space Agency. It will carry out follow-up measurements on transiting planets. This requires exquisite data that can be acquired only by a space-borne observatory and by well understood and mitigated sources of noise. Earth stray light is one of them which becomes the most prominent noise for faint stars.
 
 A software suite was developed to evaluate the contamination by the stray light. As the satellite will be launched in late 2017, the year 2018 is analysed for three different altitudes. Given an visible region at any time, the stray light contamination is simulated at the entrance of the telescope. The amount that reaches the detector is, however, much lower, as it is reduced by the point source transmittance function (PST). It is considered that the exclusion angle subtend by the line of sight to the Sun must be greater than 120\textdegree, 35\textdegree\ to the limb of the Earth and 5\textdegree\ away from the limb of the Moon. 
The angle to the limb of the Earth -- the stray light exclusion angle -- is reduced to 25\textdegree\ in a later phase. Information about the faintest star visible in any direction in the sky is therefore available and is compared to a potential list of targets. The influence of both the visibility region and the unavoidable South Atlantic Anomaly (SAA) which is dictated by the altitude of the satellite are also studied as well as the effect of a changing optical assembly. A methodology to compute the visible region of the sky and the stray light flux is described. Furthermore, techniques to prepare the scheduling of the observation as well as a possible way of calibrating the dark current and the map of hot pixels in the instrument are presented.

The simulations show that maximal stray light flux set in the science requirements of the mission is not often reached. There are seasonal variations on the amount of flux received. Lower orbits tend to present flux which exceed the requirement. The stray light increases by nearly 20\% when the altitude is reduced respectively from 800 to 700 and from 700 to 620 km. However, the South Atlantic Anomaly impacts more direly higher orbits. This high radiation region demand the interruption of the science operations. Even if the viewing zone at low altitude is smaller than at 800 km, the availability of instrument is greater. There exist two favoured regions for the observations: in April and mid-September. The field of view is the widest then as the plane of the orbit and of the terminator merge. Loosening the condition on the stray light exclusion angle allows a wider visible region for bright stars without affecting much the quality of the data.

In turn, the following recommendations can be made: (1) the stray light must be simulated to ensure that the visible region is correct; (2) The stay light exclusion angle can be reduced to 25$^\circ$ or $28^\circ$; (3) Low orbits imply that a given object can be observed more efficiently; (4) The two favoured observation periods must be used for pristine observations; (5) Short period planets should be observed in Northern winter while long period planets in the Summer.

\cleardoublepage
 \chapter*{Acknowledgements} \addcontentsline{toc}{chapter}{Acknowledgements}
  \lettrine[lraise=0., nindent=0em, slope=-.5em]{I}{} am deeply indebted to Dr. Andrea Fortier for her help, availability and continuous support throughout this master thesis. I would like to thank the Theoretical Astrophysics and Planetary Science Group (TAPS) group for welcoming me. I am grateful to the whole CHEOPS team and in particular to Udo Wehmeier, to Dr. Christopher Broeg and to Dr. Yann Alibert for their ideas and support. There are many other people external from UniBe that pointed me towards the right direction amongst them Dr. David Ehrenreich, Adrien Deline both at UniGe, Luzius Kroning at EPFL and Dr. Matteo Munari at INAF in Italy. There are people outside the CHEOPS consortium that I would like here to thank for their support -- you know who you are.
 Finally, I would like to warmly thank my advisors Prof. Willy Benz at UniBe for his trust, his remarks and his encouragements and Prof. Georges Meylan at EPFL for giving me the opportunity of working on this exciting project.
\paragraph{}
 This research has made use of the Exoplanet Orbit Database, the Exoplanet Data Explorer at exoplanets.org and of the Sloane Digital Sky Survey-III data at sdss3.org.

\twocolumn


\cleardoublepage
\tableofcontents
\clearpage

\cleardoublepage
\pagenumbering{arabic}
\chapter{Introduction}
\lettrine[lraise=0., nindent=0em, slope=-.5em]{O}{ver} the course of the last 20 years, there has been an explosion of planets detected that are orbiting other stars than our Sun -- the so-called exoplanets. Those numerous discoveries unveiled objects that are beyond what was thought to be possible. With the passing of time, the instruments became better and better yielding for the first time this year objects smaller than the Earth itself \citep{Barclay2013}. The unexpected zoo of extrasolar bodies count in their ranks planets in the range from super-Earth ($m_\oplus \lesssim m_p \lesssim 10m_\oplus$) to Neptune-like planets (for which $m_p \lesssim 20m_\oplus$). Those planets make up an interesting sample as somewhere between the two extrema lies a zone of transition from terrestrial to gaseous bodies.
There is no clear limit and the transformation can be more easily described by a blurry phasing out of the terrestrial type to a phasing in of gaseous planets. The mass-radius $m$--$R$ relationship of those objects is not yet well constrained as the dataset is sparse due to the detection techniques.
With a large sample and more precise measurement this relationship is likely to exhibit a natural scatter reflecting the largely different formation conditions that occurred in those systems. This important relationship will provide constraints on the formation of planetary companions of stars but also on the birth or abortion of stars.
 More massive planets, similar to Jupiter or even larger, also contribute to this diversity. The first discoveries in the exoplanet field were extremely surprising as they unveiled Jupiter-like bodies orbiting very close to their stars \citep{Mayor1995}. Such systems immediately prompt questions about the formation mechanisms, the migration in the system as well as heat transfer between night and day side of those planets.
\paragraph{}
The search missions such as the ground-based High Accuracy Radial velocity Planetary Search (HARPS) instrument or the space-borne Kepler satellite are finding planets and cataloguing them with the available data (\cite{Pepe2002} and \cite{Basri2005}). The amount and the quality of data that can be retrieved are limited by the performances of the instruments as well as by the physical constraints of the techniques. HARPS is based on the radial-velocity measurement which does not yield information on the dimensions other than an estimate of the mass of the object orbiting the star whereas Kepler looks for transiting planets -- planets that passes in front of their star in the line of sight of the telescope -- which gives the radius \citep{Wright2013}. Those two techniques are therefore complementary.
\paragraph{}
Follow-up missions are designed to complete the gap in knowledge of a particular system. The Characterizing Exoplanets Satellite (CHEOPS) will look for transiting systems discovered by means of radial velocities as well as transit surveys. Using precise photometric measurements of the transit, the instrument on-board the satellite will be able to determine the radii of the observed systems to a precision of 10\% which can be used to add precise points on the $m$--$R$ graph \citep{Broeg2013}. The same transit technique and photometric surveys can be applied to giant gaseous planets very close to the their host star -- the ``hot Jupiters'' -- to probe their atmosphere and gain insight into their evolution. 
\paragraph{}
CHEOPS won the call of the European Space Agency (ESA) for the first small-class mission. It is designed by a consortium lead by the University of Bern regrouping many countries with a large Swiss contribution by the Observatory of Geneva, the Swiss Space Center at Ecole Polytechnique F\'ed\'erale de Lausanne (EPFL) and Eidgen\"ossische Technische Hochschule Z\"urich (ETHZ). It will be a light small satellite launched in low Earth orbit (LEO) by the end of the year 2017.
\paragraph{}
This work focuses on one particular aspect of the instrument: the contamination of the detector -- a charged couple device (CCD) -- by stray light. Stray light can be of large impact on the image: it is a source of diffuse noise as photons emitted from the Sun bounce off the atmosphere of the Earth and may reach the telescope. As the measurement technique of the satellite is using precise photometry on slight changes in the flux of a star (down to a few tens of ppm), this source of noise is very important to study as it restricts the maximum magnitude of a target star to be observed by adding a constant noise to an exposure. The baseline orbit for this mission is a Sun-synchronous orbit (SSO) around the Earth at an altitude between 620 and 800 km above the surface. The stray light contamination depends upon several different variables : the altitude, the angle subtended by the target star and the limb of the Earth as well as optics and the choice of the baffle on the telescope. 
The objectives of this Master thesis are:
\begin{itemize}
 \item To study the behaviour of the stray light flux received by the detector at different time in the year 2018;
 \item To determine the regions where this radiation is too high;
 \item To determine the regions where the mission-wide stray light exclusion angle could be lowered;
 \item To determine the effect of the altitude of the orbit onto the amount of photons received;
 \item To study how changes in the optical design of the telescope affect the stray light;
 \item To prepare a methodology to schedule the observations.
\end{itemize}

This thesis reports the work carried out to generate observability maps for different assumptions for the altitude given a optical design and a stray light exclusion angle. It discusses also targets availability. It is divided into several chapters and sections: chapter \ref{ch:theory} gives a broad approach into the work starting for exoplanets (\S\ref{sec:theo-exo}), to detection techniques (\S\ref{sec:theo-detections}) and CHEOPS (\S\ref{sec:CHEOPS}) and its instrument as well as a theoretical introduction to the problem of stray light (\S\ref{sec:theory-sl}). Chapter \ref{ch:numerical} describes the different numerical methods and software used while chapter \ref{ch:results} presents and analyses the results. Conclusion and outlooks are drawn in chapter \ref{ch:conclusion}.
\cleardoublepage
\chapter{Exoplanets, Detection Techniques and CHEOPS} \label{ch:theory}
\lettrine[lraise=0., nindent=0em, slope=-.5em]{T}{his} chapter is intended to give a background to this project in terms of exoplanets, how they are detected and the CHEOPS mission. CHEOPS will be explained from both a scientific and an engineering point of view as this work may be used to constrain some engineering points and also mission operations. The notation $\star$ for the star and the subscript $p$ for planets are used throughout this chapter.
\section{Exoplanets} \label{sec:theo-exo}
In this section, selected parts of exoplanet astrophysics will be examined in order to present the science goals of CHEOPS as well as planet formation, evolution and detection. 

\subsection{On Planet Formation} \label{sec:theo-formation-evolution}
 In 1755, the philosopher Kant produced a model that was in some agreement with the modern ideas about how the Solar System came to be \citep{Beatty1999}. Since then, a more precise picture has emerged, but based on the only known system: ours. Exoplanets add difficulties to the picture as the range of exoplanet size extends from sub-Earth size up to gaseous giants several times the mass of Jupiter in a wide range of orbits. The existence of such a zoo starts to be explained by computer simulation of N-bodies systems, but some of the steps required are still beyond our understanding. The description of a typical planetary system around a star given here is adapted from reviews (\citet{Mordasini2010}, \citet{Armitage2007}, \citet{Lissauer1993}), some of which concentrates more on our Solar System (\citet{Beatty1999} for example) than the shear diversity out there.

The central star of a system forms from the collapse of a nebula cloud. The setting is a cloud of interstellar matter, dark and cold. This cloud does not simply sit, it is agitated by the difference of gravitational potential and expands as well as being threaded by magnetic fields. Where the concentration of material is high enough, the nebula cloud collapses under the pressure of the gravitational attraction. With the conservation of the angular momentum, the initially slowly rotating cloud starts spinning faster as the radius of the future system reduces. As some of the material has too much angular momentum, it does not fall to the protostar but orbits around it. The bubble of material left over after the formation of the protostar becomes a disk as the material collides and scatters around the equatorial plane.
At this point the planet formation starts. The formation of planets requires a growth of the particle size of at least 10 to 12 orders of magnitude. The formation process occurs simultaneously with the evolution of the protoplanetary nebula and the latest phases of star formation.

\paragraph{Dust.} It is the smallest scale of solid material present in the disk, composed of particles from sub-micron to a few centimetres. Its growth is governed by physical collisions as one particle of dust sticks to another. The gravitational attraction is not strong enough to retain the particles together and Van der Waals attraction may be a better candidate. Understanding this phenomenon of sticking dust grain together requires the modelling of both mechanical and chemical interactions and processes in play.
When reaching the centimetre-size, the dust becomes weakly coupled to the gas. This implies a drag which forces them to migrate inwards on a fast time scale. There is an unknown mechanism that make those particles grow faster than the speed at which they fall to their star. 
The nebula is considered to be turbulent, the dust cannot settle on a thin disk and therefore some other mechanism must be used, which can be described by self gravity -- this is the gravoturbulent planetesimal formation. Turbulence and instabilities in the disk can lead to over dense regions which can be sufficiently dense to collapse.


\paragraph{Planetesimals.} They are objects whose size is of the order metres to tens of kilometres. The mechanism to create planetesimal from the dust is for now unknown. The gravitational interaction between planetesimals exceed the electromagnetic forces, the gas drag as well as collective gravitational effects of the smaller bodies. To continue their growth, they accrete smaller planetesimals through collisions and merge with other planetesimals with the rule that ``the rich become richer'' as the collisional cross section depends upon the gravitational focussing which dominates and upon the intrinsic size of the object.
At this point, the largest bodies in the disk undergo a runaway growth. The latest phase of this runaway growth is a slower growth of the objects called oligarchic growth. This is due to the gravitational stirring that the largest bodies produces in the swarm of the small planetesimals, but still the largest bodies grow faster than the small ones. Planetesimals orbits and velocities are distributed over a large range of values which are not only described by the simple and perfect Keplerian velocities. Scattering -- in other words close encounters -- and other pairwise interactions interfere with the Keplerian orbits yielding more chaotic motions with the supplementary effect of the gas drag which tends to damp eccentricities and inclinations. 

\paragraph{Earth mass bodies.} Those are the future core of giant planets. At this stage, their gravitational influence is such that they re-couple with the gas disk. The gas disk is gravitationally attracted by the protoplanet which is a different process than the early interaction in the disk with rocks or dusk. Gaseous accretion onto the protoplanet continues until the disk is either depleted or is dissipated or if gravitational torques in the region have created a gap in the disk. Late formation in an almost depleted disk or close to the star where there is less material does not to lead to a gaseous planet. Therefore, the following paragraphs describes the two formation hypotheses for giant planets and the formation of terrestrial planets. 

\paragraph{Gaseous Planets.} They are more massive bodies which have a large fraction of their mass composed of gas. Two processes could explain their formation. In the first hypothesis -- the direct collapse scenario -- giant planets originate from the collapse of a part of the gas. This mechanism requires a heavy disk and a very efficient cooling. Therefore, it tends to be efficient only early on. For the second one, the beginning of the mechanism depends not on the gas disk, but rather on the mass of protoplanet. At a certain critical mass, the icy core surrounded by an envelope of gas becomes unstable and start to shrink enabling further gas accretion. The set of equations governing this growth is similar to the ones that model the formation of stars with the notable exception of replacing the source of energy: nuclear fusion by accreted planetesimals potential energy. 

\paragraph{Terrestrial Planets.} They are thought to be formed after the gas disk was dispersed. It was previously said that planetesimals had their orbit eccentricity and inclination damped by the gas disk. Once giant gaseous planets are created, all remaining planetesimals start pumping up those two same parameters by means of gravitational interaction leading to impacts.

\subsection{Evolution of Planetary Systems}
In the previous section, \emph{formation} of planets and their system was discussed. The next step is to discuss their \emph{evolution}. The boundary between these two concepts is somewhat blurry. There are two ways to look at this: either from a planetary viewpoint or from a system standpoint. The limit between formation and evolution for the former can be set at the moment when the body will have reached say 90\% of its final mass. For rocky planets, this definition is straightforward as they can only gain mass, but gaseous planets may evaporate. 
In the later view -- the system view -- the system can be considered formed once the protoplanetary disk has disappeared. In this case, the gaseous giants have a clear threshold into evolution as they accrete their gas fairly quickly. On the other hand, rocky planets accrete their mass on a much longer timescale. Hence, whatever the definition chosen, there will be uncertainties on the transition between formation and evolution. As the section on exoplanets follows the lecture notes by \cite{Armitage2007} (and references therein), the convention of the planetary point of view is chosen. 

Once the planets have formed, several phenomena can occur that will still modify the configuration of the system. Four models are well supported by observations, but none are fully understood.


\paragraph{(a) Drag by the gas disk.} During formation, the interactions between planets and the gaseous disk cause migration. This effect is important as it yields a theoretical motive for the existence of hot Jupiters. When the gas disk is still present, it is thought that eccentricities of the orbits are damped such that they become more and more circular.

{\bf Torques.}
As the orbital speed depends on the semi-major axis $a$ of a given object ($v_\text{orbital} \propto a^{-1/2}$), the planet moves faster than the gas outwards of the orbit. Therefore, the angular momentum of the gas is increased by interactions with the planet which results in a inwards migration while the gas is expelled outwards. In case the planet interacts with the interior gas, then the resulting movement are inverted. A rather complex sum of the torques implies that the gas is always expelled outwards.

{\bf Type I Migration.} For low mass planets, the exchange of angular of momentum by the planet has little effect on the gas disk and therefore, the viscous drag is more important. Thus, the net torque on the planet is simply the sum of all torques including the co-orbital resonance. The migration due to those torques is called type I. It can be shown that this migration is fast (the bigger the planet, the faster its migration) and its time scale depends upon the mass of the planet. This holds only if the influence of the planet on the gas disk is negligible. The understanding of type I migration is far from being comprehensive and the exact form of the terms of the torques is not yet well defined.

{\bf Type II Migration.} When the assumptions of negligible effect of the planet on the disk are not longer valid, \emph{i.e.}~when the planet is massive enough, the angular momentum exchange prevails over the viscous drag. The gas is repelled away from the orbit of the planet which creates a gap: a sudden drop in the surface density of the disk near the orbit of the planet. 
In most cases the type II migration causes the planet to move inwards and slowly. The effect of the gap is that it applies a barrier to the flow of gas. Accumulation of gas at the edge causes the planet to move in. Type II migration seems to be a good theoretical explanation for the existence of hot Jupiters.
\paragraph{(b) Remnant planetesimals.} Another possibility for the source of migration is interactions with remnant planetesimals especially after the dissipation of the gaseous part of the disk. Simulations point towards hints that this effect affected the early Solar System by migrating Neptune and Uranus and maybe Saturn \citep{Levison2007}. 
The cumulative effect of the remnant planetesimals start to be important when their total mass is of the same order as the mass of the planet. 
\paragraph{(c) Initial instabilities.} There is no mechanism preventing planets pairwise scattering or even collisions from happening. Massive planets tend to remain in the system while low mass planets are ejected. Initial instabilities could well explain the eccentric orbits of extrasolar bodies as the depleted gas would not damp the eccentricities. 
The outcome of planet-planet interactions can be classified into (1) stable system over a long term, (2) one planet is ejected which usually pumps the eccentricity of the remaining planet, (3) the planets collide and (4) one planet falls onto the star leaving, in general, the other in an orbit close to the star. 
Investigations on this subject demand numerous three bodies simulations. Such studies reveal that instabilities over a long term are unusual when the planets are far out. However, when the planets are located close to their star, scattering and collisions are more frequent. For larger systems, numerical computations show that there is a quantitative match to the observations.
\paragraph{(d) Tidal interactions.} Tidal interactions between the host star and a planet are important only at short distances. They have an effect on the semi-major axis and eccentricities of Hot Jupiters and other smaller very close planets. The tides arise from the gradient of gravitational forces that affect planets which have finite dimensions. Earth tides dissipate energy of the Moon which implies a slight increase of the distance between the Moon and the Earth. There is thus a torque in this system. In a perfect hydrostatic system, no torque should arise as the tides would be symmetric on the body and would be on the line joining the centres of two bodies. Due to the orbital rotation, the tides are not aligned with the centres, \emph{i.e.} creating a non-zero response time to the tidal perturbation and therefore exert a torque. 

\section{Detection Techniques} \label{sec:theo-detections}
The detection of planets orbiting stars other than our Sun can relate to the search for stellar companions. The first peer-reviewed claim (and which was actually confirmed later on) of planet discovery dates back to \cite{Campbell1988}. In this work, suspicions about the discovery of an extrasolar planet of a few Jupiter masses were raised, however the actual confirmation came in 2003. In the article, the authors hesitate indeed between a planet or a dwarf star. In this section, the detection techniques, their capabilities and their bias will be discussed. First of all, useful quantities will be defined. Unless stated otherwise, this discussion follows the book by \cite{Wright2013}. 

\begin{figure}[h]
 \begin{center}
  \includegraphics[width=1\linewidth]{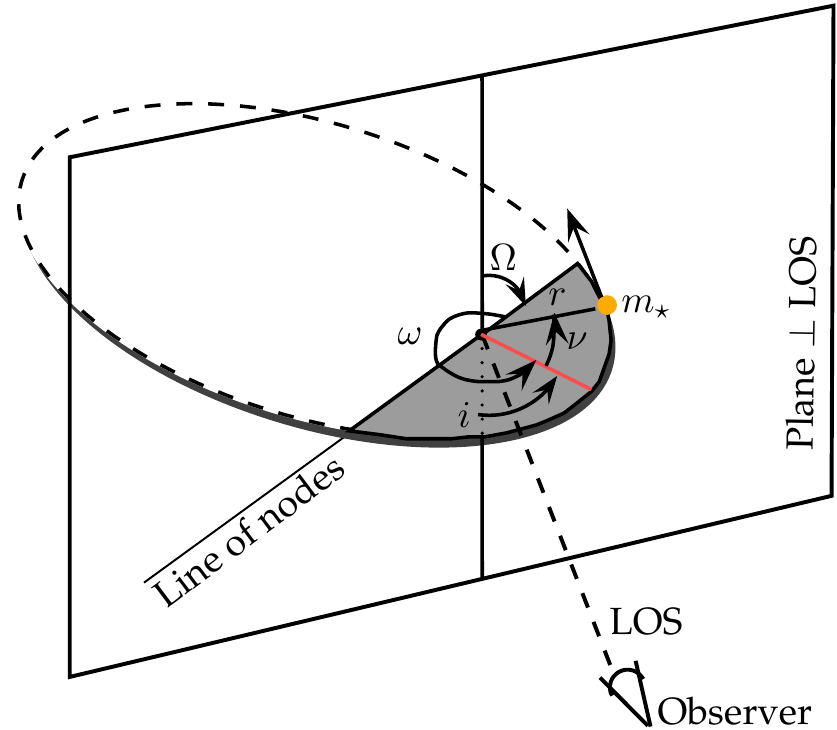}
  \caption{Depiction of the orbital elements -- The line of sight points directly to the centre of mass of the system. For clarity, the second body was omitted. The motion of the star is shown with respect to the centre of mass. The frame represents the plane perpendicular to the line of sight. The line of node is the line that is formed by the intersection of the plane of the orbit (in gray) and the plane perpendicular to the line of sight. The red line joins the periastron to the centre of mass. Both the argument of the periastron and the true anomaly increase in the direction of the motion of the object along its orbit.}
  \label{fig:exo.orbital_param}
 \end{center}
 \vspace{-1.em}
\end{figure}
Two objects orbiting each other in a two body system are doing so around the centre of mass. This point will be the origin of the coordinate system. The angles are measured from the line of nodes which is the line described by the intersection of the plane of the observer which is perpendicular to the line of sight and the orbital plane. The right ascension of the ascending node (RAAN) $\Omega$ describes the inclination of the line of nodes on the sky (see Fig. \ref{fig:exo.orbital_param}) with respect to the plane perpendicular to the line of sight (LOS). At this point, the star crosses the plane of the sky. At one of the two points, the star moves away from the Earth. The RAAN is defined at this point. The periastron defines the reference point in the orbit. Its location with respect to the line of nodes is described by the argument of the periastron $\omega$ which is in the plane of the orbit. The argument of the periastron for the planet and the star differ by $\pi$: $\omega_p = \omega_\star+\pi$. 
The position of the star is described by the true anomaly $\nu_\star$. The position of the star in this system is therefore completely defined by the coordinates $(r_\star,\nu_\star)$ where $r_\star$ describes the distance of the star to the centre of mass. The inclination of the plane of the orbit $i$ is referenced to the plane perpendicular to the LOS. The period of rotation is given by the mass of the system and the semi-major axis $a=a_\star + a_p$ where $a_i$ are defined by the distance from the centre of mass to the centre of the body:
\begin{equation}
 P^2 = \frac{4\pi}{G(m_\star+m_p)}a^3
\end{equation}
The distance from the centre of mass to the star is given by:
\begin{equation}
 r_\star(1+e\cos\nu_\star)=a(1-e^2)
\end{equation}
with $e$ being the eccentricity of the orbit. Practically, the position of the star is described by the use of another variable, the eccentric anomaly $E$ which is related to the time $T_0$ at which the body reaches the periastron through the mean anomaly~$M$:
\begin{equation}
 M = \frac{2\pi(t-T_0)}{P}=E-e\sin E
\end{equation}

\begin{figure}[h]
 \begin{center}
  \includegraphics[width=0.6\linewidth]{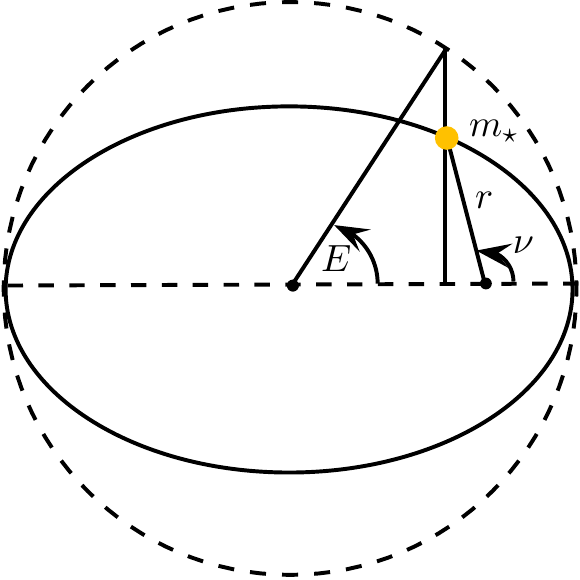}
  \caption{Depiction of the eccentric anomaly $E$}
  \label{fig:exo.eccentric-anomaly}
 \end{center}
 \vspace{-1.5em}
\end{figure}

This angle $E$ represents the angle between the periastron and the actual position of the body on a virtual circular orbit (Fig. \ref{fig:exo.eccentric-anomaly}). It allows a simple calculation of the true anomaly $\nu$ and either $a$ or $e$:
\begin{eqnarray} \label{eq:exo.E1}
\tan\frac{\nu}{2}=\sqrt{\frac{1+e}{1-e}}\tan \frac{E}{2}\\ E=\arccos\frac{1-r/a}{e} 
\label{eq:exo.E2}
\end{eqnarray}

At this point, the orbit is not fully described as the inclination $i$ is missing. It determines the angle subtend by the crossing of the plane of the sky and the plane of the orbit. It must be noted that the inclination and the RAAN are defined with regard to the observer and not to an arbitrary reference.

\subsection{Radial Velocities}
\begin{figure*}
 \begin{center}
  \includegraphics[width=0.7\linewidth]{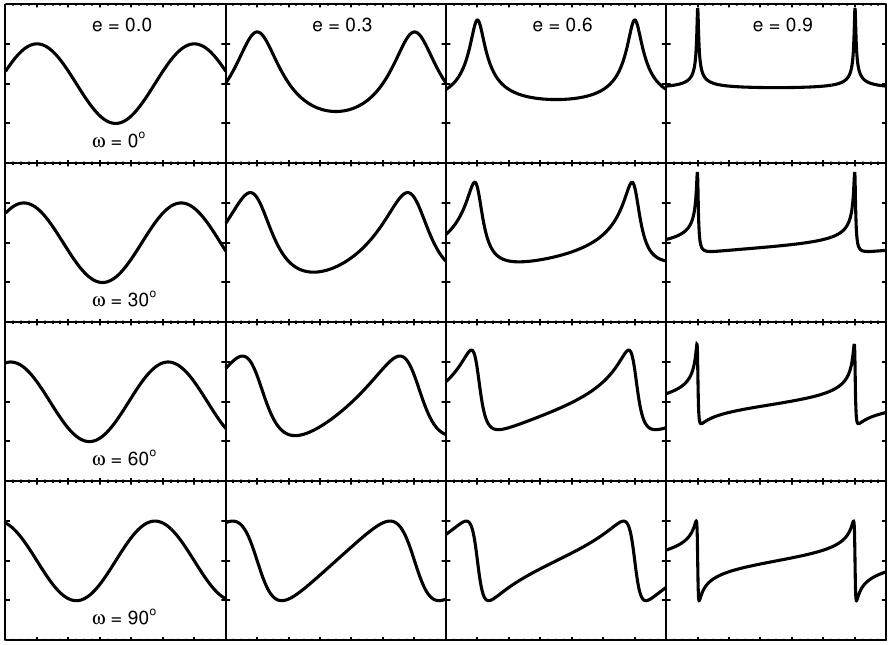}
  \caption{Different shapes of curves determined by the radial velocity detection technique as imposed by the eccentricity $e$ and the true anomaly $\omega_\star$. All curves share the same $K,P,T_0$. Taken from \cite{Wright2013}.}
  \label{fig:exo.e_omega.png}
 \end{center}
 \vspace{-1.5em}
\end{figure*}

The motion of a star around the centre of mass can be measured by precise measurements of the Doppler shift. Therefore, the technique is called \emph{radial velocities} (RV) or Doppler spectroscopy. This technique can reveal the period, some orbital parameters, the mass and semi-major axis of the planet. The Doppler measurement yield a radial velocity whose signal is given by six parameters: (1) the period of the orbit $P$, (2) the eccentricity $e$, (3) the semi-amplitude of the signal $K$ (bearing units of velocity), (4) the argument of periastron (of the star) $\omega_\star$, (5) the time for periastron crossing (for a given epoch) $T_0$ and (6) the bulk velocity of the centre of mass. The Doppler measurement yields $V_r$ is given by:
\begin{equation}
 V_r = K \left[ \cos(\nu + \omega_\star)+e\cos\omega_\star\right] + \gamma
\end{equation}
with $P,e,T_0$ are hidden inside $\nu$ and $E$ (eq. \ref{eq:exo.E1} \& \ref{eq:exo.E2}). While $K$ sets the amplitude of the oscillation of the RV curves, $P$ and $T_0$ respectively define the period and the phase of the RV curve. The two other parameters of the variation $w_\star,e$ define the shape of the signal (See Fig. \ref{fig:exo.e_omega.png}). The bulk velocity of the system is given by $\gamma$.

The inclination of the system $i$ and the RAAN $\Omega$ cannot be deduced from the RV measurements. Not knowing the inclination of the plane leads to a degeneracy of orbital parameters. Therefore, the observables are linked to the mass \emph{via}:
\begin{equation}
 \frac{PK^3(1-e^2)^{3/2}}{2\pi G}=\frac{(m_p\sin i)^3}{(m_p+m_\star)^2}\approx\frac{(m_p\sin i)^3}{m_\star^2}
\end{equation}

The right hand side of the above equation is called the mass function of the system. If the mass of the star is known, the minimum mass of the planet $m_p\sin i$ can be estimated. Therefore the true mass of the planet is larger by a factor of $1/\sin i$ which can be estimated to be 1.15 as a median if random inclinations are assumed.

The signal-to-noise ratio ($SNR$) can be approximated to:
\begin{equation}
 SNR \approx \sqrt{N} \frac{A}{\sigma}
\end{equation}
were $N$ is the number of observations, $A$ the amplitude of the signal and $\sigma$ the uncertainty. For the radial velocities techniques it can be written as 
\begin{equation}
 SNR_{RV} \sim \sqrt{N} \frac{K}{\sigma_{RV}}
\end{equation}
The shorter the period of rotation, the stronger the signal. $K$ can be written as:
\begin{equation}
 K=\left( \frac{P}{2\pi G}\right)^{-1/3} \frac{m_p \sin i}{m_\star^{2/3}} (1-e^2)^{-1/2}
\end{equation}
To ensure that the signal-to-noise ratio is enough, the precision of the measurements must be sufficiently high $\sigma_{RV} \ll KN^{1/2}$. To discover a planet with the same characteristics as Jupiter (\hbox{$K\approx 12.5 \text{ m/s}\ \cdot\sin i$}), the needed number of observations with a precision of the order of the m/s is about a few dozen. An Earth-like planet around a Sun-like star is much more difficult to detect as the amplitude induced is smaller by two orders of magnitudes. To reach this kind of precision with the instruments, bright stars or large apertures telescopes must be used. Moreover, the instruments must be exquisitely calibrated and ultra-stable. A very good example of instrument is HARPS developed by the University of Geneva (UniGe). Its accuracy is currently of about 1 m/s and is actually a spectrograph mounted on an European Southern Observatory (ESO) in La Silla, Chile.

\subsection{Transits} \label{sec:theo-transit}
Some planetary systems can be seen nearly edge-on. This peculiar geometry gives rise to eclipses of the star by the planet(s). An observer that points a telescope to this system at the moment of the eclipse sees a periodic photometric variation of the light coming from the star. This detection technique is called transit and is the method that will be used by CHEOPS to follow and characterise planets. Several space-borne missions such as Kepler or CoRoT \citep{Auvergne2009} are based on this method as well as ground-base instruments for example the Arizona Search for Planets program. This paragraph follows the paper by \cite{Winn2010}. The condition (or an approximated condition) to see a transit is that the separation between the planet and the star projected on the plane of the sky must be less than the sum of the radii, \emph{i.e.}:
\begin{equation}
 r(t_c) \cos i \leq R_\star + R_p
\end{equation}
where $r(t_c)$ is the projected separation at conjunction -- when the planet is closest with respect to the observer -- of two objects, $i$ the inclination of the orbit. $r(t_c)$ can be related to the argument of the periastron \emph{via}:
\begin{equation}
 r(t_c) = a\frac{1-e^2}{1+e\sin\omega_\star}
\end{equation}
With this definition, the impact parameter $b$ can therefore be introduced (in units of the radius of the host star) as:
\begin{equation} \label{eq:exo-transit-b}
 b = \frac{a\cos i}{R_\star}\frac{1-e^2}{1+e\sin\omega_\star} \leq 1 + \frac{R_p}{R_\star}
\end{equation}
Assuming an isotropic distribution of orbits, the transit probability is:
\begin{equation}
 P_\text{tr} = \frac{R_\star+R_p}{a}\frac{1+e\sin\omega_\star}{1-e^2}
\end{equation}

\begin{figure}[h]
 \begin{center}
  \includegraphics[width=0.9\linewidth]{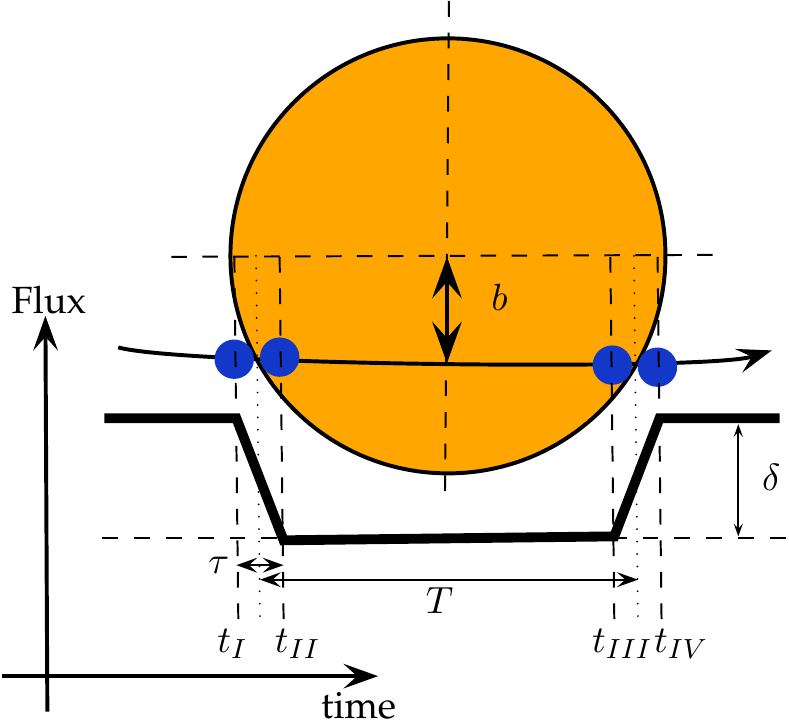}
  \caption{Geometry of transits. Adapted from \cite{Winn2010}.}
  \label{fig:exo-transit_time}
 \end{center}
 \vspace{-1.5em}
\end{figure}

In the following, a restriction is made to non-limb-darkened star, to circular orbits, and to the usual assumption on a planetary system (namely $R_p \ll a,\ m_p\ll m_\star$). The ratio of the radii is frequently used and thus named $k = R_p/R_\star$. Using those assumptions, the trajectory of the planet is a straight line in front of the star with impact parameter $b$. A sketch of the situation is represented in figure \ref{fig:exo-transit_time}.

For a non-grazing eclipse, the total duration of the eclipse is given by $T_\text{tot} = t_{IV}-t_I$ with a full duration of the eclipse $T_\text{full} = t_{III} - t_{II}$. The time associated with the dimming of the light is called ingress or egress duration $\tau$. Using the equation of motion, the time of total and full eclipse can be recovered:
\begin{eqnarray} \label{eq:exo-transit-tf}
 T_\text{full} = \frac{P}{\pi}\sin^{-1}\left( \frac{R_\star}{a}\frac{\sqrt{(1\pmb{-}k)^2-b^2}}{\sin i} \right) \\
 T_\text{tot} = \frac{P}{\pi}\sin^{-1}\left( \frac{R_\star}{a}\frac{\sqrt{(1\pmb{+}k)^2-b^2}}{\sin i} \right) \label{eq:exo-transit-tt}
\end{eqnarray}
Applying the assumption of $k \ll 1,\ R_p\ll R_\star \ll a$, the results can be simplified to:
\begin{eqnarray}
 T &\approx &T_0 \sqrt{1-b^2} \\
 \tau &\approx& \frac{T_0k}{\sqrt{1-b^2}}
\end{eqnarray}
where $T_0$ is a characteristic time scale 
\begin{equation}
 T_0 \equiv \frac{R_\star P}{\pi a} \approx 13\ \text{hr}\ \left( \frac{P}{1\ \text{yr}} \right)^{1/3}\left( \frac{\rho_\star}{\rho_\odot} \right)^{-1/3}
\end{equation}
with the mean density $\rho$.

\begin{figure*}
 \begin{center}
  \includegraphics[width=0.7\linewidth]{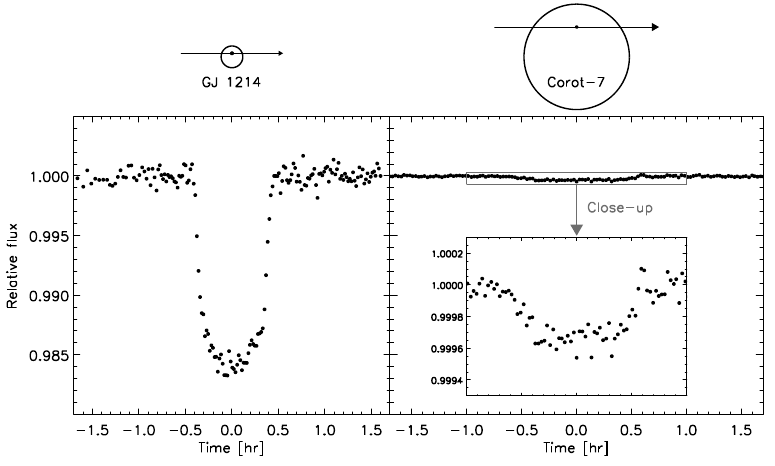}
  \caption{Example of curves for transits -- Two super-Earths of comparable sizes for stars of different spectral type. -- \emph{Left.} Spectral type of the star is M4.5V -- \emph{Right.} G9V star. Reference: \cite{Winn2010}.}
  \label{fig:exo-transit-curves}
 \end{center}
 \vspace{-1.5em}
\end{figure*}

The symbol $\delta$ in Fig. \ref{fig:exo-transit_time} represents the depth of the transit. The dimming of the light induced by the transit is proportional to the ratio of the radii $k$. The incoming flux from the system $F(t)$ is the addition of the flux of the star $F_\star(t)$ and the flux of the planet $F_p(t)$, both variable with the time. The flux of the planet depends upon the phase of the day visible to the observer. The transit will dim the total flux. Occultations (when the planet passes behind the star) dim also the flux if the observer sees the daylight part of the planet. The total flux is therefore the sum of flux of the two objects minus a correction factor due to transits and occultations:
\begin{equation}
 F(t) = F_\star(t) + F_p(t) - \begin{cases} k^2\alpha_t(t)F_\star, & \mbox{transit} \\ 0, & \mbox{no eclipse} \\ \alpha_o(t)F_p(t), & \text{occultation} \end{cases}
\end{equation}
where the factors $\alpha(t)$ are dimensionless functions of the order of one depending on the overlap area. If $F_\star$ is taken constant, then the relative flux $f(t) = F(t)/F_\star$ can be defined and therefore, the above equation take the form:
\begin{equation}
 f(t) = 1 + k^2\frac{I_p(t)}{I_\star} - \begin{cases} k^2\alpha_t(t), & \mbox{transit} \\ 0, & \mbox{no eclipse} \\ k^2\frac{I_p(t)}{I_\star}\alpha_o(t), & \text{occultation} \end{cases}
\end{equation}
where $I_\star,I_p(t)$ are the disk averaged intensities which implies that $F_p/F_\star = k^2 I_p/I_\star$. As an approximation, $f(t)$ is specified by the depth $\delta$, duration $T$ ingress or egress duration $\tau$ and a time of conjunction $t_c$.  During the transit $I_p(t)$ can be considered constant and therefore the depth of transit is roughly:
\begin{equation}
 \delta_t \equiv {f_{\text{no eclipse}}-f_{\text{transit}}}\approx k^2
\end{equation}
The approximation can be made if the visible part of the planet is in the night during the transit which, for most of the case, is geometrically favoured. For the occultation, the depth of transits solely depends on the planet averaged-disk intensity:
\begin{equation}
 \delta \approx k^2\frac{I_p}{I_\star}
\end{equation}

In the two last approximations, the variations of the ingress and egress fluxes are assumed to be linear (trapezoidal approximation). Due to limb darkening of the star, the value of $I_\star$ actually depends on the radius and therefore this trapezoidal assumption is not correction. The flux in time follows a curved line. Examples of two light curves obtained from the data is shown in figure \ref{fig:exo-transit-curves}.

Form those curves, the radii ratio $k\approx \sqrt{\delta}$ can be recovered. Therefore, the absolute planetary radius is known only if the radius of the star is known as well. Moreover, the eclipse duration can be measured in terms of $T_\text{full},T_\text{tot}$ and they can be used to measure the impact parameter $b$ and the ratio $R_\star/a$ called scaled planetary radius. Inverting equations \ref{eq:exo-transit-tf} and \ref{eq:exo-transit-tt}, one finds the lengthy expressions of those two parameters. Assuming that $\tau \ll T$ -- \emph{i.e.} small planets on non-grazing trajectories -- they can be expressed by:
\begin{eqnarray}
 b^2 &=& 1 - \sqrt{\delta} \frac{T}{\tau} \\
 \frac{R_\star}{a} &=& \frac{\pi}{\delta^{1/4}}\frac{\sqrt{T\tau}}{P}\frac{1+e\sin\omega_\star}{1-e^2}
\end{eqnarray}
Using equation \ref{eq:exo-transit-b}, the inclination can be deduced from the impact parameter. The parameter $R_\star/a$ can yield -- interestingly -- the mean density $\rho_\star,\ \rho_p$ thanks to the Kepler's famous third law:
\begin{equation}
 \rho_\star + k^3\rho_p=\frac{3\pi}{GP^2}\left( \frac{a}{R_\star} \right)^3 \approx \rho_\star
\end{equation}
The approximation is justified for $k \ll 1$.

Thus the observables of a transit yield after processing: the stellar mean density $\rho_\star$, the radius of the planet $R_p$. 
Another property of the planet can be described: the effective temperature, but with large error bars. Using the stellar radius and temperature and the planetary radius, plus assuming that the Bond albedo\footnote{For planets of the Solar System the mean albedo is 0.3 -- except for Venus.} is 0.3, one can estimate the effective temperature, \emph{i.e.} the radiative equilibrium temperature. This is highly planet- and even atmospheric-depend, hence results must be interpreted as order of magnitude estimates. To summarise, no orbital elements can be measured with this technique, it yields stellar mean density $\rho_\star$ and the radius of the planet $R_p$. Other parameters can be very roughly estimated.

The signal-to-noise ratio of the method is described by 
\begin{equation}
 SNR_{tr} \sim \sqrt{\frac{N T_\text{full}}{P}}\frac{\delta}{\sigma_{ph}}
\end{equation}
where $\sigma_{ph}$ is the uncertainty on the photometric measurements. The probability of transit for hot Jupiters ($R_p \approx R_J,\ P \sim 3$ days) orbiting a Sun-like star is $P_{tr} \approx 10$\% with a transit depth of around one percent. It can be seen on the example curves in Fig. \ref{fig:exo-transit-curves} that the transit depth for Earth-like planets orbiting a similar star as ours is of the order of $0.5\%$. The probability of detecting Earth like being low, many thousands stars must be monitored at the same time to a photometric precision of a few millimagnitudes. Quiet stars are better than active stars as small flux decrease are easier to monitor. It should be borne in mind that the main assumptions for transits detection are circular orbits and $m_p\ll m_\star$.

Ground-based instruments will operate close to $\delta / \sigma_{ph} \sim 1$ as pristine photometry is difficult from the ground. Moreover to increase the complexity further, the data cannot be acquired during the day and therefore monitoring thousands of stars becomes challenging. 
The Kepler and CoRoT missions were using the transit technique from space which is much better from a signal-to-noise ratio perspective. However, many false-positives are found in the fields due to eclipsing binaries. Follow-up missions are therefore required to better characterise and distinguish interesting targets. To confirm a transit as a planet or rather as a planet candidate, three transits have to be observed in order to show the repeatability (same depth) and the periodicity of the transit. The change that can be measured on Kepler is about 100 ppm of the photometric measurement.

A follow-up mission using the transit method, such as CHEOPS, is designed to look for transiting planets for known planetary systems. The probability that this system possesses an object that eclipse the star is fairly low if detected by RV ($\gtrsim 10-20\%$ \cite{Stevens2013}). Ephemerides of the transit can be derived from the RV-curves (see \S\ref{sec:dis-transit-detection}) and thus not the whole orbital transit must be monitored.

\subsection{Other Techniques} \label{sec:theo-TTV}
There exist several other techniques that have succeeded in discovering planets around stars \citep{Wright2013}. They some are derived from techniques used to study stellar population or from observational cosmology.
\paragraph{Microlensing.}
The General Relativity theory of Einstein predicts that a mass bends the trajectory of nearby photons. This phenomenon, called gravitational lensing, can cause the background source to change its shape, luminosity and can appear several times if the foreground mass is heavy enough. In the case of the background image and the lens being stars, the gravitational lensing is minute. Therefore only one image of the source can be seen and the only modification on the source is the luminosity as other images that could be created by the lensing effect are unresolved. 
As both the foreground and the background objects move, the microlensing is ephemeral. The photometric variability depends therefore upon the time. If the star that serve as the lens has a planetary companion, the image of the background object is perturbed further. This technique has the advantage that it can detect planet orbiting at large distances from the star or free floating planets wandering the Milky Way by a short microlensing event \citep{Sumi2011}.

\begin{figure}
 \begin{center}
  \includegraphics[width=1\linewidth]{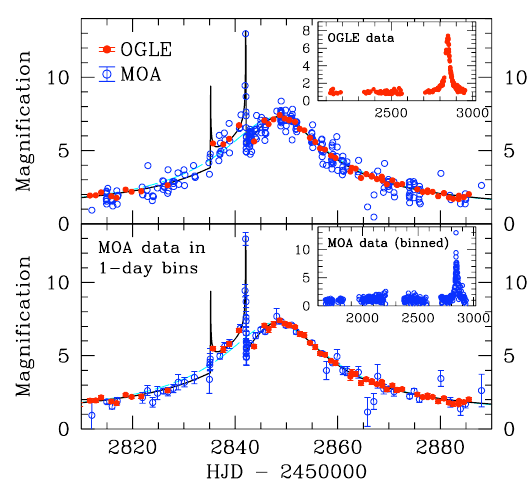}
  \caption{Gravitational microlensing typical curve -- First planet discovered by microlensing -- The OGLE-2003-BLG-235/MOA-2003-BLG-53 light curve with Optical Gravitational Lensing Experiment (OGLE) data in red and Microlensing Observations in Astrophysics (MOA) data in blue. -- \emph{Top.} -- Complete data with error bars indicated in the legend. \emph{Bottom.} -- Same data, but binned by one day. Reproduction of \cite{Bennett2008}.}
   \label{fig:exo-microlensing-HJD}
 \end{center}
 \vspace{-1.5em}
\end{figure}

The effect of a planet is to break the symmetry of the magnification curve that results from the effect of the gravitational lensing. The planet appears as a second peak on the signal. As shown in Fig. \ref{fig:exo-microlensing-HJD}, the light curves from the first planet discovered by the gravitational microlensing technique do not follow an easy analytical expression. 

The detection proportion of microlensing events is difficult to predict as it depends mostly on the angular distance lens--source. The mass ratio does play a role and a crude estimation of the probability of microlensing events is given by 
\begin{equation}
 P_{\mu l} \sim 20\%\left(\frac{q}{10^{-3}}\right)^{\sim5/8}
\end{equation}
which is determined by averaging over an uniform distribution of impact parameters.
The surveys will often give planets in the range of semi-major axis 1 to 5 AU. A Jupiter like planet has a probability of detection of roughly 30\% in the lensing zone whereas and Earth-like planet is only 1\%. A minimum mass of the Moon can even be uncovered with this technique! The difficulty is not to monitor main-sequence stars but rather to photometric-ly distinguish between background stars and the source which usually requires space-borne missions. The rate of 
occurrence of such 
events is low and therefore necessitate a huge number of stars in the field which means a large portion of the sky. The variations occur roughly on a basis of 25 days for large planets and thus daily observations are needed from the ground. For smaller planets, networks of telescope must be used to observe the sub-daily variations.

\paragraph{Astrometry.}
Astrometry uses the slight variation of the position of a given object to determine its orbital parameters. The way astrometry is used for binary stars is to measure their angular separation and their position. The problem becomes tougher for planet detection. The star is indeed visible, but it moves around the centre of mass induced by an invisible companion. Therefore, the background stars must be used to detect the variations. Those tiny perturbations must be measured on the much larger motion due to the parallax of the Earth and proper motion of the system. Astrometry is the only technique capable of yielding the true inclination and orientation of the orbit. To obtain good results with this technique, astrometric precisions of the order of the miliarcsecond (mas) for Jovian planets or even the microarcsecond (\textmu as) for Earth-like planet must be reached. This method is of course more efficient for close stars which have proper motions 
of the order of 
a thousand mas per year with a parallax motion of the order of hundreds of mas. Therefore exquisite instruments and calibrations must be used to detect the planetary signals that is several orders of magnitude smaller than the total astrometric signal. 
For comparison purposes, the Gaia mission, which aims at measuring the position of about 1 billions objects in the Milky Way and the Local Group, will have a precision of 20 \textmu as \citep{Lindegren2008}.

\paragraph{Imaging.} Direct imaging of planets is the most intuitive method -- simply put, it is to take a picture. In this frame, some photons originating from the exoplanet will hit the detector and, provided that they can be resolved from the star and that the signal of this planet is sufficiently higher than the noise, \emph{voil\`a} ! In practice, the parent star is usually much brighter than the exoplanet (a million to a billion time brighter respectively for the visible and the IR -- \citep{Kalas2008}). 
From enough measurements at different epochs, orbital parameters can be deduced. The albedo of the planet can be computed assuming the spectral type of the star is known. This means that the effective temperature can be extrapolated. From the flux of photons coming from the planet, a radius can also be estimated. The most interesting feature that can be measured is the spectrum of the planet yielding informations about the components of the atmosphere!
Distinguishing between photons from the star or from the planet is not a trivial task and depends upon the quality of the instrument. For a planet similar to Jupiter (\emph{i.e.} gas giant on a circular orbit at $r_\perp=5.2$ AU from its star) in a system at $d=20$ pc viewed face-on, the angular separation of the system is given by $\Delta \theta = r_\perp / d=250$ mas. If the planet were at $a=1$ AU, then the angular separation would drop to 50 mas only. The diffraction limit of a telescope is about $\theta_\text{diff} = \lambda / D$ with $\lambda$ being the wavelength of the observation and $D$ the aperture of the telescope. For a 8m telescope at 2 \textmu m, it corresponds to 60 mas. Therefore, this technique works only for relatively close stars and/or planets that are far away from their host star. Detectability of an object depends on many variables that are both of astrophysical ($a,i,R_p$,\dots) and engineering origin $\theta_\text{diff}$, the optics, the state of Earth atmosphere.

\paragraph{Timing.} There are several objects in the Universe that exhibit periodic behaviours by releasing energy. If the release of energy is directional and the object is rotating,  then on Earth, pulses are detected. Such objects are pulsars or pulsating white dwarfs. In this case, the timing methods is very similar to the radial velocities technique as it implies measuring Doppler shifts and gravitational perturbations on a signal with the notable difference that photons are not the object of the study. 

This photometric variability can also be seen in a different phenomenon: eclipse. Additional bodies in the system that were not detected perturb the ephemerides which can be non-negligible especially if there are resonances in the system. This technique is named Transits Timing Variations (TTV) when it is specifically used in the framework of transit measurements and is particularly useful for the Kepler mission \citep{Steffen2013}. Given enough time, timing methods can detect planets of masses smaller than the Earth.

\subsection{The Zoo of Discovered Exoplanets}
\begin{figure}[h]
 \begin{center}
  \includegraphics[width=1\linewidth]{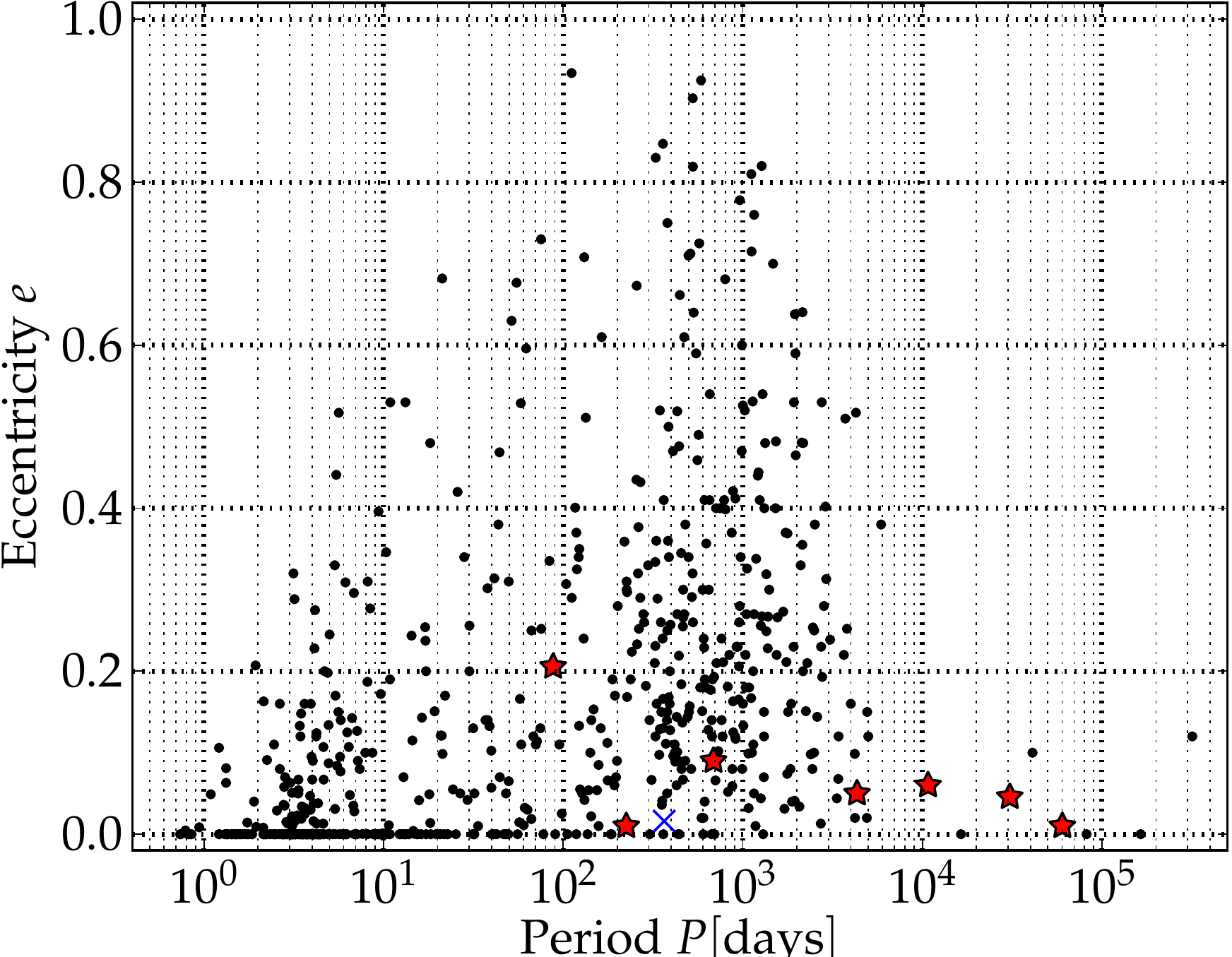}
  \caption{Diagram Period--eccentricity of a sample of 650 exoplanets. The Earth is shown by a blue cross. The red stars are the other planets of our Solar System. Source of the data: \httplink{http://www.exoplanets.org}. Note that this is a sub-sample containing exoplanet having both $e$ and $P$ defined.}
  \label{fig:exo-zoo-P-e}
 \end{center}
 \vspace{-1.5em}
\end{figure}

The Solar system upon which was based planet formation theories up to the discoveries of the first exoplanets -- and which is still used today to confront models -- is different from other known systems \citep[for example]{Baruteau2013}. This fact was realised early on and strengthened by the discoveries of many different exoplanets. The detection of hot Jupiters -- Jupiter-like planets in a very short (a few days) period orbit -- showed the importance of the migration mechanisms briefly described above. The shortest period planet detected is Kepler 42 c whose orbital period is slightly less than 11 hours \citep{Muirhead2012}. The difference to the longest period (Fomalhaut b, \cite{Kalas2008}) is 5 orders of magnitude. 
Another interesting characteristic is the distribution of the eccentricities. Indeed, in our Solar system most of the eccentricities are close to zero for planets ($e=0.21$ for Mercury) whereas the distribution is much broader in the discovered planets and in particular for giant planets. The plot in Fig. \ref{fig:exo-zoo-P-e} shows that the Solar System is not the norm and that eccentricity tends to increase for planets orbiting far away from their star. 

\begin{figure}[h]
 \begin{center}
  \includegraphics[width=1\linewidth]{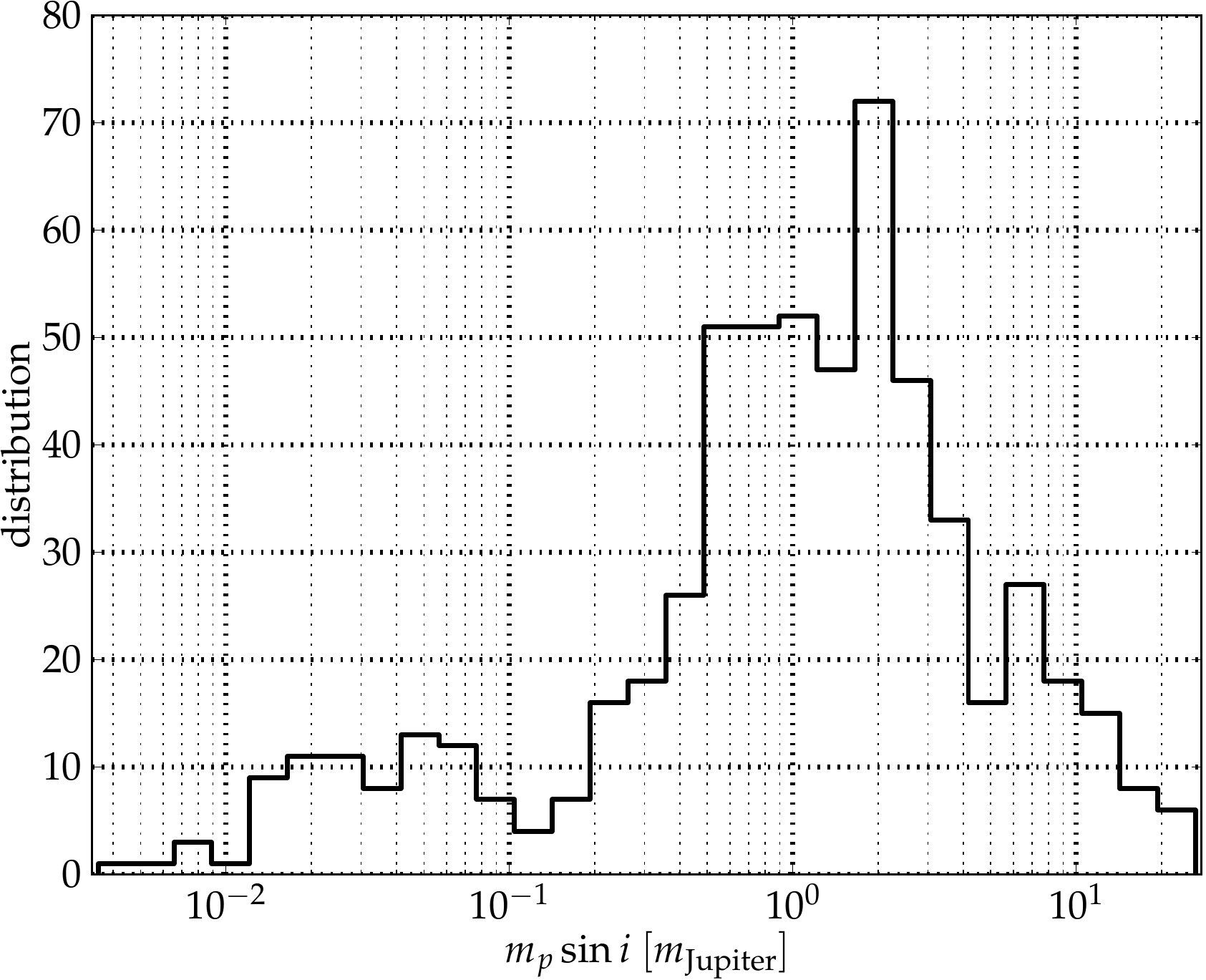}
  \caption{Distribution of mass for a sample of 590 exoplanets with 30 logarithmic bins. There is a significant peak around $1\cdot m_\text{Jupiter}$ due to detection bias. Source of the data: \httplink{http://www.exoplanets.org}.}
  \label{fig:exo-zoo-distrib}
 \end{center}
 \vspace{-1.5em}
\end{figure}

The discoveries -- or at least the confirmation -- of those exoplanets is largely due to the radial velocities techniques \citep{Mayor2012} which can determine the planet mass up to a degeneracy of $\sin i$. Fortunately, this barely impacts the statistics about planet masses with the assumption of randomly oriented orbits. In figure \ref{fig:exo-zoo-distrib}, the distribution of 590 planets is reported. A noticeable artefact in this distribution is the peak around $1\cdot m_\text{Jupiter}$. As already mentioned, most of the extra-solar planets are discovered thanks to radial velocities surveys. RV measurements are biased towards massive planets relatively close and towards massive planets as they will impact their host star. Giant planets are much easier to detect than Earth-like planets. 
However, it has been shown that those giant planets were common. This peak may shift in the future as most of the effort to find exoplanets are devoted to find Neptune masses (roughly $20m_\oplus$) down to Earth or sub-Earth masses. The lightest planet to date was discovered by \cite{Barclay2013} 
which is an object about the same size as Mercury which exemplifying the exciting search for small planets is well on its way.

The Kepler mission which was monitoring over 165'000 stars has increased the number of exoplanet discovered in the recent past by a factor of five. A number of studies (\cite{Dressing2013} and reference therein) show that planet occurrence for solar-like stars indicate that small planet radii and longer period are very common. \cite{Dressing2013} found that the occurrence of planets with radius of $0.5-4R_\oplus$ with orbital period shorter than 50 days is at least 0.5 planet per star. Giant gaseous planets ($\gtrsim 50m_\oplus$) are also common among solar-type stars \citep{Mayor2011} with an estimate of 14\%. The occurrence of planets is not a monotonous function: 
there is a gap in the range of $15-30m_\oplus$ both in observational data \citep{Mayor2011} and predictions \citep{Mordasini2009}. This fact is due to most of the planets reaching those masses would do so while the runaway gas accretion phase takes place. One hypothesis is that a migration of a gaseous giant too close to its host atmosphere would simply evaporate the atmosphere leaving only the core \citep{Udry2007}. Therefore, it is difficult to form planets that would remain in this mass range. 

Estimates vary throughout the literature and can reveal optimistic, but this demonstrates that planets around stars are common. It shows also that formation and evolution mechanisms produce a variety of different system. For example, the Kepler-11 system interestingly revealed 6 transiting planets all in the mass range of 2 to 25 Earth masses \citep{Lissauer2011}, meaning 6 planets in the same plane. Although multiple systems are unlikely to be discovered due to detection bias, there are hints such systems tend to form around solar-like stars \citep{Roell2012}.

\section{CHEOPS} \label{sec:CHEOPS}
In this section, the CHEOPS mission and its science will be described. Much information on this mission was retrieved during the different meetings -- both from the engineering and from the scientific standpoint -- that took place in the duration of this Master project. The CHEOPS mission is still in an early phase which will imply that there will be some difference to the final design. 

CHEOPS is not a normal ESA mission. Indeed, it is the first of the new small class mission meaning small budget, small size, light satellite, but still able to carry out scientific studies of great importance. The total cost of the mission should be lower than 100 M\officialeuro. The proposal of the mission was selected in fall 2012. The mission adoption will be finalised in February 2014 and the launch will take place in the late 2017. The consortium lead by the University of Bern regroups 10 nations amongst which Switzerland plays a key role.
\subsection{Science with CHEOPS} \label{sec:theo-science-cheops}
The primary objective of the CHEOPS mission is to perform high quality photometric measurements of known exoplanets. 
Those high quality measurements can contribute to different characterisation of the transiting planet. 

The science objectives of the CHEOPS mission are, as described in the \cite{CHEOPSConsortium2012}:
\begin{itemize}
\item To constrain the $m-R$ relationship for planet lighter than Saturn;
\item To identify planets with significant atmospheres;
\item To place constraints on planet migration;
\item To study energy transport in the atmosphere of hot Jupiters;
\item To provide ``golden targets'' for future mission such as the James Webb Space Telescope or the Extremely Large Telescope;
\item To provide 20\% of open time to the community.
\end{itemize}

CHEOPS will study small planets -- from ``Super-Earth'' (with masses $m\lesssim2-10m_\oplus$) to Neptune-sized bodies -- around bright stars which are well characterised. Knowing for example the activity and the spectra of a star enables to observe when the star is quiet enough to get less noise in the photometric signal and to estimate the temperature of the planet. Utilising the planet mass determined thanks to RV techniques, CHEOPS will add points in the $m-R$ diagram. Most of the known ``Super-Earths'' are thought to be rocky, but some show densities similar to Saturn's or remnants of evaporated giant planets (e.g., Kepler 10b, \cite{Wagner2012}). Neptune-sized planets will be used to place constraints on planet migration by studying their composition and thus their formation process while hot Jupiters will be used to explore atmospheres of giant planets and energy exchange. 

Another important objective of the mission is to prepare a target list for the next extremely large telescopes both on the ground and in space. The James Webb space telescope will study amongst others the birth of stars and planets as well as planetary systems \citep{Gardner2006}. 

The most obvious parameter to be measure with CHEOPS is the radius of planet from which densities can be deduced. Having the mass and the radius of a planet enable to compute the mean density and therefore to extrapolate its composition. The composition is heavily degenerated. Indeed, large error bars as well as changing ratios of elements prevents from precisely determining the composition. For example, a $5m_\oplus$ Earth-like planet composed of 50\% of rocks and 50\% of water vapour has a radius roughly twice as large a purely rocky composition \citep{Valencia2010}. 
Even if this example will be resolved by CHEOPS, there will still be a degeneracy in the composition of the planet \citep{Sotin2007337}. 
Probing the inner structures of planets is a difficult task as the structure is highly degenerate. Atmospheres compositions are slightly less difficult \citep{2011ApJ...733...65B}. 
Measurements of the flux at different time for a close in planet allow to calculate its albedo (thanks to its occultation) and even to generate a brightness map of the atmosphere (reflectivity of the high altitude clouds thanks to the phase curve). Measurements in the optical band constraint heavily the models and allow to lift degeneracy \citep{Sing2011}.

\subsection{The Instrument}
\begin{figure}[h]
 \begin{center}
  \includegraphics[width=0.9\linewidth]{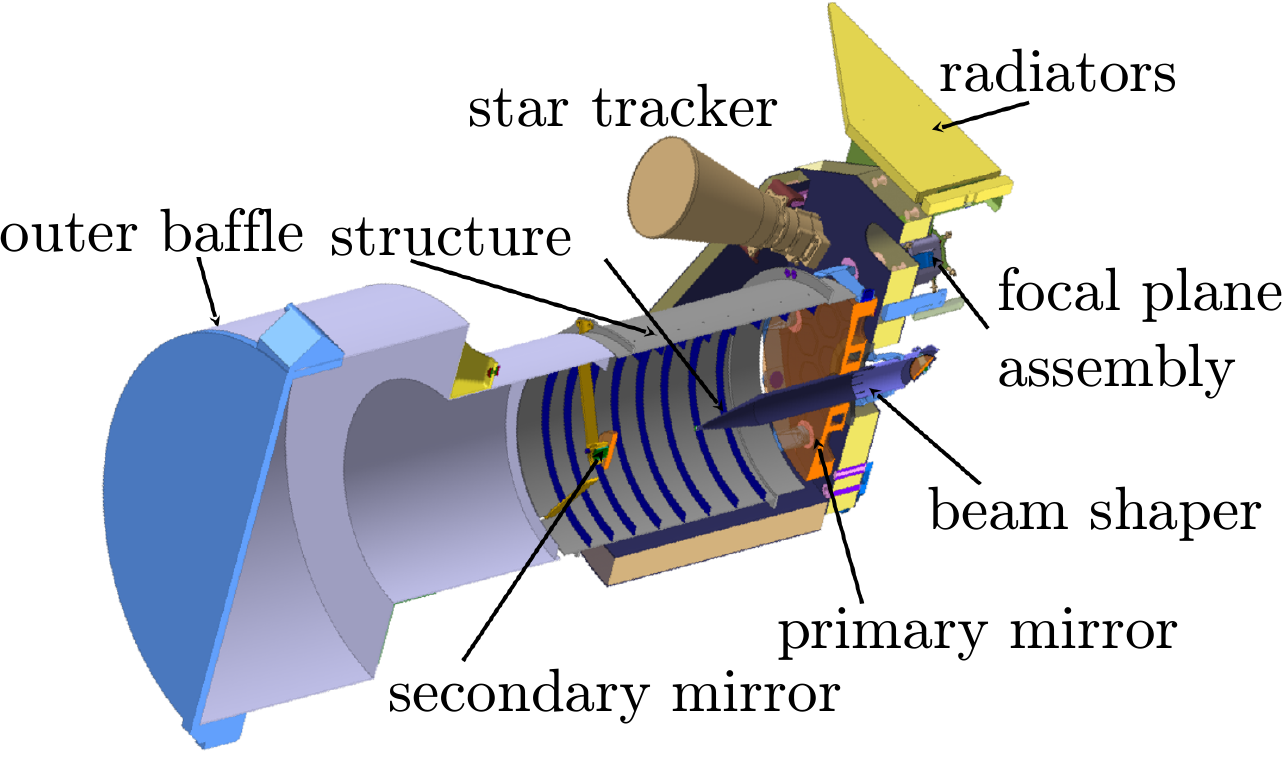}
  \caption{Cutaway view of a computer model of the instrument. The cover which is closes in this view can only be opened once and does not act as a shutter. Its goal is to protect the optical assembly during handling on the ground and during the launch. The annulus in the baffles are optical barriers for the stray light.\label{theo:instrument}}
 \end{center}
\vspace{-2.em}
\end{figure}

\paragraph{Optical Assembly.} The satellite is composed of a platform that provides everything needed (see \S\ref{sec:mission-implementation}) and one instrument: its telescope. This 30 centimetres telescope is a Ritchey-Chretien with a field of view of 0.4 degrees enclosed inside a baffle to protect its focal path from contamination (see \S\ref{sec:theory-sl} and Fig. \ref{theo:instrument} \& \ref{fig:theo-sl-analysis}). This 50--60 kg payload is equipped with a CCD composed of 1k by 1k pixels. With a size of 13 \textmu m, the angle seen by one pixel is 1 arcsec. This CCD is sensitive to wavelengths in the range 400 to 1100 nm centred on 550 nm \citep{aimo2006}. 
The useful photometric area of the image is not the whole frame, but a circle window of radius of about 100 px. Indeed, the observation technique for CHEOPS is to perform photometry on a smaller region -- the \emph{region of interest} -- than the whole field of view. The stabilisation of the satellite is such that one of the spacecraft axes always points towards the Earth -- the attitude is ``nadir locked''. Thus the image rotates about the line of sight.

\paragraph{Point Spread Function (PSF).}  \label{sec:Theory-PSF}
The \cite{SciReq} defines the PSF of the system as a flat PSF without high frequency feature (SciReq. 5.1). This $\sim700-800$ pixels wide will provide a defocused image. 
This PSF area originates from a tradeoff between the spacecraft pointing high frequency inaccuracies (jitter) and the stray light contamination. The former requires a large surface to minimise the error contribution while the latter is a diffuse source of noise and therefore is minimised when the area of the PSF is minimised. With a pixel scale of about 1 arcsec per pixel, the PSF has a radius in the range of 8'' to 9''.
\subsection{Mission Implementation} \label{sec:mission-implementation}
The satellite, which mass is of the order of 200 kg, will have a nominal lifetime of 3.5 years. The mission plans to measure radius of transiting planets to a 10\% accuracy. The magnitude limit is to target stars brighter than 12.5 in the V band. Stars brighter than magnitude 6 cannot be often observed under penalty of CCD saturation, even if the goal is to reach $V=0$. 

The choice of the orbit is driven by the fact that CHEOPS should be able to observe most of the sky while achieving thermal and photometric stability. To meet those requirements, a Sun synchronous orbit in low Earth orbit (LEO) was chosen. This nearly polar orbit is peculiar in the sense that the satellite crosses the equator each day at the same (solar) local time. 
As the Earth is not perfectly spherical, its oblateness causes the line of nodes\footnote{the line of nodes is the line joining the plane of the orbit to the plane of reference (the ecliptic).} of the orbit to precess around the rotation axis. Placing the satellite at the right inclination for a given altitude compensates exactly the change of relative position of the Sun in the sky; in turns, the Sun remains at a constant angle with respect to the plane of the orbit. Another added advantage is that eclipses of the Sun by the Earth are uncommon in comparison to a low inclination LEO orbit.
The local time of ascending node (LTAN) of 6 am preferred implies that the viewing zone will be in the South during the maximal observation time. As CHEOPS is likely to be a passenger on a dual launch, a LTAN of 6 pm is also considered. The altitude of the spacecraft is not completely defined yet, but will be in the range of 620 and 800 km translating into orbital periods of 97 to 101 minutes. As it is studied in this project, the influence of the altitude is great over several important quantities such as the stray light, but also the amount of radiation received.

As stated earlier, the satellite will be three-axis stabilised (as for most of space-borne observatories), but nadir locked. The performance of the attitude determination and control subsystem is increased by adding the instrument in the control loop. Two star trackers acquire the attitude ten times a second while reaction wheels alter the orientation of the spacecraft. The advantage of reaction wheels over thrusters are two-fold: they do not need consumables and they do not threat to contaminate the instrument. The spacecraft will provide about 50W of continuous power to the instrument and be able to downlink at least 1 Gbit (using S-band) of data per day down to ground. Communications to and from the satellite will use the S-band which has a low data rate, but is easy to use.

\subsection{From the Scientific Objectives to Accuracy and Constraints} The requirements are defined in the \cite{SciReq}. The purpose of this section is to give a motivation for those concerning closely this project. Most of the requirements for this mission were inherited from the CoRoT mission. To get sufficiently high signal-to-noise ratio (SNR) for the transit of an Earth equivalent before a solar-like star dictates that the noise over 6h (the duration of the transit plus margin) to be of the order of 20 ppm assuming a 50 day orbital period for the planet. Similarly, for Neptunes orbiting in 13 days around K dwarfs, the noise should be no greater than 85 ppm. The pointing accuracy yields a rms jitter of $\sim 8$ arcsec RMS over 10 hours. 
This very low noise imposes to have a very well calibrated flat field, an optimised read-out (a tradeoff between read-out noise and dark current noise). CCD operate better -- meaning less noise -- at low temperature. The CDD is therefore foreseen to be cooled down to about 233 K ($-40^\circ$C)\footnote{Spatial environment does not imply cold temperatures. Moreover, with direct Sun light or the heat generated by the electronics, the temperature inside the spacecraft can easily reach room temperature.}, stabilised to an accuracy of 10 mK.  

In this 20 ppm noise budget, there are several incompressible items such as the read-out noise, the jitter or quantum efficiency changes. From the allocations of the other noises, the stray light noise was set to be maximal 1 photon per second per pixel. 
In the discussion (\S\ref{ch:results}), it is shown that this requirement is easy to reach with the current performance. A minimal angle from the line of sight of the telescope to the limb of the Earth (\S\ref{sec:numerics-visibility-constraints}) can also be defined. One of the goals of this work is to see whether it can be lowered. This angle also defines the baffle of the satellite and the rejection factor for the stray light (see next section). 

\section{Stray Light Contamination} \label{sec:theory-sl}
Unfortunately for astronomers, there are sources of light in our local neighbourhood: The Sun of course and other objects that reflect its radiation. For a space-borne observatory in LEO, another great source of light pollution is this radiation emitted by the Sun which is reflected by the surface (and by the atmosphere) of the Earth. The wavelengths of interest here are in the range of 400 to 1100 nm which implies that the contaminating radiation is reflected by the Earth at the surface before reaching the telescope. This contamination is called Earth stray light. 
Its effects are the focus of this study. The amount of stay light received at a given point in LEO depends of course if the satellite is mostly over a region in the daylight or in the night and also on the altitude of the satellite. How much of this unavoidable flux actually reaches the detector is up to a skilled optics manufacturer to decide. 
At the end of the day, the noise is like any source of noise: it degrades the signal to noise ratio. This section gives an introduction to stray light analysis and mitigation techniques. 
\subsection{The Problem and its Mitigations}
A good practice in telescope design is the early study of stray light as it is a telescope-wide design and fabrication issue \citep{Pompea1995}. Late changes caused by an overlook of this issue can result in immense difficulties, in delays and in the explosion of cost if not in the scraping of the program altogether. Stray light analysis must be systematic and take into account (1) the optical design, (2) the mechanical design, (3) the thermal model, (4) the scattering and reflectance characteristics of every surfaces involved. Any photon originating outside the field of view of the telescope which does not use the optical path is considered as stray light. Thus, this noise is generated by the actual optics. Another type of stray light not considered in this study is noise generated by thermally emissive objects close to the optical system. These photons can reach the detector through diffraction and scattering due to micro-roughness and dust on all surfaces inside the telescope.

All the surfaces are not analysed to the same depth. Indeed, baffle surfaces that can be seen from the focal plane are consider critical. The existence of direct paths depends upon the nature of the telescope and must be minimised by blocking the path, shifting mechanical parts or changing if possible the surface. A careful study of a perfect system may not be sufficient as in every instrument there exist misalignments. Mechanical stresses applied to instruments in space are more dire during the launch than any moment of the life of a ground-base instrument. Thus misalignments and misplaced surfaces arise during production and subsequently during handling or launch.

To reduce the flux of stray light on the detector, several techniques are used. They fall into mechanical solutions or material science. Of course, the optimisation of the system does not mean that the image is free of unwanted radiation. In most of the programs, there is not enough time to carry out a thorough design process as the focus is set to reaching the best performances -- or at least the \emph{required} performances -- with as less resources (\emph{e.g.} mass, volume, complexity, cost, \dots) as possible as soon as possible. 

A very visible feature of stray light mitigation is the presence of the baffle -- the tube around the telescope. The installation of vanes in the baffle is another measure. Those annulus are placed in the baffle with a certain height to block the path of the light coming from given off-axis angles. The types and shapes of vane can be various, even if conservative designs offer usually good results for a relative low cost. The use of the so-called aperture, field and Lyot stops are also common and consist in placing disk with a small central aperture. Their names depend upon their location with respect with the optical path. Aperture stops reduce the size of the bundle of radiation capable of reaching the focal plane. Field stop is an aperture at the intermediate images to limit the field of view of the optics to the one of the detector. Lyot stops prevent the detector from directly seeing the baffle.

Preventing stray light means also to treat black surfaces as important optical elements. In optimised systems it can make tremendous differences. Black surfaces are surfaces of low reflectivity and that are ``black'' for a given wavelength. They are used to attenuate transmission along existing paths. Relevant surfaces are carefully studied, coated and calibrated. The selection of these materials can be painful and tedious especially in space where outgassing and shaking at launch play a important role.

The use of a baffle, vanes and stops are the most efficient techniques to prevent stray light contamination. Reduction of the noise due to stray light can be significant.

\subsection{Analysis \& Characterisation}
To ensure a low noise in the detector, the best way is to simulate the contamination. In a nutshell, the commercial \verb=Zeemax= code traces a very large number of rays to detect the paths to the detector. This very large number is at least of the order of a few dozen millions rays to ensure a sufficient resolution. For CHEOPS for example, $50\cdot 10^6$ rays were computed for the preliminary study. The observational constraints on many systems defined by the angle from the line of sight to bright objects (the Sun, the Earth, the Moon, Venus, \dots) are often dominated by stray light considerations.

The most prominent output of such simulations -- especially in the scope of this project -- is the Point Source Transmittance (PST). This function describes the ability of the system to reduce the flux of stray light depending of the off-axis angle $\alpha$. Hence the PST describes the rejection of stray light. The PST is defined as the irradiance at a reference plane -- usually at the detector --  divided by the input irradiance:
\begin{equation}
 PST(\alpha) = \frac{P_\text{det}(\alpha)}{P_\text{inc}(\alpha)}=\frac{E_\text{det}(\alpha)}{E_\text{inc}(\alpha)}
\end{equation}
where $P_\text{inc}$ is the total incident power from external stray light sources and $P_\text{det}$ the total stray light arriving at the detector. The PST does not contain any information about the distribution of stray light across the detector \citep{Neubert2011}. 
The PST can be described by a two argument function or as in this study by the approximation of a axisymmetric function about the line of sight. The behaviour of such a curve should be a rapid decrease in the energy reaching the detector with growing off-axis angles. If spikes are seen in the PST, the angles around which the spikes are located represent significantly worse area in term of contamination. Those must be studied in order to find their source and change the design to mitigate them.

Another output is the distribution of stray light on the focal plane for a given off-axis angle. This prediction is not part of this work as it was not available. This distribution is useful in the image reduction process as it affects background subtraction. However, this necessitate a design which is close to a final design to start discussing how stray light is distributed across the field. The emphasis must be on preventing the formation of bright spots in the focal plane rather than on a long study with many maps of the distribution of stray light at the detector \citep{Breault2010}.

The simulations are of great help in the design of a complex system. What about tests in laboratory with real equipment? While tests of prototypes have many advantages, the weaknesses of this techniques are large for the stray light characterisation. Indeed, the construction of a realistic enough prototype requires resources and time. Moreover, the design of the instrument must be almost finished which means that if a bad surprise arise during testing, the consequences can be dire. 

\subsection{CHEOPS} \label{sec:theo-sl-CHEOPS}
The first analysis of stray light as described in the previous section was performed by RUAG during the pre-study phase of the mission. The baffle has undergone since then an extensive redesign in order to reduce the dimension of the satellite such that its envelope would fit in different launchers such as Vega or Soyuz. Indeed, the distance between the mirrors was reduced by 10 cm to 40 cm and the baffle is now shorter. The management of the stray light was therefore adjusted by changing the shape and location of the vanes placed in the baffle. However, during this preliminary phase, the design of the stray light contamination was optimised and the current goal is to reproduce the results obtained by RUAG as they allow to reach the scientific objectives of the mission.

A constraint on the quality of the rejection (\emph{i.e.} the PST) was imposed to be lower than $3\cdot10^{-12}$ for angles larger than $35^\circ$. This number is also inherited from previous missions. For angles smaller than $35^\circ$, the PST can increase dramatically therefore observations for low angles could not take place. This obviously reduces the amount of visible sky. The goal is to reduce this angle to gain visibility.

\begin{figure}
 \begin{center}
  \includegraphics[width=.8\linewidth]{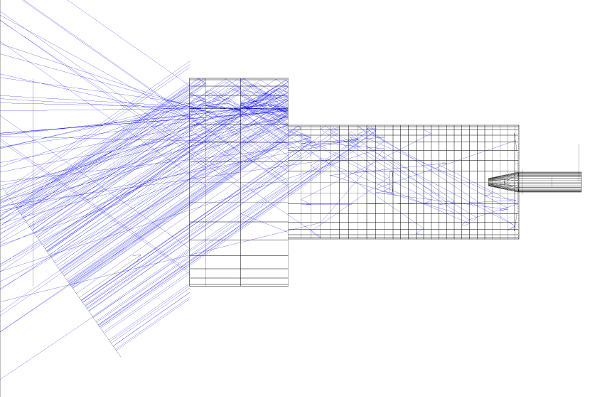}
  \caption{Schematic view of the telescope including the back-end optics in current design at the time of the study. The blue line represent a few of the 50 millions rays traced. \cite[private communication]{Munari2013}}
  \label{fig:theo-sl-analysis}
 \end{center}
 \vspace{-1.5em}
\end{figure}

The stray light analysis is now carried out by the Italian members of the CHEOPS consortium (INAF) (Fig. \ref{fig:theo-sl-analysis}). They have separated the optics into two parts: (1) the telescope and (2) the back-end optics. This has the advantage to be able to modify the design of one of the two without affecting the other analysis. The back-end optic model is currently basic and can therefore be simulated on a short time scale. Future work with a more realistic model will show -- or at least the team claims -- that the most prominent radiation is coming from the secondary mirror. 

The number of rays simulated is of the same order as for the back-end optics. The simulation of the telescope is much more complex as the geometry is more complicated. To speed up the whole process only ``important'' scattering paths are taken into account, meaning that a photon that scatters not towards the back-end optics or other high valued part of the telescope are discarded. The resulting PST at the time of writing of this thesis was in agreement with the PST from RUAG with the advantage that it extends down to $5^\circ$.

The choice of the PST for this work is the RUAG one. Indeed, the state of the optical design was far from being frozen with preliminary results that show ``reasonable'' agreement with previous work. More work was required to eliminate a spike that appears in the PST as well as to increase the sample of rays for large angles. A discussion of the effect of those differences is proposed in section \S\ref{sec:dis-PSTs}. 


\cleardoublepage
\chapter{Numerical Methods} \label{ch:numerical}
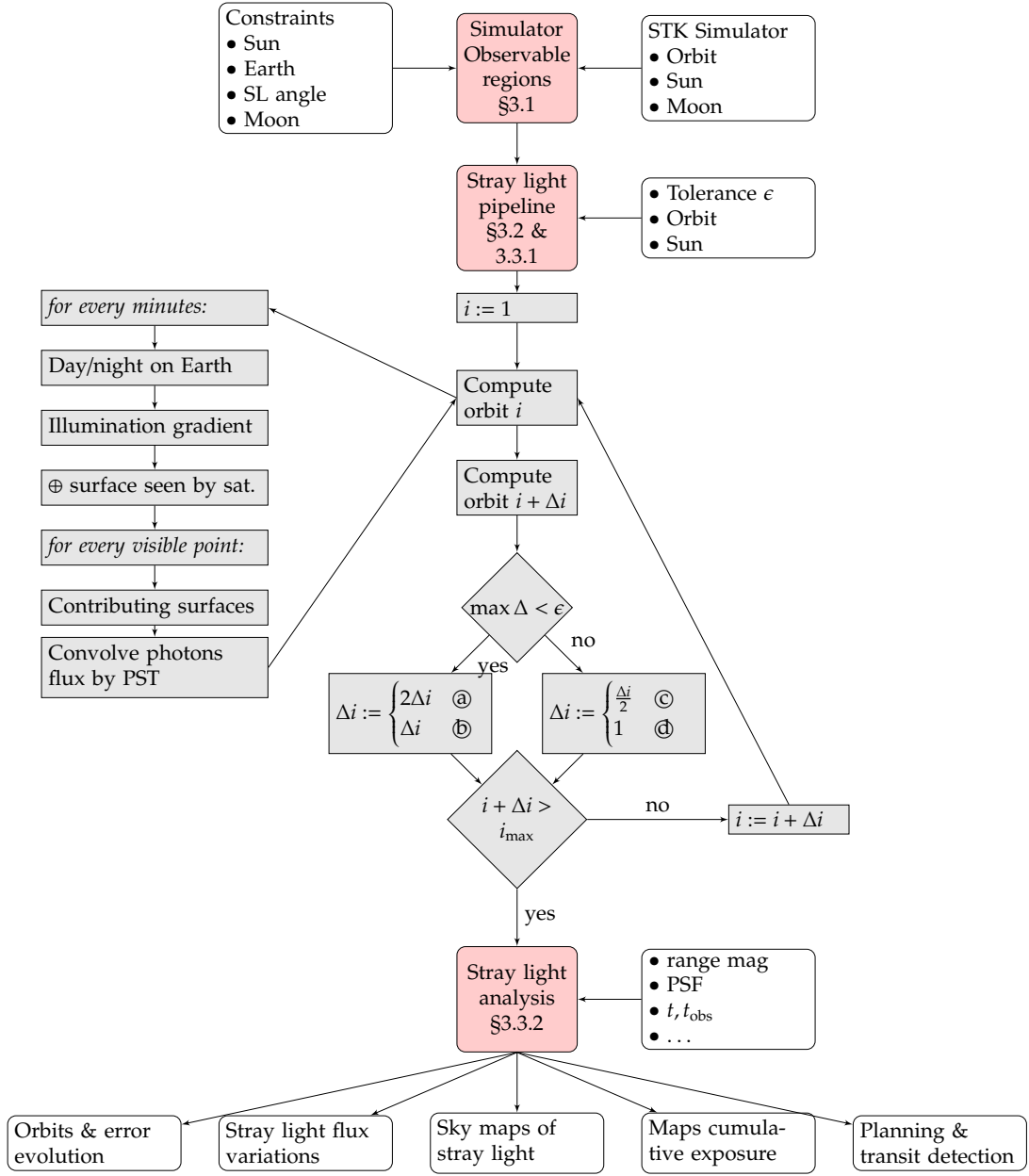
\begin{figure*}
\begin{center}
\tikzstyle{decision} = [diamond, draw, fill=gray!20, text width=4.5em, text badly centered, node distance=2.cm, inner sep=0pt]
\tikzstyle{block} = [rectangle, draw, text width=7.5em, rounded corners, minimum height=2em, node distance=3.5cm]
\tikzstyle{algo} = [rectangle, draw, fill=gray!20, text width=5em, node distance=1.5cm]
\tikzstyle{algolarge} = [rectangle, draw, fill=gray!20, text width=7em, node distance=2.5cm]
\tikzstyle{algoverylarge} = [rectangle, draw, fill=gray!20, text width=10em, node distance=2.5cm]
\tikzstyle{code} = [rectangle, draw, fill=red!20, text width=5em, text centered, rounded corners, minimum height=5em, node distance=2.5cm]
\tikzstyle{line} = [draw, -latex']
\tikzstyle{cloud} = [draw, ellipse, text width=5em]
   
\resizebox{!}{1\textwidth}{
\begin{tikzpicture}[scale=2, node distance = 2.5cm, auto]
  \node [code] (raw_maps) {Simulator\\Observable regions\\\S\ref{sec:numerics-obs}};
  \node [block, left of=raw_maps] (constraints) {Constraints\\\textbullet\ Sun\\\textbullet\ Earth\\\textbullet\ SL angle\\\textbullet\ Moon};
  \node [block, right of=raw_maps] (stk) {STK Simulator\\\textbullet\ Orbit\\\textbullet\ Sun \\ \textbullet\ Moon};
  \node [code, below of=raw_maps] (sl) {Stray light pipeline\\\S\ref{numercis:stray_light.f} \& \ref{sec:numerics-pipeline}};
  \node [block, right of=sl] (tolerance) {\textbullet\ Tolerance $\epsilon$\\\textbullet\ Orbit\\\textbullet\ Sun};
  \node [algo, below of=sl] (orbit_0) {$i:=1$};
  \node [algo, below of=orbit_0] (orbit_i) {Compute orbit $i$};
  
  \node [algoverylarge, left of=orbit_0, node distance=6cm] (for) {\emph{for every minutes:}};
  \node [algoverylarge, below of=for, node distance=1cm] (c1) {Day/night on Earth};
  \node [algoverylarge, below of=c1, node distance=1cm] (c2) {Illumination gradient};
  \node [algoverylarge, below of=c2, node distance=1cm] (c3) {$\oplus$ surface seen by sat.};
  \node [algoverylarge, below of=c3, node distance=1cm] (c4) {\emph{for every visible point:}};
  \node [algoverylarge, below of=c4, node distance=1cm] (c5) {Contributing surfaces};
  \node [algoverylarge, below of=c5, node distance=1cm] (c6) {Convolve photons flux by PST};

  \node [algo, below of=orbit_i] (orbit_next) {Compute orbit $i+\Delta i$};
  \node [decision, below of=orbit_next] (step) {$\max \Delta <\epsilon$};
  \node [algolarge, below left of=step] (go) {$\Delta i:=
  \begin{cases}2\Delta i&\textcircled{a}\\
	      \Delta i&\textcircled{b}
	      \end{cases}$};
  \node [algolarge, below right of=step] (nogo) {$\Delta i:=\begin{cases}\frac{\Delta i}{2} & \textcircled{c}\\1& \textcircled{d}\end{cases}$};
  \node [decision, below left of=nogo, node distance=2.5cm] (imax) {$i+\Delta i>i_{\max}$};
  \node [algo, right of=imax, node distance=4.5cm] (iplus) {$i:=i+\Delta i$};
  \node [code, below of=imax, node distance=3cm] (analysis) {Stray light analysis\\ \S\ref{sec:numerics-analysis}};
  \node [block, right of=analysis] (params) {\textbullet\ range mag\\\textbullet\ PSF\\\textbullet\ $t,t_\text{obs}$\\\textbullet\ \dots};
  \node [block, below of=analysis, below of=analysis, node distance=1.2cm] (A) {Sky maps of stray light};
  \node [block, left of=A] (B) {Stray light flux variations};
  \node [block, left of=B] (C) {Orbits \& error evolution};
  \node [block, right of=A] (D) {Maps cumulative exposure};
  \node [block, right of=D] (E) {Planning \& transit detection};

   \path [line] (constraints)--(raw_maps);
   \path [line] (stk)--(raw_maps);
   \path [line] (raw_maps)--(sl);
   
   \path [line] (sl) -- (orbit_0);
   \path [line] (tolerance)--(sl);
   \path [line] (orbit_0) -- (orbit_i);
   \path [line] (orbit_i) -- (orbit_next);
   
   \path [line] (orbit_i.west) -- (for.east);
   \path [line] (for) to (c1);
   \path [line] (c1) to (c2);
   \path [line] (c2) to (c3);
   \path [line] (c3) to (c4);
   \path [line] (c4) to (c5);
   \path [line] (c5) to (c6);
   \path [line] (c6.east) to (orbit_i.west);
   
   \path [line] (orbit_next) -- (step);
   \path [line] (step) -- node {yes} (go);
   \path [line] (step) -- node {no} (nogo);
   \path [line] (go) -- (imax);
   \path [line] (nogo) -- (imax);
   \path [line] (imax) -- node {no} (iplus);
   \path [line] (iplus.north) to (orbit_i.east);
   \path [line] (imax.south) to node {yes} (analysis);
    \path [line] (params) -- (analysis) ;

    \path [line] (analysis.south) to (A);
    \path [line] (analysis.south) to (B);
    \path [line] (analysis.south) to (C);
    \path [line] (analysis.south) to (D);
    \path [line] (analysis.south) to (E);
\end{tikzpicture}
}
\caption{Flow chart of the project. Red blocks represent the different code suites used in this project. Gray cells depict algorithmic instructions and white either inputs or outputs of the project. \label{fig:data-flow} STK is the commercial software used to generate the orbit of the satellite. $i$ is the orbit number, $\Delta i$ represent the orbit step size and $i_{\max}$. $\max\Delta$ is the maximal (relative) difference in the equivalent magnitude. The conditions on the orbit step size are $\textcircled{a}$ if $\Delta i$ is smaller than a maximum orbit size; $\textcircled{b}$ else; $\textcircled{c}$ if $\frac{\Delta i}{2}\geq 1$; $\textcircled{d}$ else.}
\end{center}
\end{figure*}

\lettrine[lraise=0., nindent=0em, slope=-.5em]{T}{his} chapter on numerical techniques presents the algorithms and the softwares used and prepared during this work. The presentation starts by discussing the calculation of the visible area that was coded by Luzius Kronig from EPFL/Swiss Space Center, continues by describing the stray light code of Andrea Fortier at UniBe and myself and ends by the description of the various algorithms and techniques developed to derive stray light maps and tables. A flow chart summarises the numerical codes used in Fig. \ref{fig:data-flow}.

\section{Observable Area} \label{sec:numerics-obs}
An visible region in the sky is an area that the satellite can observe at a given time. The geometry of this zone is primarily defined by the position of the satellite relative to objects that could contaminate the image either by direct imaging or by diffuse light such stray or zodical lights. Most of the constraints are therefore derived from the pointing direction -- the line of sight -- of the telescope. 
\subsection{Constraints} \label{sec:numerics-visibility-constraints}
\begin{figure}
 \begin{center}
  \includegraphics[width=0.8\linewidth]{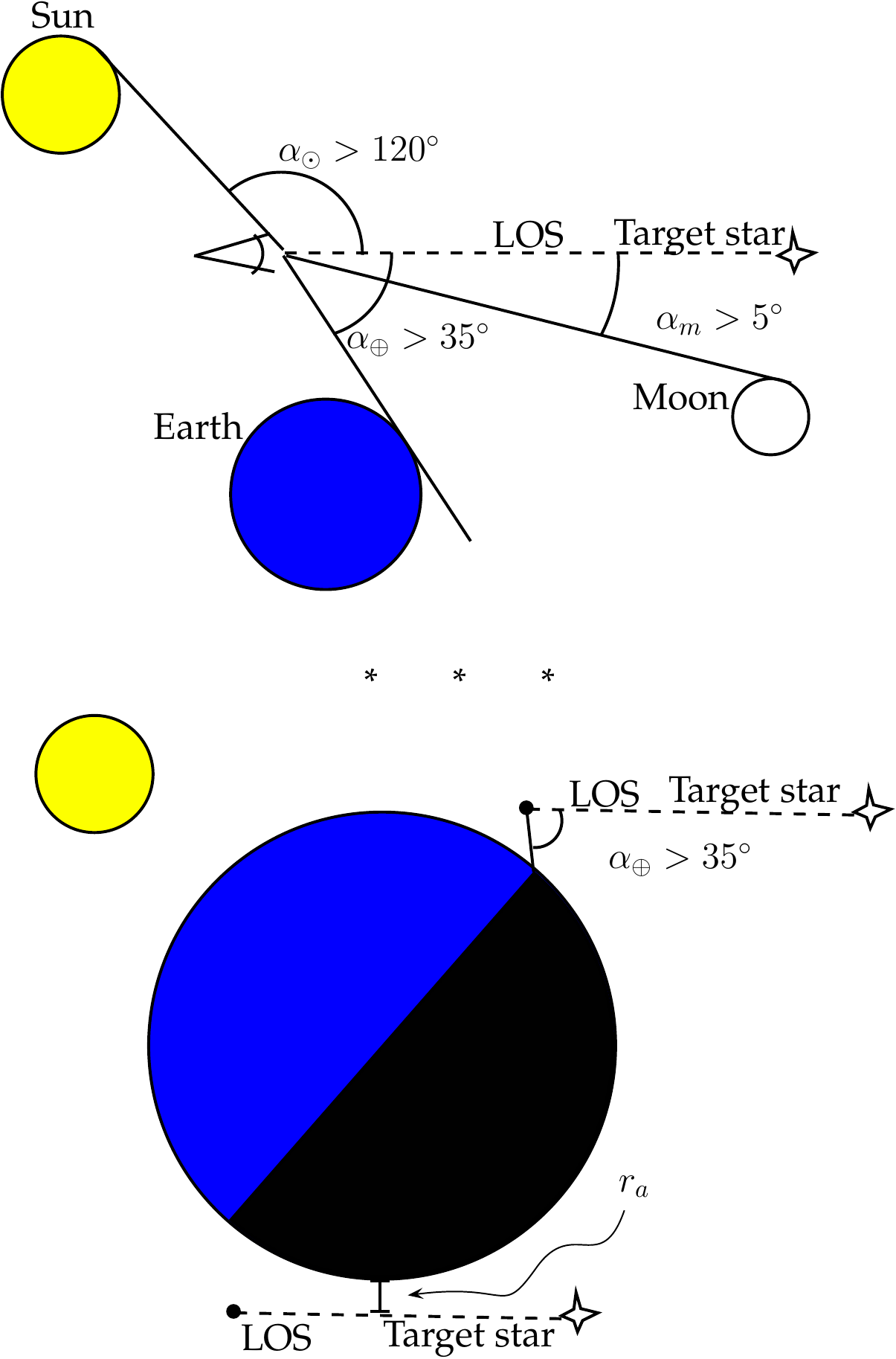}
  \caption{\emph{Top}. Exclusion angles from the line of sight (LOS) of the telescope. \emph{Bottom.} Close-up for different cases for observations above the dark side of the Earth.}
  \label{fig:exclusion-angles}
 \end{center}
\end{figure}

The first constraints on this area are that it is forbidden to look directly at or close to bright objects namely the Sun, the Earth and the Moon. Moreover, to reduce the contamination by diffuse light of the Sun or Earth and Moon stray lights, there are conditions on the minimum angles between the line of sight (LOS) of the telescope and the limb of those objects. The restrictions are as described by the \cite{SciReq}:
\begin{description}
 \item [SciReq. 3.1: Earth Occultation] The telescope shall not point to a target, which projected altitude from the surface of Earth is less than 100 km;
 \item [SciReq. 3.2: Earth stray light exclusion angle] In order to limit stray light contamination, the minimum angle allowed between the line of sight and any (visible) illuminated part of the Earth limb, the so-called Earth stray light exclusion angle shall be $\alpha_\oplus=35^\circ$ (goal: 28$^\circ$). This angle could be adapted (lowered) as a function of the target magnitude;
 \item [SciReq. 3.3: Sun exclusion half-angle] The Sun must be outside the cone around the line of sight (LOS) of the telescope having a half-angle, the so-called sun exclusion angle, of $\alpha_\odot=120^\circ$;
 \item [SciReq. 3.4: Moon exclusion half-angle] The bright Moon must not be inside a cone around the line-of-sight of the
telescope having a half-angle, the so-called moon exclusion angle, of $\alpha_m=5^\circ$.
\end{description}
The constraints shape the observable regions; figure \ref{fig:exclusion-angles} depicts those exclusion angles.
The values of the exclusion angles are widely in use for space-borne observatories with similar orbits such as CoRoT. The stray light exclusion angle depends upon the altitude of the satellite. The higher the altitude, the less important the stray light contamination. In addition to the altitude, another important parameter -- as discussed in section \ref{sec:theory-sl} -- is the PST which is further discussed in section \ref{sec:dis-PSTs}.
There exists an additional condition that arises from the thermal regulation of the satellite. Indeed, there is going to be a radiator on the satellite to dissipate the extra heat produced by the on-board electronics. The radiator must point towards the so-called ``deep space'' meaning away from the Earth, the Moon and the Sun. The satellite has a given orientation in space (the attitude of the spacecraft). This constraint is however already met by the attitude control of the satellite -- which is in a ``nadir-locked'' state -- and the exclusion regions concerning the Earth.

The South Atlantic Anomaly (SAA) plays also an important role. As the plane of the orbit is much inclined, most of the latitudes are spanned by the satellite. The trajectory will cross very often the SAA. This region above the South Atlantic ocean experiences a much higher flux of protons and electrons than above other locations on the Earth due to a perturbation of the Earth's magnetic dipole (see \S\ref{sec:dis-SAA} for further information). The applicable requirement for the SAA is the radiation flux received by the telescope which is:
\begin{description}
 \item [Radiation flux] The radiation flux for the detector to operate shall be less than 2 protons of 50 MeV per second per cm$^2$.
\end{description}
The maps of the visible regions in the sky are generated without taking the SAA into account and the restriction is applied afterwards during the analysis. Otherwise, if the SAA would already be included in the observability maps, it would be impossible to perform an adaptive orbit step as described in the next section. Indeed, the crossing of the SAA does not always occur at the same time in the orbit. Moreover, the time spent over the SAA changes as well (for 800 km, between 0 and 20 minutes).
\paragraph{}
To compute the region of the sky that can be observed by the satellite a \verb=MATLAB= code was written. This program was created before the selection of the satellite by ESA to show the feasibility of this mission. The orbits used, which are described in a previous section \ref{sec:mission-implementation} and in Table \ref{tab:numerics-table}, are generated by the specific commercial software \citeauthor{STK2013} (See references). This tool enables an user to simulate in a realistic way any -- civilian or military -- satellite mission. The position of the Sun and the Moon are predicted by the \verb=STK= software as well.

\begin{table}
 \begin{center}
  \begin{tabular}{r|rrr} \toprule
   $h$	&	$i$	&	$P$	&	RAAN	\\	\midrule
800 km	&	98.6$^\circ$	&	100.87 min	&	190.4$^\circ$	\\	
700 km	&	98.2$^\circ$	&	98.77 min	&	190.4$^\circ$	\\	
620 km	&	97.9$^\circ$	&	97.04 min	&	190.4$^\circ$	\\	\bottomrule
  \end{tabular}
  \caption{Considered Orbits -- $h$ is the altitude of the orbit (above the Earth mean equatorial radius of $R_\oplus = 6378.14$ km), $i$ the inclination with respect to the equator, $P$ the period and RAAN the right ascension of the ascending node. Reference frame: Mean Earth Equator of J2000. They are all circular.\label{tab:numerics-table}}
 \end{center}
\vspace{-1.5em}
\end{table}

\subsection{Coordinate System \& Discretisation}
\paragraph{Coordinate System.} The observable region in the sky must be in reference to the satellite. To describe the system (Sun, Earth, Moon, satellite) a common coordinate system must be chosen. The chosen frame is the International Celestial Reference Frame (ICRF). This inertial frame is centred on the barycentre of the Solar system and is defined by 608 extragalatic sources mostly quasars and active galactic nuclei \citep{McCarthy2004}. The reference axes are defined with the observed objects. The realisation of this coordinate system has its principal plane as close as possible as the mean equator at J2000. The origin of this plane is also as close as possible to the equinox at J2000. This system was adopted by the International Astronomical Union (IAU) in 1997 and is therefore widely used -- which makes this choice natural.

In order to perform the computations, the centre of this coordinate system is translated into the satellite. Therefore, all positions are translated to the satellite and not the Earth.
\paragraph{Discretisation in Space.}
The sky is simulated by a grid of points described by the following relationships:
\begin{eqnarray}
\alpha_g(i) = \frac{\Delta \alpha}{2}+i\Delta\alpha,\ i=\{0,\dots,n-1\} \nonumber\\ 
\delta_g(j) = -\frac{\pi}{2} + \frac{\Delta \delta}{2}+j\Delta\delta,\ j=\{0,\dots,m-1\} 
\label{eq:grid}
\end{eqnarray}
where $n=40,m=20$ describe the resolution of the gird, $\Delta \alpha =\frac{2\pi}{n}$, $\Delta \delta =\frac{\pi}{m}$ and therefore the grid spans $\alpha_g \in (0,2\pi)$ and $\delta_g \in \left(-\frac{\pi}{2},\frac{\pi}{2}\right)$. This grid yield cells on the sky of $\Delta \alpha \times \Delta \delta$ or $9^\circ \times 9^\circ$. In order to be consistent throughout this work, the grid defined by $\alpha_g, \delta_g$ was used at each step of the generation of the maps. The fairly large zone of one cell is mainly a constraint that arises from the computation time of the different steps which scales at least as $\mathcal{O}(n\cdot m)$.

\paragraph{Discretisation in Time.}
The orbital period of the satellite depends of course on the altitude following the famous third Kepler law:
$$ P = 2\pi \sqrt{\frac{a^3}{Gm_\oplus}} $$
The observability maps will be used to plan the observation and therefore, it is better to have a time in minute rather than the position in a given orbit. The time $t_0=0$ min is chosen to be on January 1st 2018 at midnight -- \emph{i.e.} 2018-01-01 00:00. The code allows to speed up the calculation by repeating the results obtained for a given orbit a certain number of time as the map changes slowly over time. For the generation of the maps used in this work, this option was not considered and a visible region was computed for every minute in 2018.

\subsection{Computational Details \& Outputs} \label{sec:numercis-MATLAB-details}
The algorithm is fairly straightforward. For every cell and for every minute, the code tests whether the angular distance between the Sun, the Earth and the Moon are those required in \S\ref{sec:numerics-visibility-constraints}. It can also compute whether the satellite is in the SAA (see \S\ref{sec:dis-SAA}). Then, the computation for the stray light is a simple function which computes the angle from the LOS to the horizon. If the terminator is seen by the satellite, from the LOS to the terminator. This method of computing the stray light angle is fast as it considers only one direction for the stray light. Sadly, it is not sufficient to ensure that no point on the Earth will be seen with an angle smaller than the requirement of 35$^\circ$. A proof of this is given in appendix \S\ref{app:terminator-angles}. In the current version of this software, some targets are therefore claimed ``visible'' when they are actually forbidden. 
These targets are discarded by the stray light simulator and therefore this is not an issue that is present in the result of this study.
\paragraph{}
The output files of the \verb=MATLAB= code returns a list of the right ascension $\alpha$ and the declination $\delta$ of the grid points which are observable. In order to reduce the number of files generated, the visible points are all grouped in one single file for one orbit. Hence, the final output files are given in the following format : \verb=t, ra, dec= in, respectively, minutes and radians.
\section{Stray Light Code} \label{numercis:stray_light.f}
\begin{figure}
 \begin{center}
  \includegraphics[width=1\linewidth]{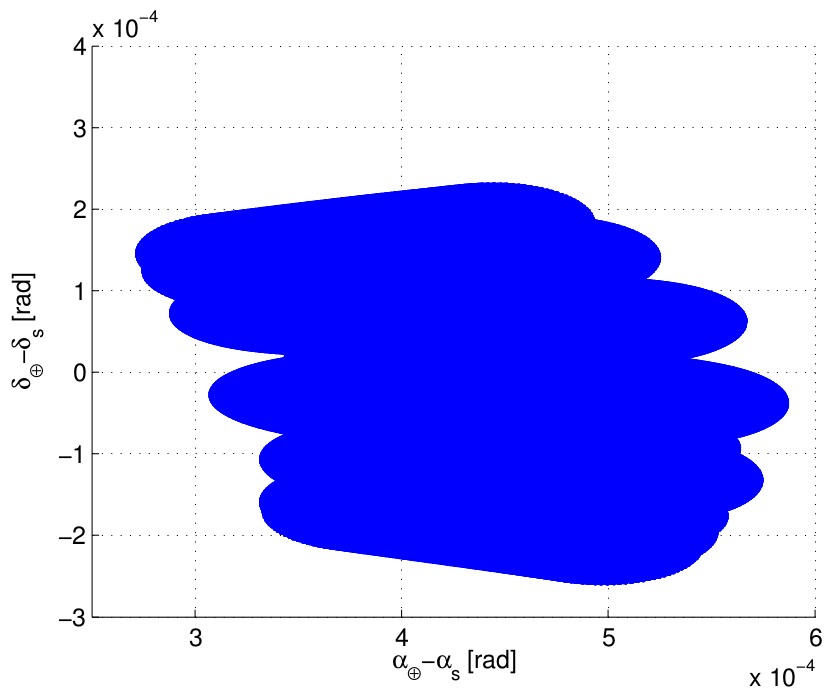}
  \caption{Dispersion of the difference in the position of the Sun. The coordinates referenced to the centre of the Earth $\alpha_\oplus,\delta_\oplus$ are plotted against the position of the Sun with reference to the satellite $\alpha_s,\delta_s$.}
  \label{fig:numerics-sun-pos-dispertion}
 \end{center}
\end{figure}

To compute the Earth stray light that enters the telescope and hit the detector, a dedicated code in \verb=Fortran= was designed by Andrea Fortier from UniBe and corrected during this work \citep{Fortier2013}. This program named \verb=stray_light.f= can compute independently of the observability map code, the stray light contamination at any time and in any direction. Ephemerides predictors for the Sun \citep{SunAlmanac2012} and later for the Moon \citep{meeus1988astronomical} provide the information necessary to operate in independence of the observability map software\footnote{The orbit must be however the same for both codes.}. During this project, the stray light code underwent extensive recoding to remove bugs and optimise the simulations. 
\paragraph{}
The coordinate system used for the stray light analysis is the same as used for the visible area computations: \emph{i.e.} the ICRF. The position of an object in the sky further away from the Sun therefore it is defined by two angles: the right ascension ($\alpha$ or RA) and the declination ($\delta$ or DEC) measured with respect to J2000. The position of the objects in the observability code are referenced to the satellite. 
This is not the case for the stray light code where the centre of the coordinate system is placed at the centre of the Earth. This does not impacts the computations of the position of the Sun and objects further away, but it greatly changes the position of the Moon and of course of the Earth. The error made at this point was evaluated to check that it would not impact the results of the study. 
As shown in Fig. \ref{fig:numerics-sun-pos-dispertion}, the dispersion of the errors made on the position of the Sun are at maximum $\Delta \alpha_\odot =\pm2\cdot 10^{-4}$ in right ascension while the error in declination is $\Delta \delta_\odot=(4.2\pm 1) \cdot 10^{-4}$. This is sufficient to compute with enough precision the relative alignment of the Sun, the satellite and the Earth. 
The former -- the Moon -- is not of great interest because the stray light due to the presence of the Moon is not included and because the Moon plus an angle of 5\textdegree\ from its limb is considered as ``not visible'' in the observability map code. The later -- the position of the Earth -- is of course of the foremost importance. The difference in the representation of the position with the two reference points is actually only the sign and therefore it can be trivially recovered. 
\hfill
\paragraph{The Steps to Stray Light Evaluation.}
The stray light computation is divided in several steps:
\begin{enumerate}
 \item Compute the illuminated part of the Earth;
 \item Compute the gradient of the illumination;
 \item Compute which parts of the illuminated Earth are visible by the telescope;
 \item For every visible region or ``targets'', do the following:
 \begin{enumerate}
  \item Compute the flux of light bouncing off Earth's surface that reaches the telescope;
 \item Convolve this signal by the PST to get the flux of Earth stray light at the detector.
 \end{enumerate}
\end{enumerate}
Those steps are explained in further details below. It is necessary to caution the reader that the description of the process is very linear here and does not exactly mirror the implementation in which numerical optimisation was planned. 
\paragraph{}
In the software, a few simplifications are made: Firstly, only the \emph{Earth} stray light is computed. Contamination from the Moon is not considered as it is assumed that its exclusion is sufficient to remove its contribution to the noise. Secondly, the Earth is considered to be a perfect sphere thus neglecting its bulge and hence slightly miscalculating the off-axis angle of an incoming photon. The surface of this sphere is supposed to be uniform: no distinction is made between oceans or land, the weather is not simulated (therefore thunderstorms and aurorae are not considered) and cities not modelled. To get a upper estimate of the stray light, the albedo of the Earth is set to 1. The albedo is a multiplicative factor in the calculation of the flux (See \S\ref{numerics:functions} for more details). The value of the averaged albedo ($\sim$ 0.3 -- 0.5) on the Earth could therefore be used. 
However, it was chosen to set the albedo to 1 to compute a worst case scenario, but with little difference to a finely tuned value because of the logarithmic behaviour of the final output of the project (as discussed in \S\ref{sec:dis-sensitivity}). Lights that could be observed in LEO are supposed not to contributed to the stray light due to the distance LOS--Earth's surface which must be larger than 100 km. This minimum distance is the thickness of the atmosphere.

\paragraph{Illuminated Earth.}
The first step is to describe with enough precision the illuminated part of the Earth. The only interesting part of the Earth at this point is the one in daylight. Indeed, it is considered that half of the Earth which is the night does not contributed to stray light.There are thunderstorms, aura or even city lights that could contaminate the signal if the LOS were too close to the limb of the Earth. 
Those effects which are extremely complex to simulate are not integrated in the code, but the constraint to point above the atmosphere with a minimal altitude of 100 km is supposed to ensure that there is no contamination.

The surface of the Earth is represented by a grid for which every cell is determined to be either in the day or in the night depending on the dot product of its normal and the direction of the Sun. The minimum grid that can achieve sufficient accuracy, \emph{i.e.} that the flux of stray light has converged, is quite large with 1000 by 500 cells.

\paragraph{Illumination Gradient.}
For the cells that are in the sunlight, the intensity $I$ of the light is computed by determining the cosine of the angle at which the light impacts the cell $c$. This depends upon the relative position of the cell and the Sun and can be computed using the following formula:
\begin{equation}
 I_c \propto \cos \varphi
\end{equation}
where $\varphi$ is the angle between the direction of the Sun and the normal vector of the cell (See \S\ref{app:great-circle-distance}).
The amount of radiation that arrives at a particular cell is also given by the energy received by the Sun at the distance of the orbit of the Earth. Here, the common value of about 1360 W/m$^2$ is not considered as it takes the whole spectrum into account. The instrument is sensitive to the range of wavelengths from 400 to 1100 nm and the solar ``constant'' for this interval can be computed to be 880 W/m$^2$. 

\begin{figure}[!h]
 \begin{center}
  \includegraphics[width=1\linewidth]{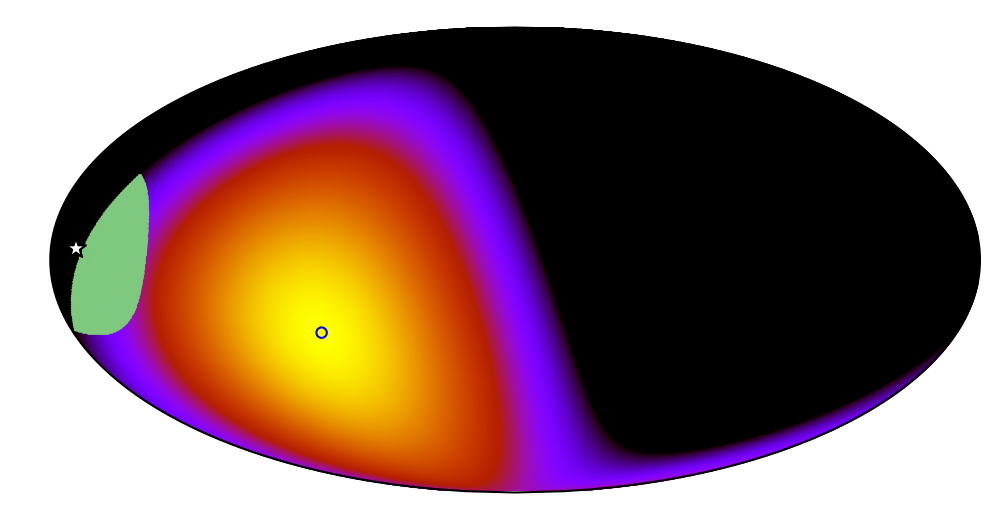}
  \caption{Mollweide projection of the illuminated Earth. The circle on top of the maximal illuminated region is the projection of the Sun, the white star-shaped marker symbolises the satellite and the green region is the illuminated Earth seen by the satellite.\label{fig:numerics-illuminated-earth}}
 \end{center}
\end{figure}

\paragraph{Changing Viewpoints.}
The two previous parts can be carried out without the knowledge of the orbit at that particular time. Indeed, the satellite plays no role. This step determines which illuminated cells can be seen by the satellite at any given time. Assuming a position for the satellite\footnote{Such files exists for the different orbits used throughout this project. The orbital parameters used are expressed in \S\ref{tab:numerics-table} and then fed to the advanced commercial software STK that computes the position of the satellite with time.}, the code finds all cells within a cone defined by all visible tangent directions on the surface of the Earth and flags them as relevant to the stray light computation. See figure \ref{fig:numerics-illuminated-earth} for an example.

\paragraph{Flux of Photons at the Telescope.} The flux of photons at the entrance of the telescope (not at the detector or somewhere else in the optical path) depends upon the direction of the line of sight. Let $\theta$ be the angle between the LOS and the photon path. The target star is denoted by $\pmb{r}_\star$, the position of the satellite by $\pmb{r}_s$ the cell position by $\pmb{r}_c$ all with reference to the centre of the Earth. The direction to which the photon bounced off the cell to the satellite can be defined by the vector $\pmb{v} = \pmb{r}_s-\pmb{r}_c$. If the angle $\theta$ is larger than $\pi/2$, then no photons will hit the telescope:
\begin{eqnarray}
\pmb{r}_\star\cdot \pmb{r} =  |\pmb{r}_\star|\cdot |\pmb{r}|\cos\theta \\
\nonumber\\ 
 \cos \theta \begin{cases} >0&\text{No hit} \\ \leq 0 & \text{photon hit} \end{cases}
\end{eqnarray}
To obtain the value of the total flux of photons all that is left is to integrate on all possible directions. The surface of the Earth is assumed to be Lambertian\footnote{A Lambertian surface is a surface whose radiance (power per unit solid angle per unit projected source area) is the same independently of the angle of view. Its apparent brightness is the same.}. This is justified as the albedo is assumed to be constant and because the radiant intensity (power per unit solid angle) depends upon the cosine of the angle described by the direction of the radiation leaving the point. The same cosine factor is applied to the surface that emits the radiation and therefore they cancel out.

\begin{figure}
 \begin{center}
  \includegraphics[width=1\linewidth]{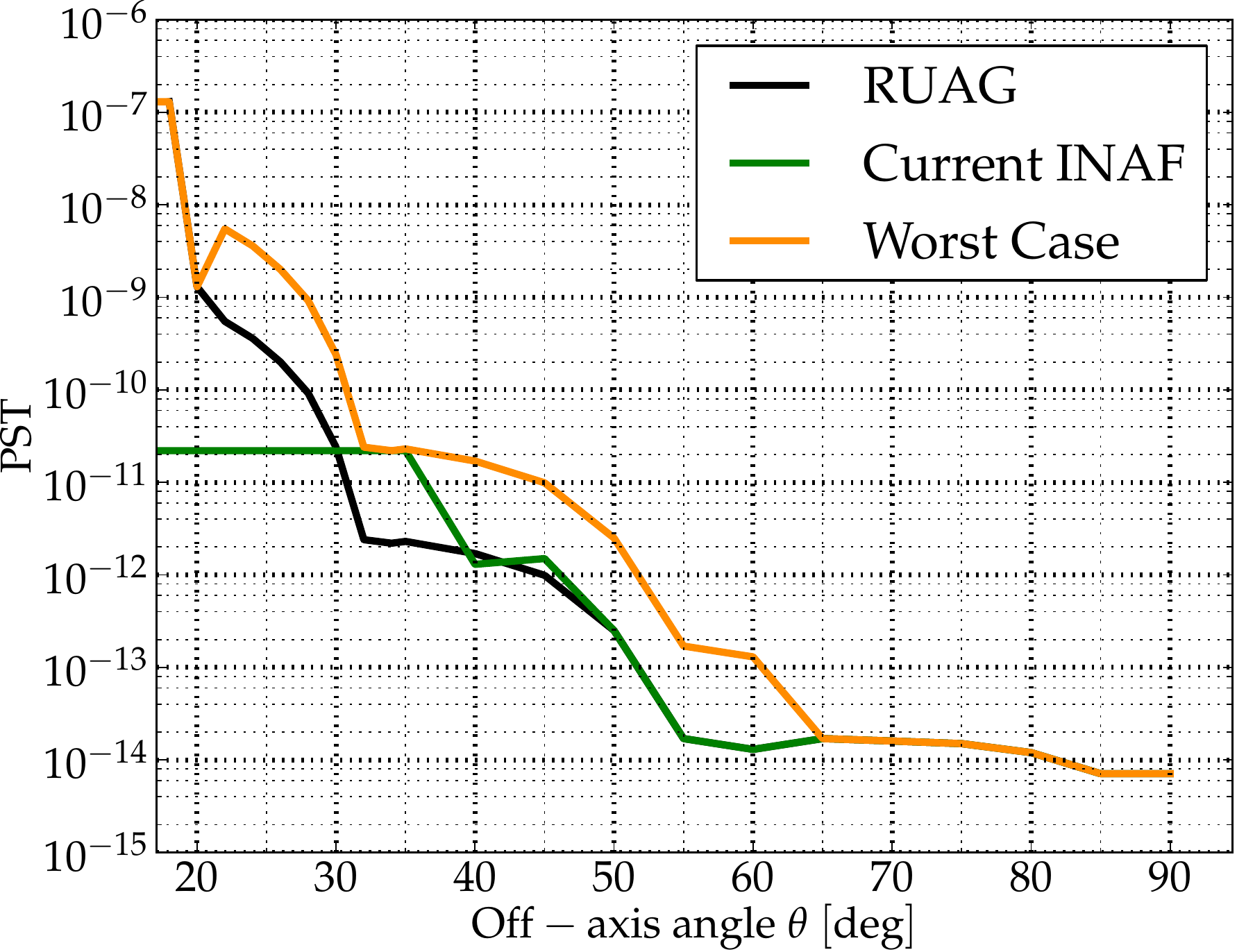}
  \caption{The PST as modelled by RUAG, INAF the latest PST and the worst case as a function of off-axis angle.}
  \label{fig:numerics-ruag-pst}
 \end{center}

\end{figure}

\paragraph{Accounting for the Photon Rejection.} Up until this point, the considerations were mostly geometrical. The Earth is assumed to be a perfect sphere which reflects light with the same radiance in any direction. The cells that can be seen by the satellite are defined by its orbit. Hence computing the angles from the LOS to them is easy.
So, at this point in the code, the amount of photons reaching the pupil of the telescope is computed each with an off-axis angle $\theta$. The discussion about stray light analysis in section \S\ref{sec:theory-sl} says that the transmission of the incoming photons through the telescope can be approximated by a axisymmetric function that depends upon the off-axis angle $\theta$ of the radiation : the point source transmission (PST). 
As explained in the cited section, the determination of this function is a tedious process. Due to this, one of the PST used in the project is the PST issued by RUAG in the preliminary phase of the project as shown in  figure \ref{fig:numerics-ruag-pst}. The energy that arrives at the detector is multiplied by the value of the PST as function of the off-axis angle of the photon. 
An interpolation of the PST is performed to fill the gaps in the knowledge of this function which is sampled \emph{typically} every 5 degrees from 20\textdegree\ up to 90\textdegree. The flux is also converted to photons per second per pixel at this point. As the stray light is scattered by different parts of the telescope and not through the optical path of the mirrors and focal planes, its signal is not convolved by the PSF and hits a single pixel.
Others PST are also used: a worst case scenario (basically RUAG PST times 10) and the latest PST produced by the consortium (section \ref{sec:theo-sl-CHEOPS} \& \ref{sec:dis-PSTs}).

\paragraph{Practicalities.} The code used in this project was developed for the purpose of this work and is therefore optimised to perform stray light computations for known observability maps and ephemerides for the Sun. It takes as input parameters a PST (\verb=theta [deg],PST=), the position of the satellite (in the frame of reference of the Earth) and the Sun in separate file containing \verb=x,y,z,ra,dec,r= with distances in km and angles in radians. Finally two files describe the target list: one containing the number of targets to optimise the \verb=Fortran= memory management and file handling and a list of targets with two columns: the right ascension and the declination both in radians. 

Given a target list and all other input files associated with it, the time required to compute the stray light depends on the season as the number of illuminated Earth cells seen by the satellite vary with time (\S\ref{sec:dis-sl}). The number of targets is between 20 and 160 depending on the position in the orbit: when the satellite is over the day side, less targets are available due to the stray light exclusion angle. Targets are cells of $9^\circ \times 9^\circ$ as described in \S\ref{sec:numercis-MATLAB-details}.
To complete the computation of one orbit at 800 km (100 or 101 minutes), it takes between 2 and 4.5 minutes. As the altitude of the satellite is decrease, this computation time increases due to the growing number of seen illuminated cells on a single 3.3 GHz CPU.

\section{Combination}
The two code described earlier must be used together to generate stray light maps for each minute. For this purpose, several \verb=Python= scripts were designed to handle every step of the generation from the input map to the different plots. This pipeline has several interfaces: the data generated by the \verb=MATLAB= code and the ones generated by \verb=stray_light.f= must be treated. Due to these interfaces constraints, a modular approach was chosen. It allows improvements of part of the pipeline or the data as well as iteration on parts of the data flow easily.

\subsection{A Pipeline to Compute the Stray Light Flux} \label{sec:numerics-pipeline}
To generate a stray light map of the observable sky at a given time, a \verb=Python= script looks for the corresponding observability map, creates the list of visible points which are then given as inputs to the stray light calculator along with a few other variables (position of the Sun and the Earth, \dots) as described in section \S\ref{numercis:stray_light.f}. To speed up computation, an adaptive time step is implemented. An orbit is computed every minute, however as there are around 15 orbits per day, consecutive orbits are very similar in most of the cases. Hence the capability of skipping orbits that are too similar. This is done by computing orbit $i$ and then orbit $i+10$ ($i+8$ for the 620 km altitude orbit). If the comparison of the two orbits $(i, i+\Delta i)$ reveals a difference of more than 5\% in stray light flux, then the orbit $i+\Delta i/2$ is computed. 
The algorithm always tries to increase the number of step and doubles the step size if it is (1) smaller than the optimum (10 or 8) and (2) the error to the previous orbit is less than 5\%. The maximum bearable error to the next step was chosen such that it would ensure that the \emph{maximal} error in the orbit in term of equivalent magnitude (see \S\ref{numerics:functions}) is about 0.05 magnitude. The maximum step size is set to ensure that the number of orbit for a given day in the year is at least one. Especially for low orbits, it may happen that the difference from one orbit to the next is still larger than 5\%. This is handled by making sure every single orbit is computed in those excessively changing moments.
\begin{figure}[!h]
 \begin{center}
  \includegraphics[width=1\linewidth]{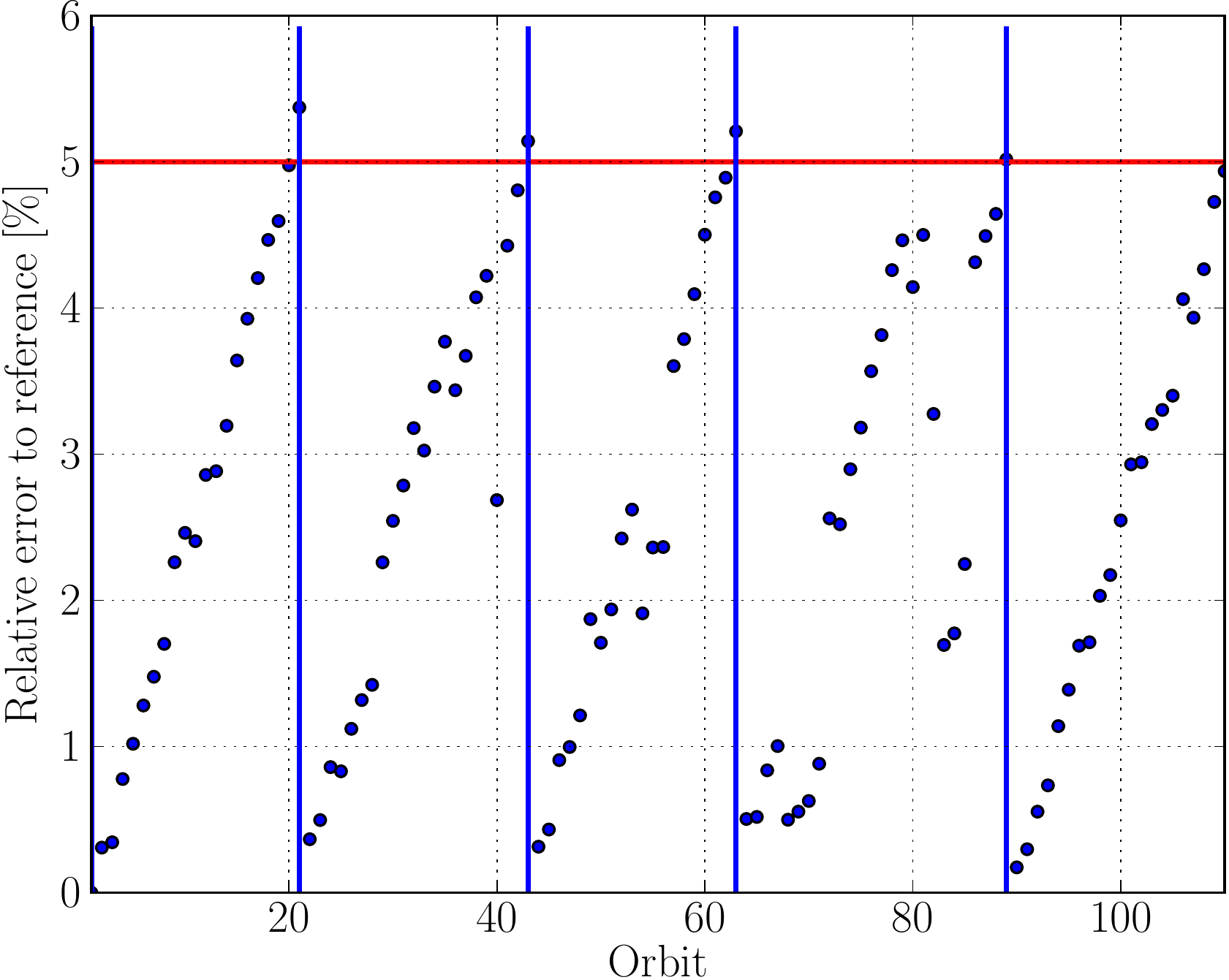}
  \caption{Example of errors relative to the previous reference orbit for orbits 1 through 110 at 800 km. The red line represents the 5\% limit while the blue vertical lines are orbits whose difference to the previous reference is larger than 5\%.\label{fig:numerics-error-evolution}}
 \end{center}
\end{figure}

Figure \ref{fig:numerics-error-evolution} shows an example of evolution of the error when computing stray light for orbits 1 to 110 for 800 km altitude. If the upper limit of the step size of 10 was not enforced, step size would be equivalent of computing one orbit every two days approximatively. The temporal resolution would not be large enough to detect errors in the observability maps or to follow, for example the Moon in the observable region. As it can be seen in the figure, the number of orbits which can be skipped vary on a short timescale due to numerical errors and differences in the target list. 
This scheme yields orbit step size of 1, 2, (3), 4, 5, (6), (8) or 10 orbits (The numbers in parenthesis are rare occurrences -- see \S\ref{sec:code-efficiency} for a discussion of the orbit step efficiency).

This step of generating stray light maps for the year 2018 is the most lengthy one. The time needed to compute a whole year with the adaptive orbit step described in the above depends on the altitude as the stray light is less prominent at high altitude. It takes for between 30 and about 100 hours to finish all computations on a single CPU running at 3.3 GHz. The outputs are one file per minute containing the following characteristics: time, grid point position in radians (right ascension, declination) and the flux of stray light in ph px$^{-1}$ s$^{-1}$. The relative high number of output files ($\sim 5-10\cdot 10^5$) weight around 1--2 GB and will be analysed in the next steps.
The temporal resolution is chosen to be the same as the observability maps. There is another deeper reason to this 60 seconds resolution: the baseline of frame exposure is 60 seconds. This leads to difficulties when patching two orbits together. For 800 km altitude, the orbital period is 100.87 minutes and thus there will be orbits consisting of 100 minutes and others of 101 minutes. This is corrected in the code by assuming that this extra minute is an integer. To compare two orbits one of 100 minutes and the other of 101 minutes, the longer orbit can be shifted by minus one minute to find the best fit to the shorter one. 

\subsection{Analysis of Stray Light Data} \label{sec:numerics-analysis}
To interpret the data, several analysis tools were developed. The scripts range from the very simple counting of computed orbits to a much more elaborated probability of observing a transit estimation. A description of some of those tools is provided here. A discussion on all of these different steps is proposed in chapter \ref{ch:results}.
\paragraph{Computed Orbits.} This is the first logical step -- compulsory in order to proceed to other functions --: check which orbits have been computed. This seems redundant as this list can easily be retrieved from logging at the previous step when computing the stray light maps. In order to speed up the computations, most of the stray light data was generated using 3 CPUs: not in parallel, but working on different periods in the year. Moreover, the stray light simulation handler can ask to compute orbits in the order for example $i,i+10,i+5,i+2,i+8$ depending on the maximal difference between the orbits and therefore the list of orbits is not straightforward to establish. To avoid any mistake when merging back the results of the different CPUs, this scripts tests if minutes 0, 20 and 60 were computed. As the SAA is not considered, each of those minutes had to be computed. If any fails, then it is considered that this orbit was not computed.

\paragraph{Analysis of the Error Evolution.} The second step is to check that the error from one orbit to the next is not larger than the 5\% limit or if it is, then that every orbit is computed. This operation is rather slow as the code must compare the minutes pairwise in each orbit and due to non-integer period, several combinations must be made to find the best fit. 

\paragraph{Analysis of the Stray Light Flux.} From this point forward, the data is considered as robust if the two previous steps did not pointed out to erroneous behaviours. The stray light flux depends upon the amount of Earth illuminated cells seen and their illumination. In order to characterise those changes the stray light flux is analysed by four different quantities:
\begin{enumerate}
 \item The \emph{maximal value}: the maximal value of the flux for a given orbit;
 \item The \emph{maximal mean}: taking the maximum value of the flux for one minute, repeating for every minute in the orbit and averaging;
 \item The \emph{twice-averaged mean}: taking the average of the flux at a particular moment in the orbit, repeating for every minute in the orbit and taking the average on those averages per minute;
 \item The \emph{evolution in one orbit} of the stray light in one direction.
\end{enumerate}

\paragraph{Stray Light Flux Maps.} The list of the visible targets is completed by the stray light flux. Therefore, a projection of the targets can be made on a map with a colour assigned to the stray light. A conversion to limiting magnitude can be made to represent how faint the observed stars can be given a certain SNR (see \S\ref{numerics:functions}). Those maps have the advantage -- beside being aesthetically pleasant -- of showing the varying visible regions in the sky and the effect of the relative alignment of the Earth and the Sun. 
\paragraph{Cumulative Observation Maps.} The cumulative observation time is a value that is computed on a long period of time (in this case, the year 2018). There is a minimum of observation time to consider the observation worth it. On top of this minimal observation time, the telescope takes time to align its LOS to the star; this acquisition time is set to 6 minutes\footnote{The value of 6 minutes is given by the capabilities of the rotation wheels of the attitude and determination control subsystem.}. Those maps give informations about which regions of the sky are invisible which can be observed the longest and the effects of the constraints and the SAA. Those maps are relatively long to establish as the appearance and disappearance times -- which occurs many times over the course of the year -- of every grid point must be computed.

\paragraph{Scheduling the Observations and Probability of Detection.} To prepare the scheduling of the observation, a catalogue of interesting objects must be established. This catalogue is read by the code and then, a chart of when this object can be observed is computed for a short period of time (say one week for clarity of the chart). However, this chart can be computed for a long period of time and store into the memory. This is very useful to see when a target is visible and if its visibility coincides with the predicted transit time.

On a similar note, the probability of detecting a transit is a function of how often and how long the instrument is able to stare at it. To get an estimate of this probability for the whole sky, the sky grid of $40\times 20$ cells is used as if the grid points were targets. This kind of sky map reveals what kind of target can be observed and when, which is a very important information.

\subsection{Computational Details}\label{numerics:functions}
\paragraph{From Stray Light Flux to Maximum Magnitude Visible.} 
The question asked here is: given the stray light flux, what is the faintest target visible? The function \verb=flux2mag= (and the reciprocal \verb=mag2flux=) yields this limiting magnitude (or inversely the flux) of the target star. The noise represented by the stray light to the flux of the star must be at maximum $T=10$ ppm. The flux is related to the magnitude $m$ by:
\begin{equation}
 F = 10^{-2/5\cdot m}
\end{equation}
which has units of ergs s$^{-1}$ cm$^{-2}$. The flux of stray light is given in units of ph s$^{-1}$ px$^{-1}$. Moreover, only photons with a certain energy matter and not all the spectrum of light. Therefore the flux is now given by:
\begin{equation}
 F_\text{sl}(m_V) = \underbrace{F_V J_V\frac{\Delta \lambda}{\lambda}}_{\equiv A}\cdot\underbrace{\left(\frac{R_\text{tel}}{R_\text{PSF}}\right)^2}_{\equiv B}\cdot10^{-2/5\cdot m_V}
\end{equation}
where $F_V = 3640$ Jy cm$^2$/ergs converts from ergs to Jansky for the band $V$, $J_V$ transforms from energy to the flux of photons for the band $V$: $J_V=1.51\cdot10^7$ photons s$^{-1}$ m$^{-2}$ / Jy. $\frac{\Delta \lambda}{\lambda}=0.8$ is the bandwidth. Those three parameters grouped in $A$ depend upon the choice of the band. $R_\text{tel}$ is the radius of the aperture in metres (hence $R_\text{tel}=0.15$ m). $R_\text{PSF}=15$ px is the radius of the PSF in pixels. The PSF is modelled here as an axisymmetric step function (top hat function) such as 
\begin{equation} \label{eq:modelled-PSF}
 \text{PSF}(r) = \begin{cases} 1 & r\leq 15\text{ px}\\
			     0 & r>15\text{ px}
              \end{cases}
\end{equation}
Those two radius are geometrical parameters and their ratio squared transforms from m$^2$ to pixels and is noted for convenience $B$. As discussed in section \S\ref{sec:Theory-PSF}, the goal to reach for this mission is close to a top hat function and hence reasonably well represented by the very simple function in eq. \ref{eq:modelled-PSF}. Of course, the real PSF will affect the number of photons received from a star. A discussion on the sensitivity on the limiting magnitude on the PSF is provided in section \S\ref{sec:dis-sensitivity}.

The SNR for the stray light is given by $SNR=1/T$ and therefore, the maximum stray light flux tolerable for a given magnitude is:
\begin{equation} \label{eq:num-mag2flux}
 F_\text{sl}(m_V) = T\cdot AB\cdot10^{-2/5\cdot m_V}=\frac{AB\cdot10^{-2/5\cdot m_V}}{SNR}
\end{equation}
This computation is implemented in \verb=mag2flux=. The reciprocal function uses equation \ref{eq:num-mag2flux} to express $m_V(F_\text{sl})$. This implies that there is a logarithmic of base 10 function. In order to avoid numerical errors while performing this computation, the stray light flux is clipped to a minimal value of $10^{-40}$ ph s$^{-1}$ px$^{-1}$. For future references, Tab. \ref{tab:sl-to-mag} presents the the limiting star magnitude for a given stray light flux without considering any correction on the images.

\begin{table}
 \begin{center}
  \begin{tabular}{l|r} \toprule
  $F_\text{SL}$ & $V_\text{lim}$ \\
  $\left[\frac{\text{ph}}{s\cdot\text{px}}\right]$ & [mag] \\ \midrule
$10^{-9}$ & 26.5 \\
$10^{-8}$ & 24.0 \\
$10^{-7}$ & 21.5 \\
$10^{-6}$ & 19.0 \\
$10^{-5}$ & 16.5 \\
$10^{-4}$ & 14.0 \\
$10^{-3}$ & 11.5 \\
$10^{-2}$ & 9.0 \\
$10^{-1}$ & 6.5 \\
$1$ & 4.0 \\
\bottomrule
  \end{tabular}
\caption{Comparative of the stray light flux and the equivalent limiting magnitude of an observed star.\label{tab:sl-to-mag}} 
 \end{center}

\end{table}

\paragraph{Comparing Two Orbits.} This function is important as it decides whether the orbit step size in the calculations of the stray light flux must be reduced. It must, therefore, read the data from the two orbits and compare them. This function is quite conservative in the sense, that it returns the \emph{maximal} difference in the orbit and not the \emph{mean}. The two orbits (the \verb=reference= and the \verb=current= as defined in the code) are compared pairwise: minute by minute. As evoked before, this is a none trivial problem as the orbital period is not an integer of minutes. Therefore, three combinations are tested using a shift $s=\{0,1,2\}$ in time between the reference and the current orbits.

\paragraph{Great Circle Distance.} \label{app:great-circle-distance}
The angular separation of two points is required several times in the different steps of this project. This separation must be computed reliably and robustly otherwise several numerical issues arise. Those numerical issues are due to two things: (1) singularities of the inverse trigonometric functions at multiples of $\pi/2$ and (2) the inverse cosine which handles badly two close-by points \citep{Vincenty1975}. Therefore, to compute the angular distance between two points 1 and 2, the following formula was used:
\begin{equation}
\begin{aligned}
 &\varphi =\arctan\left(\frac{\sqrt{a^2+b^2}}{\sin\delta_1\sin\delta_2+\cos\delta_1\cos\delta_2\cos\Delta\alpha}\right) \\
 &\text{where}\ a = \cos\delta_2\sin\Delta\alpha, &\\
 & \ \ \ \ \qquad b = \cos\delta_1\sin\delta_2-\sin\delta_1\cos\delta_2\cos\Delta\alpha
 \end{aligned}
\end{equation}
Similarly, the angular separation cannot be simply computed by the scalar product, but by the equivalent, but always correct cross product:
\begin{equation}
 \varphi =\text{arctan}\left( \frac{\left| \pmb{r}_1\times \pmb{r}_2 \right|}{\pmb{r}_1\cdot \pmb{r}_2} \right)
\end{equation}
where $\pmb{r}_i$ are vectors that point to the position of the two coordinates.
\cleardoublepage
\chapter{Results \& Discussion} \label{ch:results}
\lettrine[lraise=0., nindent=0em, slope=-.5em]{I}{n} this chapter, the results of the simulations are presented and discussed. The focus is first on the stray light flux and its behaviour and then shifts to the interpretation of the stray light constraint on the visible regions. All references to winter or summer are to be understood with respect to the Northern hemisphere. All maps of the sky (or of the Earth) are plotted using a Mollweide projection.
\section{Stray Light}
\subsection{Code Efficiency} \label{sec:code-efficiency}
\begin{figure}
 \begin{center}
  \includegraphics[width=1.\linewidth]{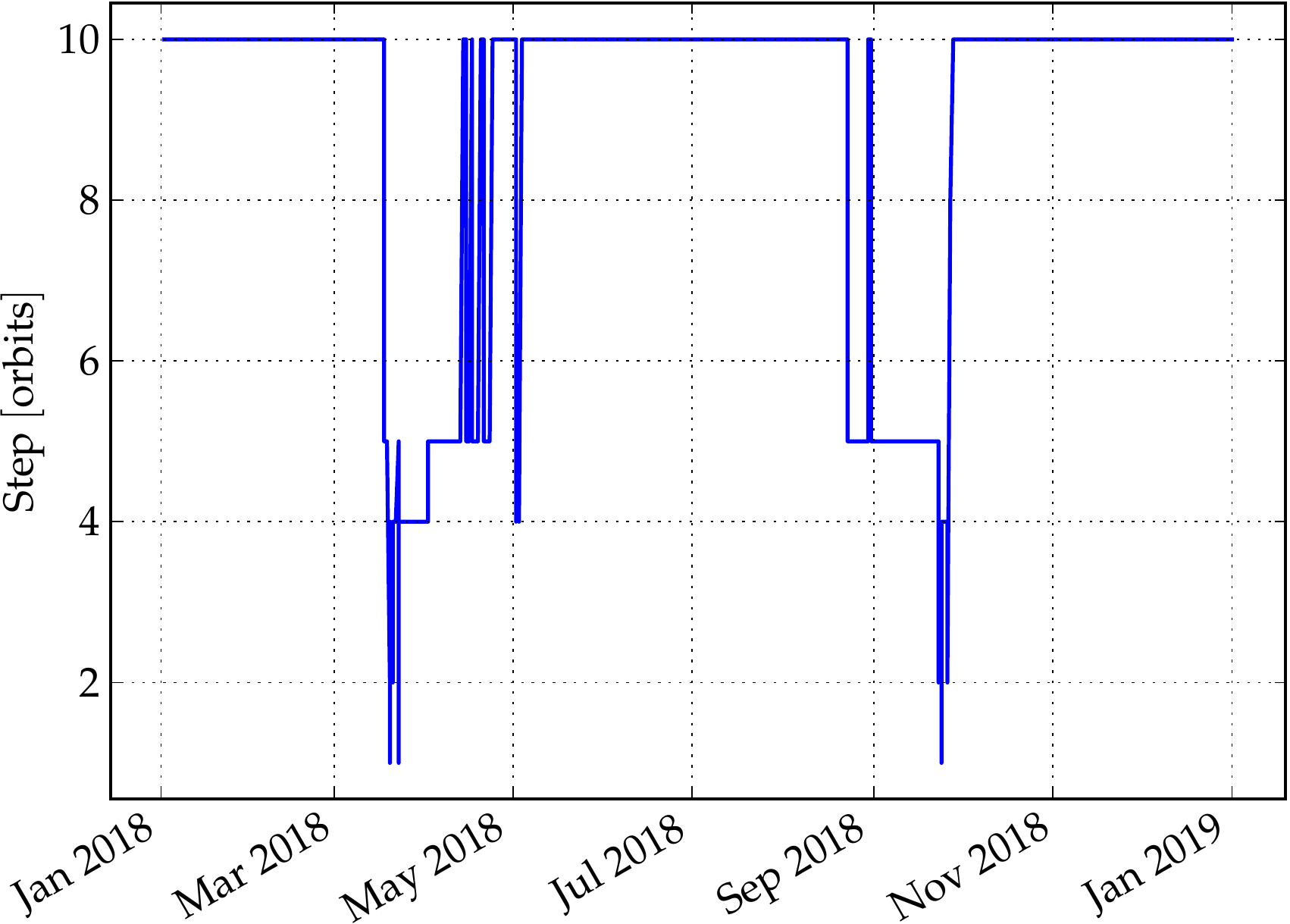}
  \caption{Orbit step applied in function of time. 800 km and a stray light exclusion angle of 35\textdegree. \label{fig:error-evolution-step-800}}
 \end{center}
 \vspace{-2em}
\end{figure}

This section treats the technical performance of the code(s) and not of the data. The \verb=MATLAB= code that generates the {\bf observability maps} was not altered other than the way it stores the data. To complete the calculations during the whole year, about 1.5 day (on a single 3.3 GHz CPU) is needed for a low resolution (the whole sky is represented by the $40\times 20$ cell grid). Being now aware that this code does not ensure that all regions are actually observable (see \S\ref{sec:numerics-visibility-constraints} and \S\ref{app:terminator-angles}), this part should be rewritten in a compiled language to allow the computations for a finer grid in a reasonable amount of time.
\paragraph{Stray Light Flux Pipeline.} The computation of the stray light is performed by the pipeline which decides which orbits to compute and which are too redundant to matter. The 5\%\ maximal error tolerated from one orbit to the next can be hard to reach for low altitude orbits. As a consequence, a tremendous number of orbits -- and in particular for the orbit at 620 km -- are computed with respect to the minimal number of orbits that could be simulated (roughly 1.5 per day instead of 15). 
At high altitude, it can be noticed by studying Fig.~\ref{fig:error-evolution-step-800}, that the step size reduces  around two particular dates in the year: April and October. The underlying reason is revealed in \S\ref{sec:dis-sl}. For the Sun-synchronous orbit at 800 km (SSO800) orbits, the combined \verb=Python= and \verb=Fortran= algorithm completes the computation in about 1.5 day. As the orbit is lowered, the error evolution becomes more erratic and the step size becomes smaller and smaller. Using a SSO620 orbit implies a computation time close to 7 days.
The peaks occur when the plane of the orbit merges with the plane of the day-night terminator. Indeed, during this period, a lot of point are visible with a very low stray light flux. From one orbit to the next, stray light fluxes can fall to zero thus may imply great changes. The number of orbits that need to be computed may not vary much with another more sophisticated implementation. The time spend to compute the stray light for a given list of points and a given time was already drastically reduced by the updates and corrections to the \verb=Fortran= \verb=Stray_light.f= code. A tradeoff was made in the optimisation of the coding between speed and the effort that should be made to understand it as this project is not the final software that will produce real maps of the sky.

\paragraph{Stray Light Flux Analysis.} The tools developed to analyse the flux from different standpoints are all written in \verb=Python=. The complete analysis of the output from the pipeline is modular. It cannot be performed completely automatically: the user must still check the parameters before running the analysis. All actions that require treating the data of the complete year or at least a significant amount of time -- cumulative maps, scheduling , probability of transit -- require between half an hour and 2 hours of CPU time.

\subsection{Time Evolution of the Flux} \label{sec:dis-sl}
\begin{figure}[!h]
 \begin{center}
 \includegraphics[width=1\linewidth]{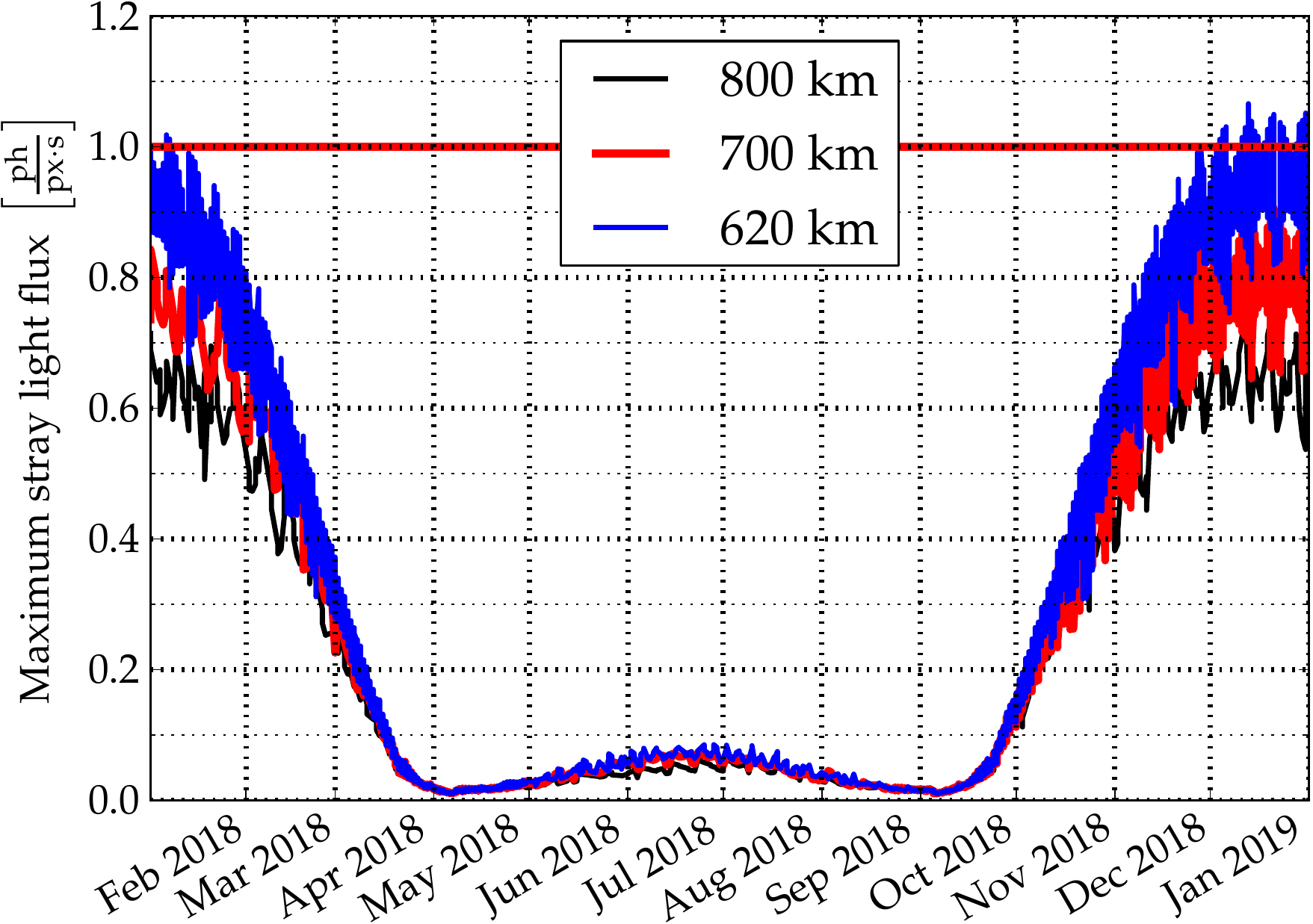}
 \includegraphics[width=1\linewidth]{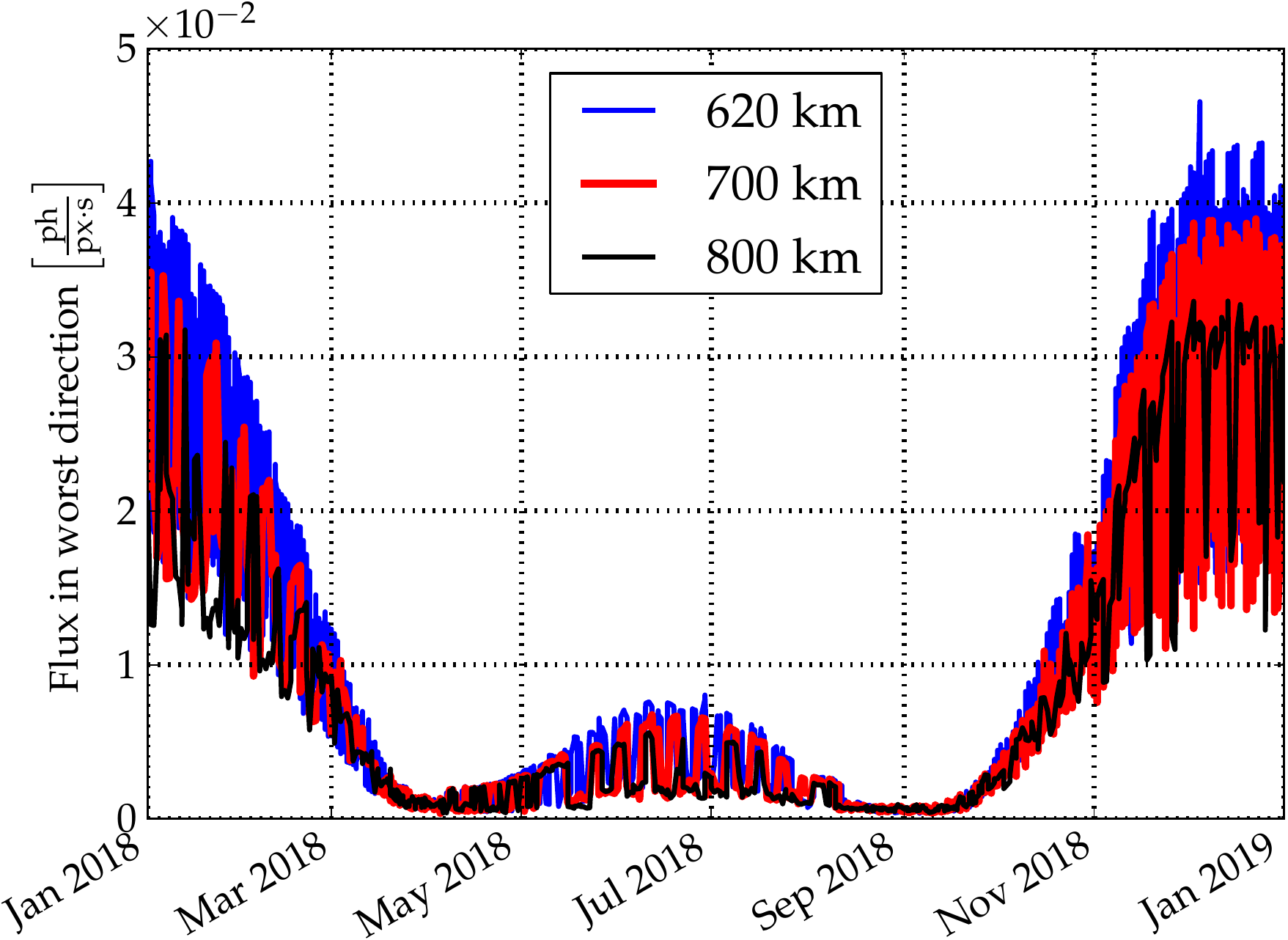}
  \includegraphics[width=1\linewidth]{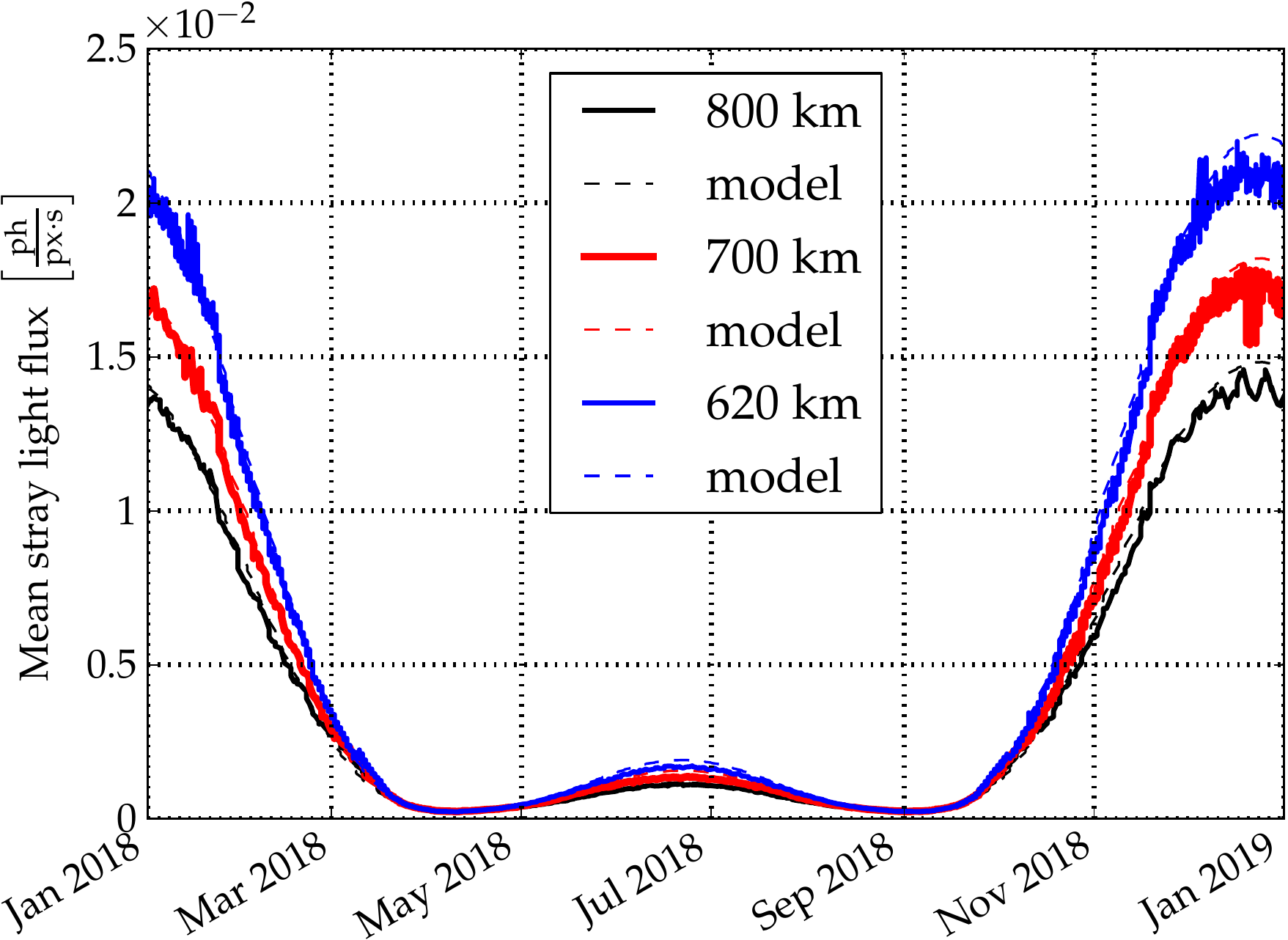}
  \caption{Flux of stray light as function of time for three different altitude -- \emph{Top.} Maximal value of the flux for a given orbit -- \emph{Middle.} Maximum value of the flux for one minute, repeating for every minute in the orbit and averaging. -- \emph{Bottom.} Average of the flux at a particular moment in the orbit, repeating for every minute in the orbit and taking the average on those averages per minute.\label{fig:all-altitudes-fluxes}}
 \end{center}
 \vspace{-1.5em}
\end{figure}

Fig. \ref{fig:all-altitudes-fluxes} show the the seasonal behaviour of the stray light.
The key point to explain the smooth-W shape of the stray light fluxes is to remember the parameters of the orbit. The orbit is inclined by 98.6\textdegree\ (with respect to the equator) at 800 km which does not correspond to the inclination of the axis of rotation of the Earth (see modelling the behaviour \S\ref{sec:dis-modelling}). Thus the angle between the terminator of light on the Earth and the plane of the orbit changes throughout the year.

This implies the existence of the two minima and maxima in the amount of Earth illuminated surface seen by the satellite. The minima take place when the plane of the orbit coincides with the plane of the terminator. At those moments, only a small part of illuminated Earth can be seen by the satellite. These dates do not coincide with the spring and fall equinoxes as there is a shift between the plane of the orbit and the rotation axis of the Earth. However, their positions are symmetric around the summer -- or winter -- solstices. 
\paragraph{}
A Sun-synchronous orbit means to pass the equator at the same solar time at each orbit. In the case of interest, this time is known as local time of the ascending node (LTAN), which is chosen to 6 am. 
It was chosen to favour coordinates visible from the Southern hemisphere due as there are more ground instrument there. Due to the choice of the LTAN, the overlapping of the planes of the orbit and of the terminator happen soon before and after the summer solstice. 
If this was to change and the LTAN were to be set to 6 pm\footnote{because CHEOPS is in a dual launch, a 6 pm LTAN might be better in the search for a co-passenger, although it would change the geometry of the visible regions in the sky.}, then it the plot would be shifted by 6 months.

The maxima are centred around the solstices of the Earth/Sun system. The summer solstice has a smaller effect in term of stray light than the winter solstices even if it is the local maximum. Again, the reason is to be found when observing the geometry of the eclipse. In the summer, the angle from the plane of the orbit to the terminator is about 15\textdegree\ while in the winter it increases up to $\sim$30\textdegree. This means that there is more illuminated surface that can be seen by the satellite in the winter and hence more stray light.
\paragraph{}
Looking at the maximum or the mean value of the stray light, one can see a similar increase due to the altitude. The maxima and minima happen at the same time regardless of the altitude chosen for the satellite. Interesting points to notice are (1) the difference between the mean and maximum stray light (Top and bottom of Fig. \ref{fig:all-altitudes-fluxes}) which is of a factor about 100 and (2) the curves for the mean value in the summer are smoother. The first point can be explained by the fact that the decrease in stray light flux when increasing the angle from the line of sight and the target star is fairly rapid. Indeed, the surface of an illuminated cell decreases as the cosine of the angle and so does the stray light flux which means a global decrease in $\cos^2$. 
As the observable region is not limited to targets close to the Earth's limb (where the cosine takes its maximal value), most of them will have contributions of stray light which are low. 
The second point is due to the fact that taking the maximum of the stray light flux is much more sensitive to changes and extreme values than the mean.

The differences between the three plots of Fig. \ref{fig:all-altitudes-fluxes} are due to the method by which they are computed (see \S\ref{sec:numerics-analysis}): the top plot presents the maximal value of the flux for a given orbit. The middle one yields the maximum value of the flux for one minute, repeating for every minute in the orbit and averaging. Finally, the bottom one depicts the average of the flux at a particular moment in the orbit, repeating for every minute in the orbit and taking the average on those averages per minute.
The noise of the signal decreases as the quantity is more and more averaged. Taking the average of the maximum stray light over the orbit yields the highest flux value. It gives an indication of whether a post-treatment of the observable region is necessary to ensure meeting the requirement and also reflects the overall quality of the observation: 
the more stray light, the more degraded the SNR. The flux in the direction of the maximal flux is biased by the fact that the maximal flux direction does not ensures that the point is visible for all the orbit. The points that were not available during the whole orbit were restricted to their period of visibility. It can be seen that the average in the worst direction is about twice the value of the mean of the mean and very noisy. However, the interesting information of this graph is that the during the alignment of the planes of the orbit and the terminator, the worst direction yield stray light fluxes similar to the mean of the mean.
The average of the average yield a rather clean signal that follows a law that can be derived from trigonometry (\S\ref{sec:dis-modelling}). It shows the global tendency of the stray light better than the two others as it takes all points into account.
\subsection{Modelling the Behaviour of the Flux} \label{sec:dis-modelling}
The shape and the behaviour of the stray light can be modelled to obtain an approximation quickly. A flux of photons in a solid angle $\Omega$ and of intensity $I$ is described by:
\begin{equation} \label{eq:dis-defn-flux}
 F = \int_{\text{d}\Omega}I\cos\vartheta\text{d}\Omega
\end{equation}
where $\vartheta$ is the angle between the normal of the surface through which passes the flux and the direction of the flux.

\begin{figure}[h]
\begin{center}
\includegraphics[width=0.85\linewidth]{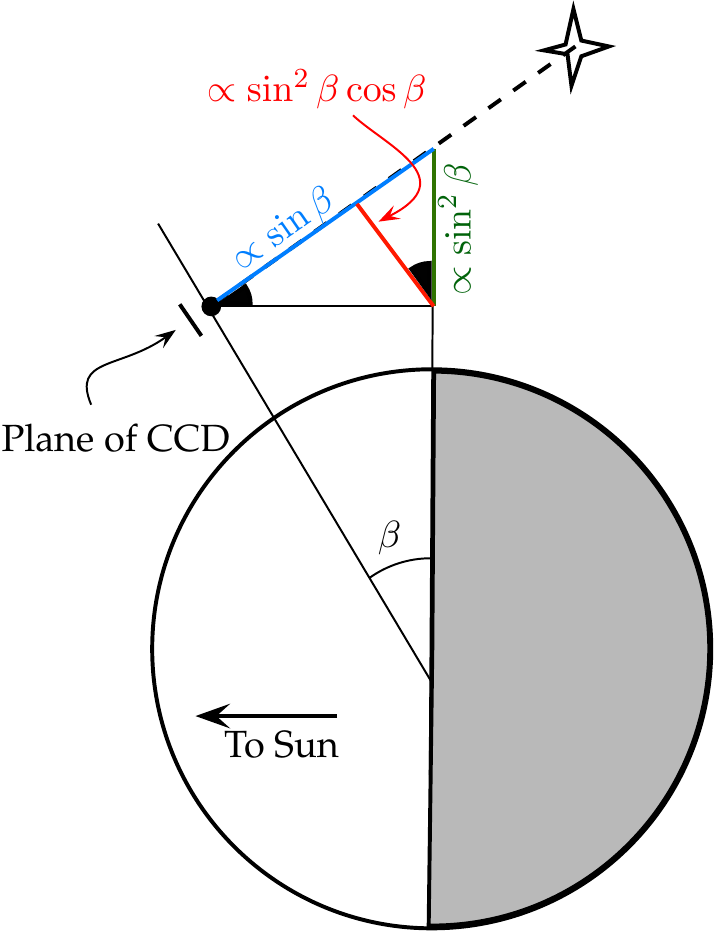}
\caption{Sketch the solid angle of the illuminated Earth seen by the satellite. The black wedges represent the angle $\beta$. The angle between the LOS and the line joining the centre of the Earth is assumed to be 90\textdegree. The blue line is a projection of the illuminated Earth, the green line is a projection of this blue line on the plane of the terminator and the red line is the projection of the green line onto the plane of the CCD. It is assumed that the angle between the LOS (broken line) and the line determined by the angle $\beta$ is always 90\textdegree. \label{fig:dis-model}}
\end{center}
\vspace{-1.5em}
\end{figure}

To recover the shape of this curve, the angle between the plane of the terminator and the plane of the orbit must be computed. At the equinoxes, and by definition, the axis of rotation of the Earth is contained within the plane of the terminator. During the solstices, the terminator and the rotation axis subtend an angle of 23.5\textdegree\ \citep{meeus1988astronomical}. 
In between those two extrema, this angle $\phi$ is given by a sinusoid function of the form $\phi = A \sin\left(t\cdot\frac{\pi}{2T}+\varphi_0\right)$ where $T\sim92$ days is the duration of one season, $A=23.5^\circ$ and $\varphi_0$ changes the date of reference. Sun-synchronous orbits are very highly inclined. They are retrograde orbits ranging between 620 and 800 km show inclinations of respectively 97.5\textdegree\ to 98.6\textdegree . Hence the plane of the orbit and the terminator are separated by (for 800 km) 14.9\textdegree\ during the Northern summer and 32.1\textdegree\ in the winter. This angular separation $\beta$ is null during the minima of the stray light flux which yields a behaviour of the form:
\begin{equation}
 \beta(t) = -A\sin\left( t\cdot\frac{\pi}{2T}+\varphi_0 \right)-\left(i-\frac{\pi}{2}\right)
\end{equation}
where $i$ is the inclination of the plane of the orbit with respect to the equator.
The angle $\beta(t)$ is assumed to be equal to $\vartheta+\pi/2$ hence equation \ref{eq:dis-defn-flux} is now:
\begin{equation}
  F(t) = \int_{\Omega}I\sin[\beta(t)]\text{d}\Omega
\end{equation}
The solid angle d$\Omega$ remains to be computed. 
To do so, it is assumed that the averaged angle between the plane of the terminator and LOS is $\pi/2-\beta$ (see Fig. \ref{fig:dis-model}). The two latest assumptions can be justified by asserting that (1) there is more stray light when $\beta$ is large and (2) the observable region is away from the seen illuminated part of the Earth. 

\begin{figure}[h]
\begin{center}
\includegraphics[width=1\linewidth]{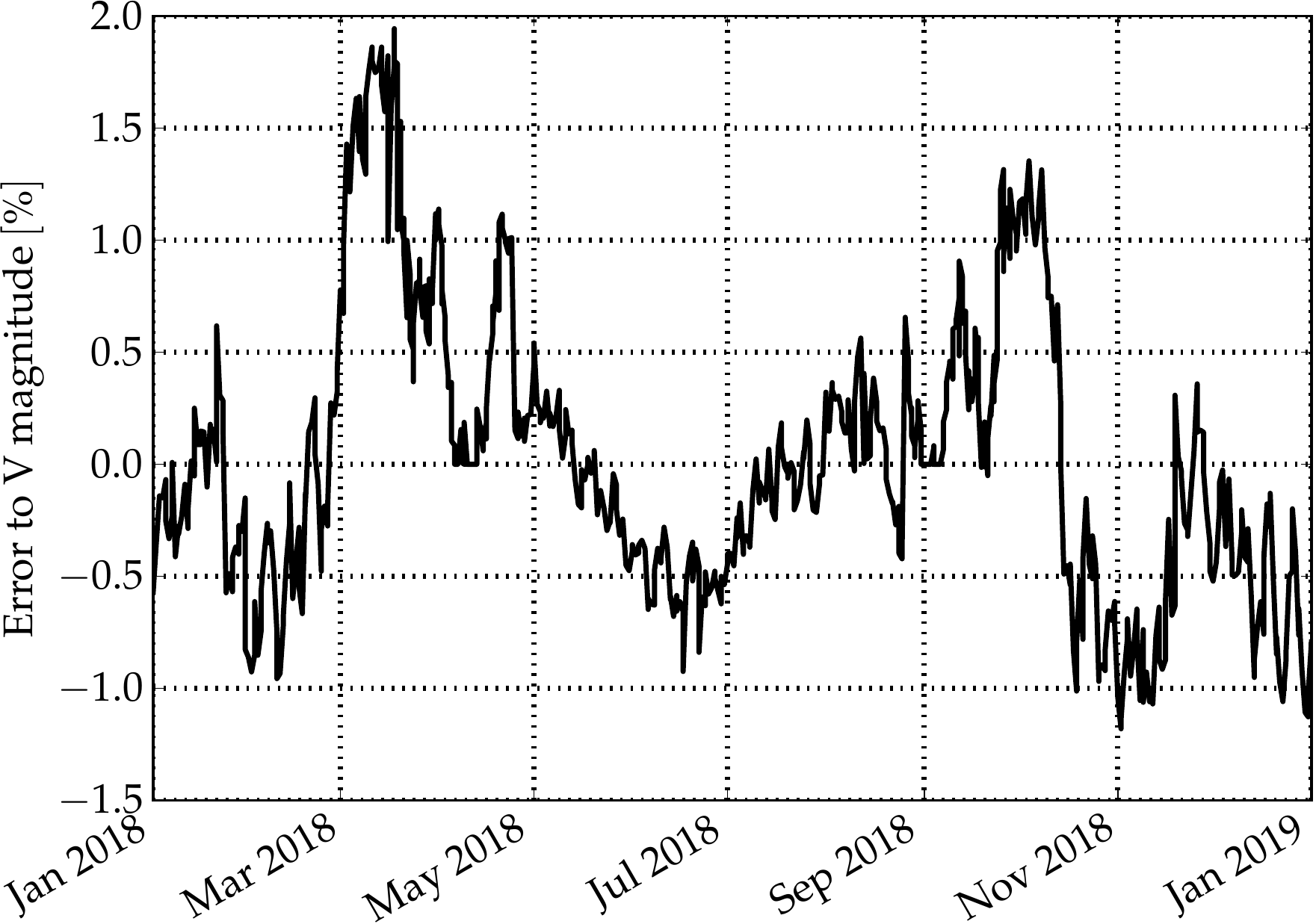}
\caption{Residuals in percentage of equivalent magnitude from the fit of the data for a 35 degree stray light exclusion angle, orbit of 800 km and RUAG PST. Negative values mean overestimation of the stray light flux.\label{fig:dis-residuals}}
\vspace{-1.5em}
\end{center}
\end{figure}

Only the behaviour of the flux will be modelled, not its value which depends upon many different variables. Taking a fixed direction for the LOS is equivalent of doing the integral on all infinitesimal solid angles:
\begin{equation}
  F(t) \propto \int_{\text{d}\Omega}\sin[\beta(t)]\text{d}\Omega\approx\sin[\beta(t)]\overline{\Omega}
\end{equation}
where the symbol $\overline{C}$ must be understood as the average of the quantity $C$. The (averaged) solid angle represents the amount of illuminated Earth as seen by the satellite. 
The surface of the Earth is projected onto a flat surface. It is $\propto \sin\beta$ (the time argument is dropped for clarity; the blue line in Fig. \ref{fig:dis-model}). This surface has to be projected to a surface which is perpendicular to the LOS (green and then red lines on the sketch) yielding an averaged solid angle:
\begin{equation}
 \overline{\Omega} \propto \sin^2\beta\cos\beta
\end{equation}
which, combined with the expression of the flux, gives
\begin{equation}
 F \propto \sin^3\beta\cos\beta
\end{equation}
This expression still needs to be scaled. In this work, the maximum value of the flux is used. The date of this maximum is easily determined to be at the winter solstice. The scaling depends upon the season to account for the deviation from the averaged LOS. Indeed, the Sun exclusion angle is further away when over the terminator hence a wider observable region. 
A simple model in the form of a single multiplicative factor of 1 in the winter (September to April) and 1/2 in the summer (April to September) is used such that the following expression fits the data reasonably well (less than 5\% -- Fig. \ref{fig:dis-residuals}):
\begin{equation}
 F(t) \propto \sin^3\beta(t)\cos\beta(t)\times\begin{cases}1\ \text{winter}\\\frac{1}{2} \ \text{summer}\end{cases}
\end{equation}
It has to be noticed that this function fits the \emph{mean} flux and not the maximal flux or the flux to a certain star throughout the year.


\subsection{Different Point Source Transmittance Functions} \label{sec:dis-PSTs}
The PST used for the previous section -- the RUAG PST -- does not reflect the latest optical design. Indeed, this PST was computed before CHEOPS was selected by ESA. In the mean time, another team took over the optical design (INAF, see \S\ref{sec:theo-sl-CHEOPS}). The PST emerging form the altered optical design is somewhat similar to RUAG's (Fig. \ref{fig:numerics-ruag-pst}). The design is not finalised yet and thus its PST can still evolve. Thus the RUAG PST was used as a baseline in this work.

To understand the effect of a changing PST, two different functions were used: the latest INAF PST and a worst case scenario suggested by a member of the INAF stray light team. 
The former, while being quite similar to the estimation of RUAG, has a peak around the off-axis angle of 35\textdegree\ which may well disappear in the final design while the later -- the ``worst case'' is the RUAG's PST multiplied by 10 in the range from 35 to 60\textdegree. This effect is difficult to predict as the PST is not scaled by the same factor for every angle. The dependence of the flux on the PST is of 
course great as it is a multiplicative factor. Low Earth-LOS angles are certainly more common than high angles due to the presence of the Sun exclusion angle which limits how the satellite can look away from the limb of the Earth.
The two above different PST -- the INAF and the worst case ones -- are both equal to the original PST at high angles.

\paragraph{INAF PST.} The simulations yield that, indeed, INAF PST yield stray light fluxes that are quite similar to RUAG's while being slightly worse than RUAG. 
At an altitude of 800 km for the orbit, the simulation look like the 620 km case for the mean stray light flux. However, maximum values reach up to 1.4 photon per pixel per second around the Northern winter solstice, again it is most certainly due to the spike around 35\textdegree.

\paragraph{Worst Case Scenario.} The worst case simulation yield fluxes that are close to 10 times as high as in the initial simulation. Maximal values of the stray light flux over one orbit are below the threshold value only during a 6 months period centred around the June solstice. A telescope with this PST could observe only very bright star if a noise of 10 ppm is wanted!
                                                                                                                                                                                                                                                                                                                                            
\paragraph{} Altering the PST has a great impact on the results. The closer the changes are to the 35\textdegree\ minimum exclusion angle, the greater the effects. Even though the latest estimation is close to the original PST, it would restrict further the visible zone in the sky during winter as it implies fluxes higher than the requirements for the 800 km case already. The worst case scenario yield maximal value of the stray light which are much higher than what could be tolerated. The observable zone would be very narrow in winter which may force to observe only very bright targets or/and short period planets.
\subsection{Smaller Stray light Exclusion Angles}
A way to increase the visible region without an altitude increase is to reduce the stray light exclusion angle. It is clear that this would not increase the observation capabilities for faint stars, but rather for the bright and very bright stars ($V<9$ and $V<7.5$). In order to study the effect of the stray light angle, this exclusion angle is reduced by slightly more than the equivalent angular separation of the cell. The new stray light exclusion angle is set to $25^\circ$.

It is clear that the different means used will increase as the added points are close to the limb of the Earth and thus higher stray light flux. An analysis must be performed \emph{a posteriori} to ensure that the restriction on the stray light flux -- namely the 1 photon per second per pixel or any other constraint on the equivalent stray light magnitude -- is respected and that CHEOPS is allowed to look at those regions. This is particularly useful in winter for the maximal value of the stray light. Such post-treatment of the data is actually required for every orbit to consolidate the observable zone or simply to construct the schedule of observations. This aspect is described and discussed in some details in \S\ref{sec:dis-targets-available}. 
It must be borne in mind, that reducing the stray light exclusion angle increases the viewing zone, but not necessarily the amount time it is visible. What changes most is that the visibility region for bright stars increase. Indeed, a lower exclusion angle allows to observe in regions that will produce more stray light.

Even if the effects are very clearly seen when computing the different maps, it can be said that increasing the number of visible points does increase the computation time as well as the treatment time. The computation time is affected in two different ways: (1) the number of points being larger, it takes more time to compute all of them and (2) the software uses the maximal difference between two orbits (see \S\ref{numerics:functions}). As the stray light exclusion angle is lower, the value of the stray light flux from one orbit to the next computed can be larger than the threshold value. Therefore, as the variability of the stray light flux is higher, more orbits are computed which also increases the computing time.

\subsection{Sensitivity to Other Parameters} \label{sec:dis-sensitivity}
A question about the sensitivity to the parameters (other than the PST and the altitude) used in this project can be paused to evaluate the robustness of the results. This section is divided in two parts: first the variation of the parameters used for the stray light flux calculations and then for the analysis. To evaluate the difference in limiting magnitude $\Delta m$ that two different fluxes would produce, the Pogson law is used. It indicates that (given two fluxes $f_1,f_2$):
\begin{equation} \label{eq:ratio-flux2mag}
 \Delta m = -\frac{5}{2}\log_{10}\left[ \frac{f_2}{f_1} \right]
\end{equation}

\paragraph{Stray Light Computations.} 
\subparagraph{Albedo.}\vspace{-1.2em}
In the stray light code, the albedo $A$ is taken to be one to yield a worst case scenario. This factor is a multiplicative factor in the computation of the stray light as:
\begin{equation}
 F_\text{sl} = S_V\int_{\Omega({\text{i,s,}\oplus})} A\cos\vartheta\cos\varphi\text{d}\Omega
\end{equation}
where $\Omega({\text{i,s,}\oplus})$ is the illuminated Earth visible from the satellite, $S_V$ the Solar flux in the relevant band, $\vartheta$ the angle LOS--illuminated surface element, $\varphi$ angle direction of Sun--normal of surface element (See \S\ref{app:great-circle-distance}). Hence using eq. \ref{eq:ratio-flux2mag}, it reveals that changing the albedo to an averaged value of 0.35 \citep{Pater2001} would push the limiting magnitude by about 1.1 magnitude fainter. Around the poles, the albedo is higher due to the ice and the snow (around 0.9) while oceans or continents have a much lower albedo (down to 0.1).
\vspace{-1.2em}
\subparagraph{Number of Cells.}
To chose the minimal number of cells that would be sufficient to represent the Earth, a conversion study was performed. $1000\times 500$ proved sufficient and the expected changes in the flux if the number was to be increased would not be significant.
\vspace{-1.2em}\subparagraph{Finer Orbital Steps.} As stated before, the maximal difference in flux is ensured to be less than 5\% from one computed orbit to the next. This 5\% represent a change of magnitude of about 0.05 which is completely negligible with respect to the albedo.
\subparagraph{Finer Sky Grid.}\vspace{-1.2em} The number of points at which to look in the sky will not change the flux received at the satellite, but it will smooth the curve of the stray light flux (Fig. \ref{fig:all-altitudes-fluxes}) and also be more precise in the limits of the zone where faint stars can be observed (see next section on cumulative observations).
\paragraph{Stray Light Analysis.} In the conversion of the stray light flux to the limiting magnitude (eq. \ref{eq:num-mag2flux}), several parameters from the design of the instrument are taken into account.
\vspace{-1.2em}\subparagraph{PSF \& Aperture.} The most important one, as the band is clearly defined, is the aperture and the PSF. Changing the aperture of the telescope by 1 cm with a constant PSF increases the flux by $\sim13\%$ or a 0.13 mag brighter. Of course, the PSF is not finalised yet and therefore it is hard to predict the direction of the changes. The current trend is to increase the size of the PSF and have it as a top-hat function with edges not exactly perpendicular in order to take into account the jitter of the instrument. To get a more precise value therefore of the equivalent magnitude, a convolution of the signal is to be made. Given a change in what would be the equivalent PSF radius for a top-hat function of $\pm 20\%$ would lead to a change of $^{-0.20}_{+0.25}$ respectively in the magnitude.

\subsection{Stray Light Flux \& Short Term Variations}
As demonstrated in the previous sections, the stray light flux presents a seasonal variation. To show this, different means have been used; all of them taking the average over one orbit. How does the flux varies in one orbit? To answer this, it must be noticed that these infra-orbit variations are also subject to seasonal variations. However, it is still important to study them as the baseline time of an exposure is 60 seconds. Therefore, the background stray light flux will not be the same for the whole orbit. It can be reminded at this point that CHEOPS will be in a ``nadir-locked'' attitude (\S\ref{sec:CHEOPS}), which will effectively imply that the satellite always present the same side to the Earth. Therefore the gradient of the stray light flux will remain in the same direction.
\paragraph{}
To predict the behaviour of the stray light flux, the coordinates of a target must be known. The assumption is made that it is visible for the whole orbit. Also, the Earth is close to the December solstice (where the stray light flux is the highest). There are therefore two clear distinctions that will arise: (1) the satellite is above the night side and looks to a target whose projection onto the Earth is also in the night and (2) the satellite is above an illuminated portion of the Earth.
The stray light flux will therefore peak when the satellite is the furtherest above the day light. The increase or decrease of the flux may not be continuous as the relative alignment Earth--LOS--Sun may imply local maxima of the stray light flux. 
The range of flux that can be reached by the stray light can span by several equivalent star magnitudes. This implies that (1) the whole orbit must be assessed with a reasonable discretisation in time (60 seconds as for the baseline frame exposure) and (2) this study on the stray light must be done in order to assess the observability -- or rather the quality of the SNR -- for a given object. 

\section{Cumulative Observations} \label{sec:dis-yearly}
The cumulative observations maps represent the amount of time a ``worthy'' observation can be made in a given direction over the year. What does ``worthy'' mean actually depends upon the targets observed. For example, the observation of a bright ($V<9$) target could require observations of at least 50 minutes per orbit. On top of this, during acquisition process -- the time required to point the satellite and start the observation --, it is assumed that the object must be visible. Any other shorter observations that are not taken into account.

The constraint on a minimum observation time is depicted in figures \ref{fig:results-800-50-1} through \ref{fig:results-800-others}. Potential observations of less than the set observation time plus the acquisition delay (fixed to $t_\text{ac} = 6$ min throughout this project) were discarded. Observations that meet the requirements are summed to obtain a map of the regions that can be observed most during 2018. Three different criteria are applied: (1) the altitude, (2) the presence of the SAA and (3) the constraint on the magnitude of the target. 
\subsection{Shapes \& Observation Time of the Cumulative Maps}
\begin{figure*}[!h]
\begin{center}
\includegraphics[width=0.49\linewidth]{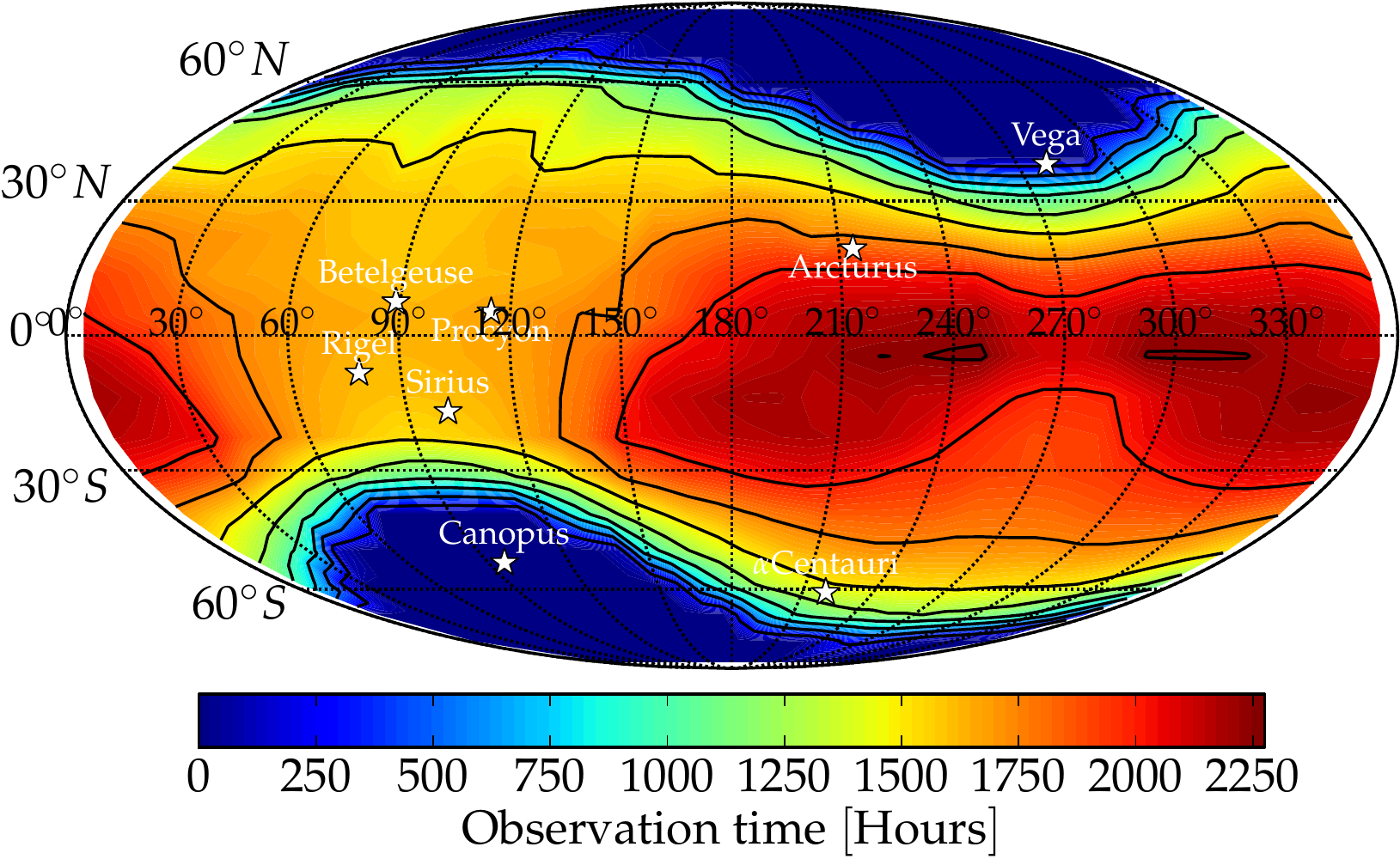} 
\includegraphics[width=0.49\linewidth]{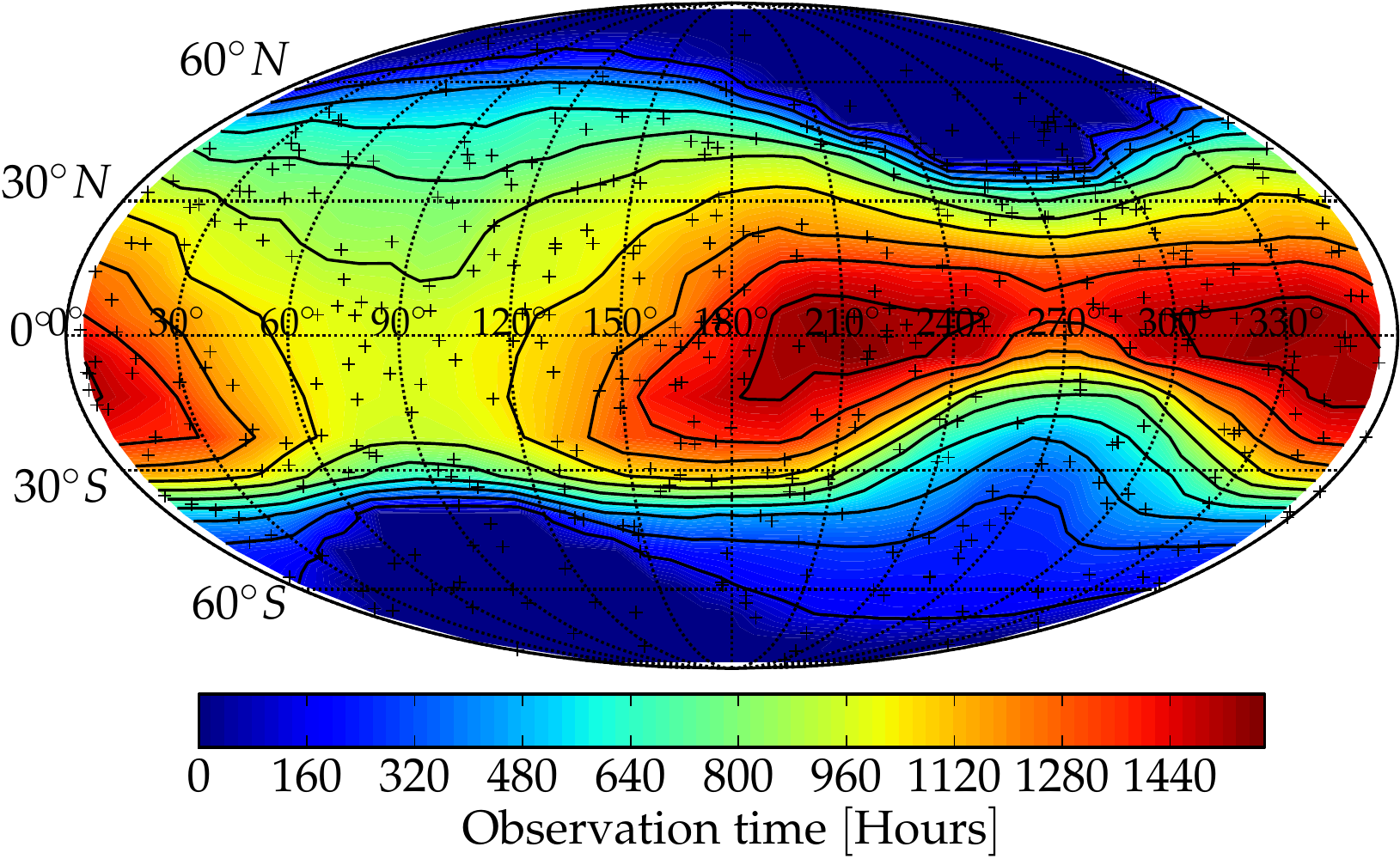}  

\caption{Cumulative observation time at 800 km. Acquisition time set to 6 minutes. -- \emph{Left.} 50 minutes of observation without any other constraint. -- \emph{Right} 50 minutes of observations with SAA interruptions. There is a global loss of observation time as well as regions that become barely visible. The crosses represent the currently known exoplanets.\label{fig:results-800-50-1}}
\vspace{2em}
\includegraphics[width=0.49\linewidth]{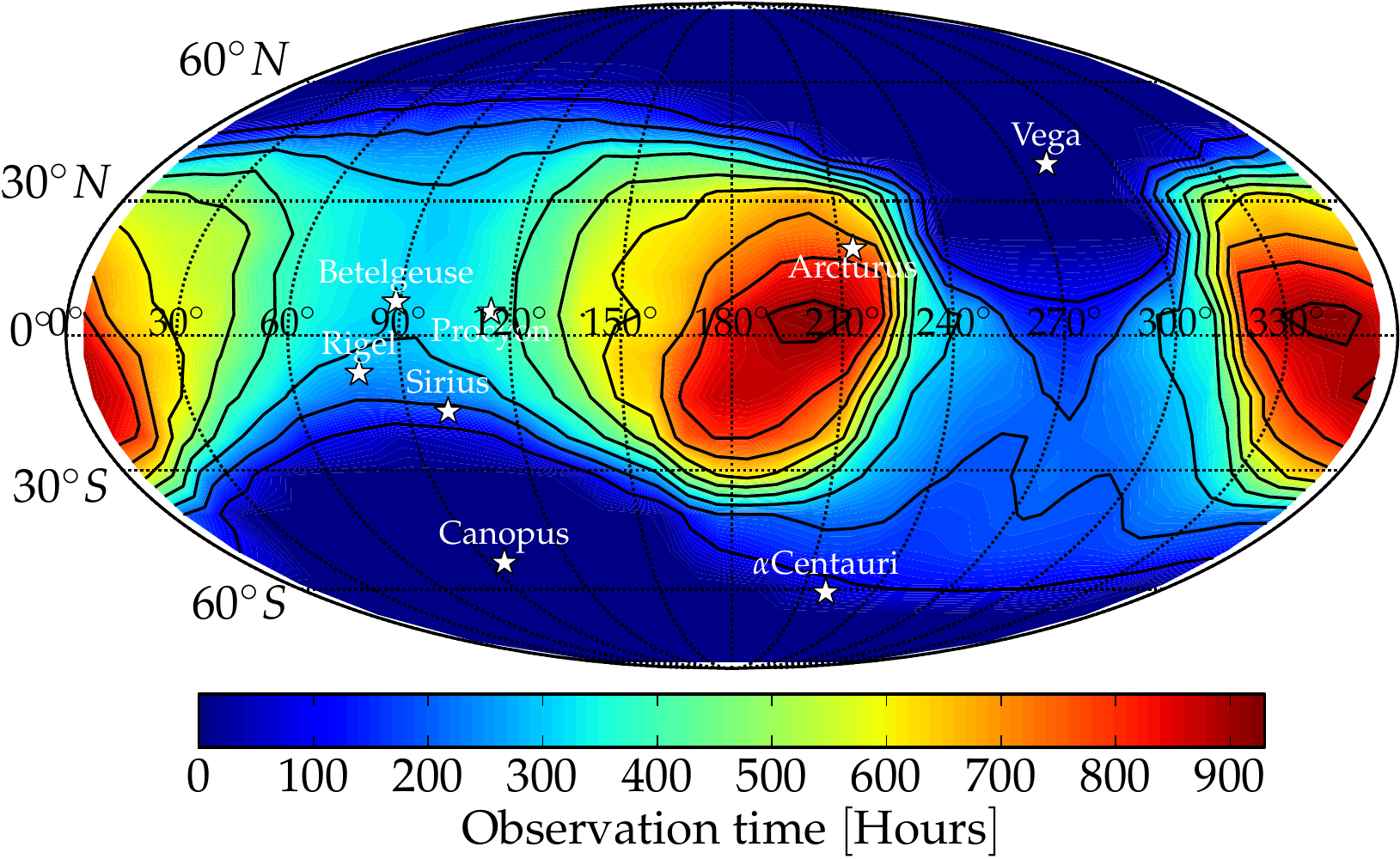} 
\includegraphics[width=0.49\linewidth]{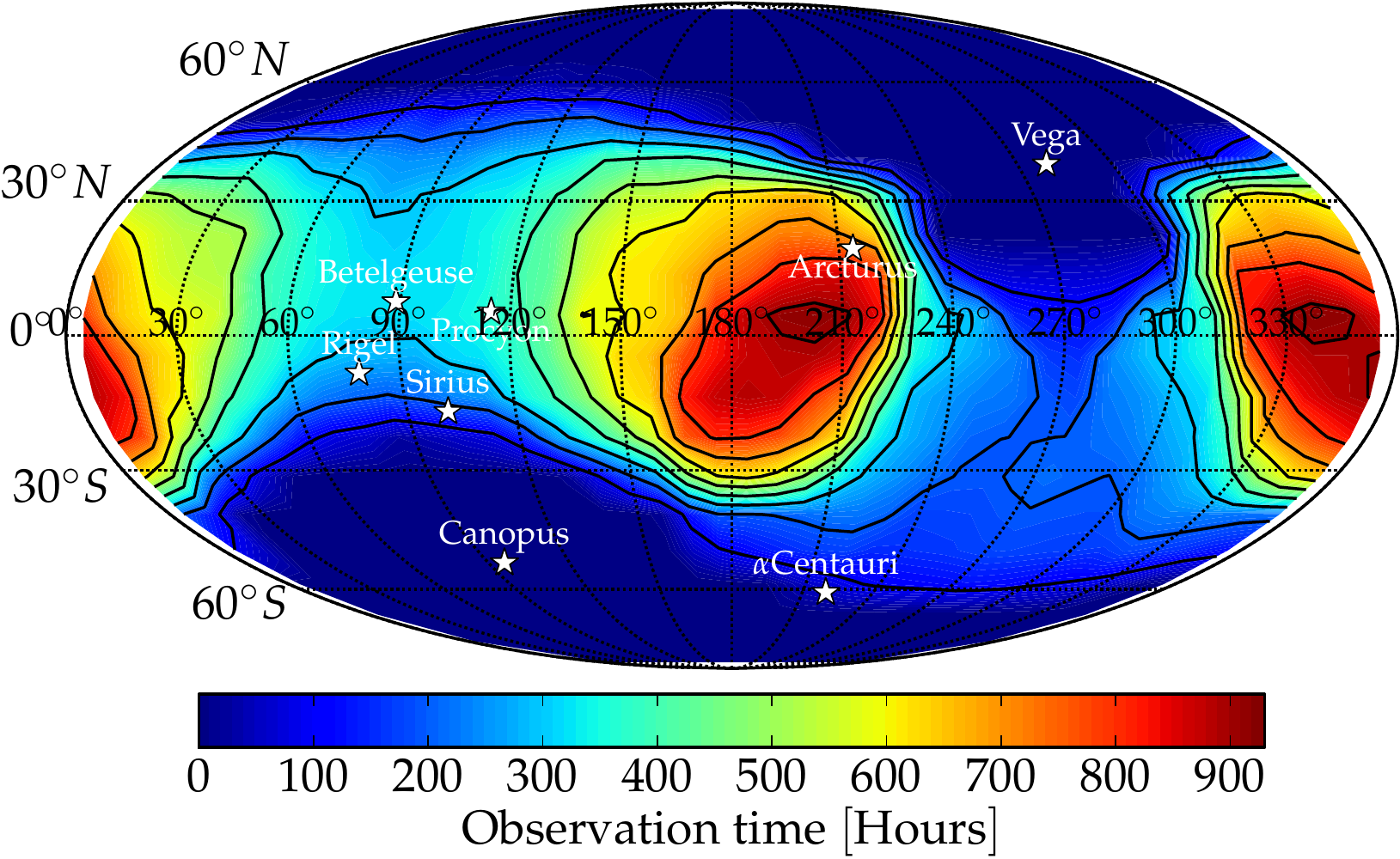} 

\caption{Observation maps including SAA interruptions. -- Cumulative observation time at 800 km. Acquisition time set to 6 minutes. -- \emph{Left.} 50 minutes of observation of faint stars (12.5 mag). -- \emph{Right} INAF PST for faint stars (12.5 mag).\label{fig:results-800-50}}
\vspace{2em}

\includegraphics[width=0.49\linewidth]{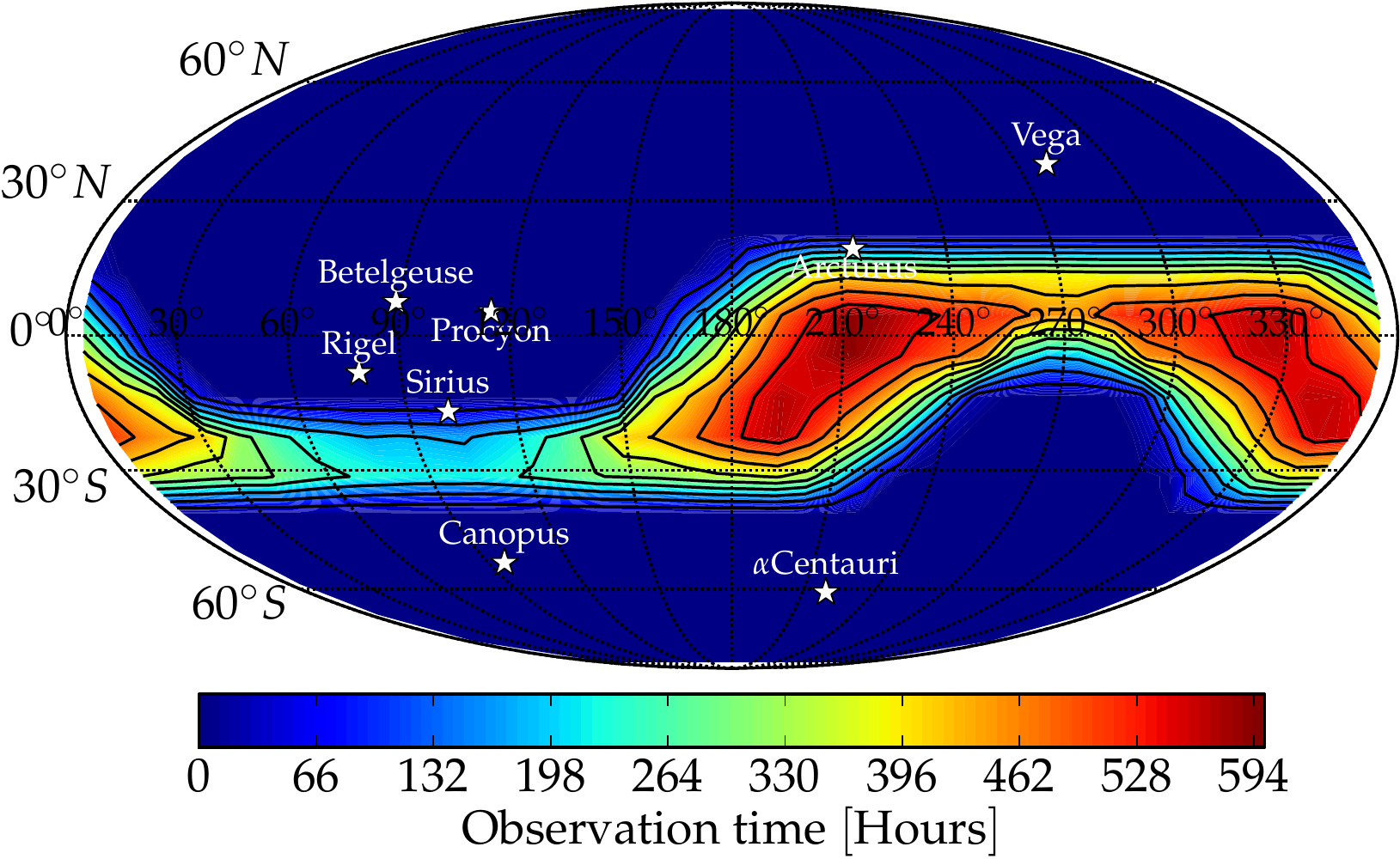} 
\includegraphics[width=0.49\linewidth]{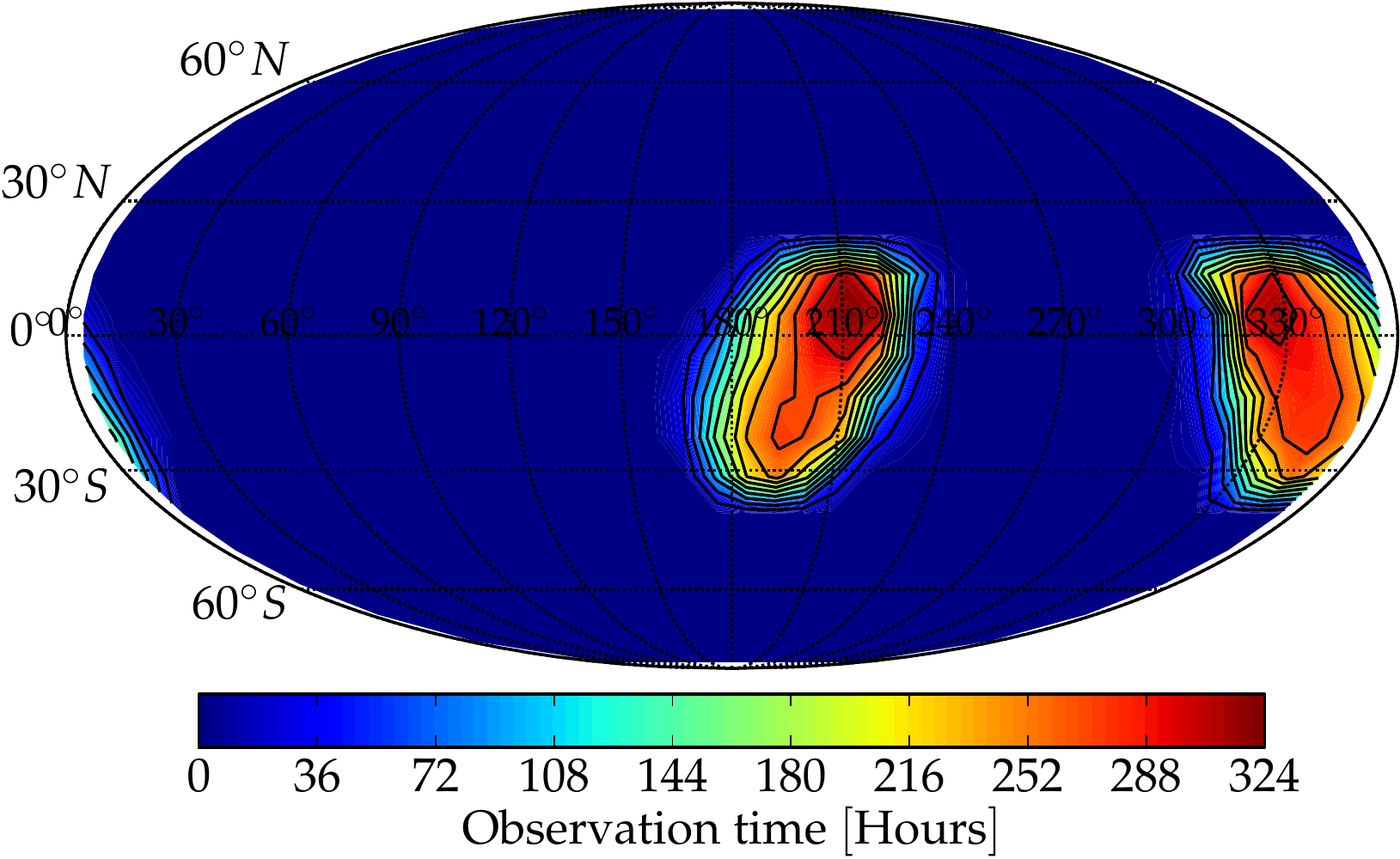} 
\caption{Long observations of 80 minutes minimum with SAA interruptions -- Cumulative observation time at 800 km. Acquisition time set to 6 minutes. -- \emph{Left.} No restriction for the magnitude of the target star. -- \emph{Right} Observation of faint stars (magnitude 12.5). Most of the observation time is in the Southern hemisphere.\label{fig:results-800-80}}
\end{center}
\end{figure*}

In figures \ref{fig:results-800-50-1} through \ref{fig:results-800-others}, the observable zone will move from left to right with time (during the winter solstice, the observable zone is centred around $\alpha = 90^\circ$). The cumulative observation time varies throughout the year as the position of the satellite relative to the day--night terminator changes. As explained in the previous section, there are moments during which the planes of the terminator and the orbit coincide. It is during this time that the cumulative observations are the longest. 
This is due to the fact that the constraint due to the Earth stray light exclusion angle is not applicable as the satellite orbits right above the closest illuminated part of the Earth and observes above the night side. Even if the coincidence of the two planes does not last long, as the alignment is getting better, the observation time increases over a rather long period (at least three to four weeks). As the Sun exclusion angle is a constraint that bears the shape of a sine, it is symmetric. Indeed, the summer solstice is flanked by two regions of favoured visibility: the ``butterfly wings''. 
They can clearly be seen in Fig. \ref{fig:results-800-50} and they are also represented in the stray light minima in figures \ref{fig:all-altitudes-fluxes}. These yearly cumulative maps are biased by the observation time: for long observation times, those butterfly wings migrate towards positive declinations. The SAA has the same effect on the position of the wings. Shorter observation times can be achieved before this optimum and therefore the wings extend further away in right ascension. During this optimum, the stray light is less of a constraint as the telescope can always look away from the day side of the Earth. Similarly, the summer and winter solstices are maxima in the stray light which are mirrored in the cumulative observations by the minima (around $90^\circ$ and $270^\circ$). 

\subsection{Effects of the Altitude}
As the satellite is placed on a higher orbit, (1) the viewing zone widens roughly\footnote{This behaviour is derived from trigonometric considerations of the satellite in orbit, its inclination and the Sun exclusion angle.} as $\sim\arccos(1/(R_\oplus+h))$ and (2) the orbital period increases as $P \propto (R_\oplus+h)^{3/2}$. Placing CHEOPS on a higher altitude has also a higher cost in terms of energy at launch and during orbit insertion. 
This implies crossing a more active and expanded SAA. The observable region clearly increase in size as the satellite is placed at higher altitude. The region that benefits the most of this increase in altitude is between $\alpha \sim 60^\circ$ to $120^\circ$. 
This region is observed around the winter solstice and therefore the constraint on the position of the Sun does not allow for observations at low declination angles while the constraint on the Earth eclipse remains the same. The overlapping of those two constraints results in this region being poorly observable.
\paragraph{}

The South Atlantic Anomaly impedes the observation capabilities for durations that vary between 0 and 20 minutes (at 800 km). The duration of the cut-off will be longer at higher altitude as the satellite enters the inner Van Allen Belts further. The shortest time between two crossings of the SAA is an orbital period minus the maximum duration of the SAA crossing. This yields typically 80 minutes. Therefore for an observation time of minimum 100 minutes, the observable regions are restricted to a few ``islands'' of little amplitude (at 800 km: maximum 600 hours against 1600 for a 50-minute long observation). However, the shape of the observable regions is strongly related to the SAA: some targets appear shortly before crossing the SAA and cannot be observed for the required amount of time. Hence they do are not count as one observation.

\subsection{Effects of the Stray Light}
The last constraint that is added on those observations is the stray light flux. As discussed before, the stray light flux was computed using a dedicated code and uses as an input any region of observation for a given minute. For the CHEOPS mission, the magnitude of the stars to be observed is in the range of 6 to $\sim12.5$ in the V band. The $35^\circ$ stray light angle represents an exclusion for the obviously too high fluxes of stray light and does not take into account the magnitude of the target. 
The purpose of this work is to compute the value of the stray light at any time for any allowed direction. 
The impact of the stray light for magnitude 9 or brighter is negligible for most of the cases. However when approaching the observation limit, the stray light may become prominent (\emph{e.g.} Fig. \ref{fig:results-800-80}). 
It is important to bear in mind that what is measured here is the photometric flux of the target to the best accuracy possible to detect minute flickering of light. An Earth-like planet around a Sun-like star presents a dimming of a few tens of parts per millions (ppm). The maximum tolerated contamination due to stray light is set to 10 ppm (see \S\ref{numerics:functions}). 
This contamination is maximum at winter and summer solstices. This effect is strong and reduces for an orbit of 800 km, the cumulative observation time by nearly half with the favoured butterfly region (for 50 minute observations). The region of the summer solstice is also much impacted. If the interruptions due to the SAA are added, it becomes nearly invisible. If the constraint on the observation period is increased to 100 minutes, no targets can be observed in this area. 
\paragraph{}

\begin{figure*}[!h]
\begin{center}
\includegraphics[width=0.49\linewidth]{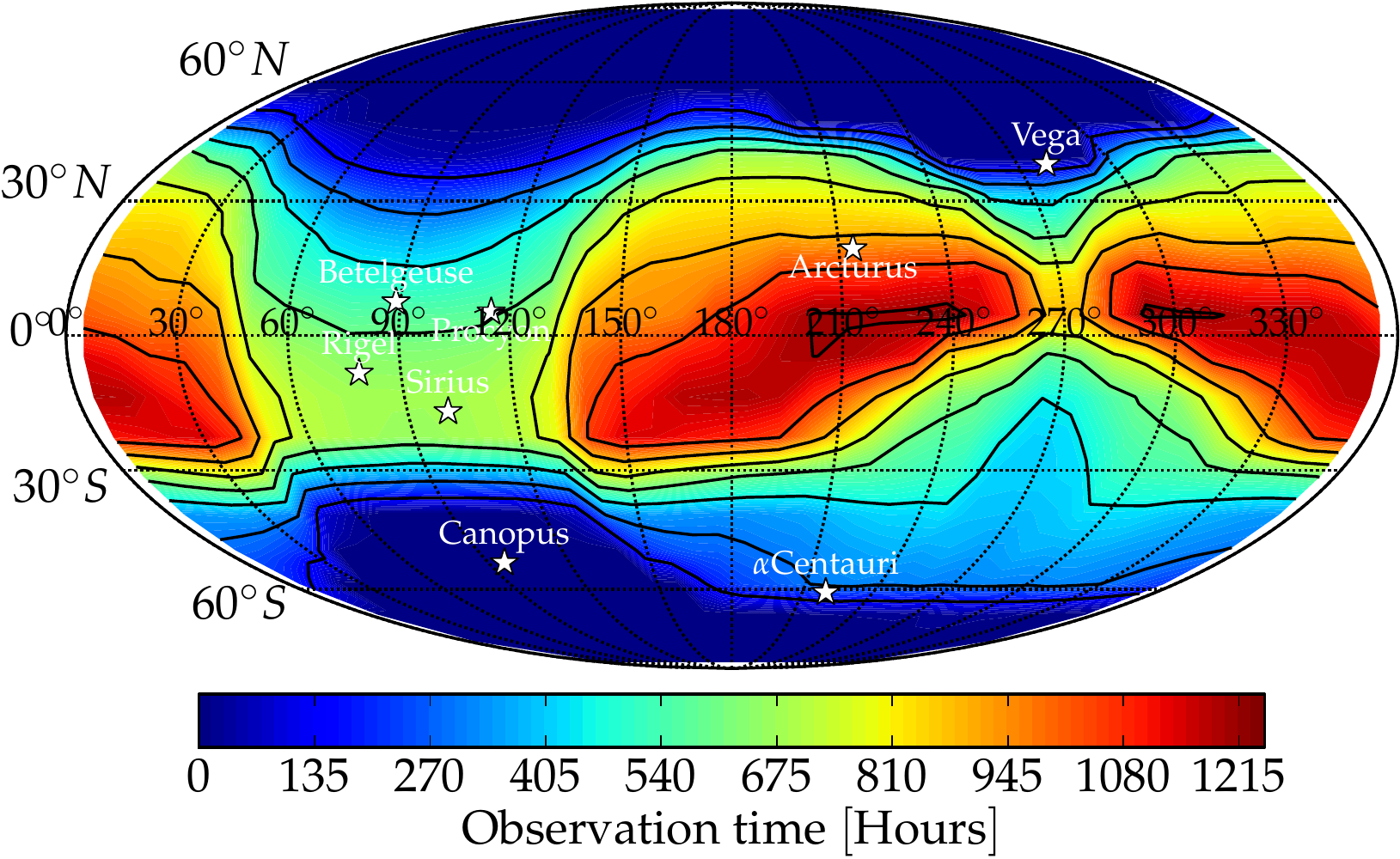} 
\includegraphics[width=0.49\linewidth]{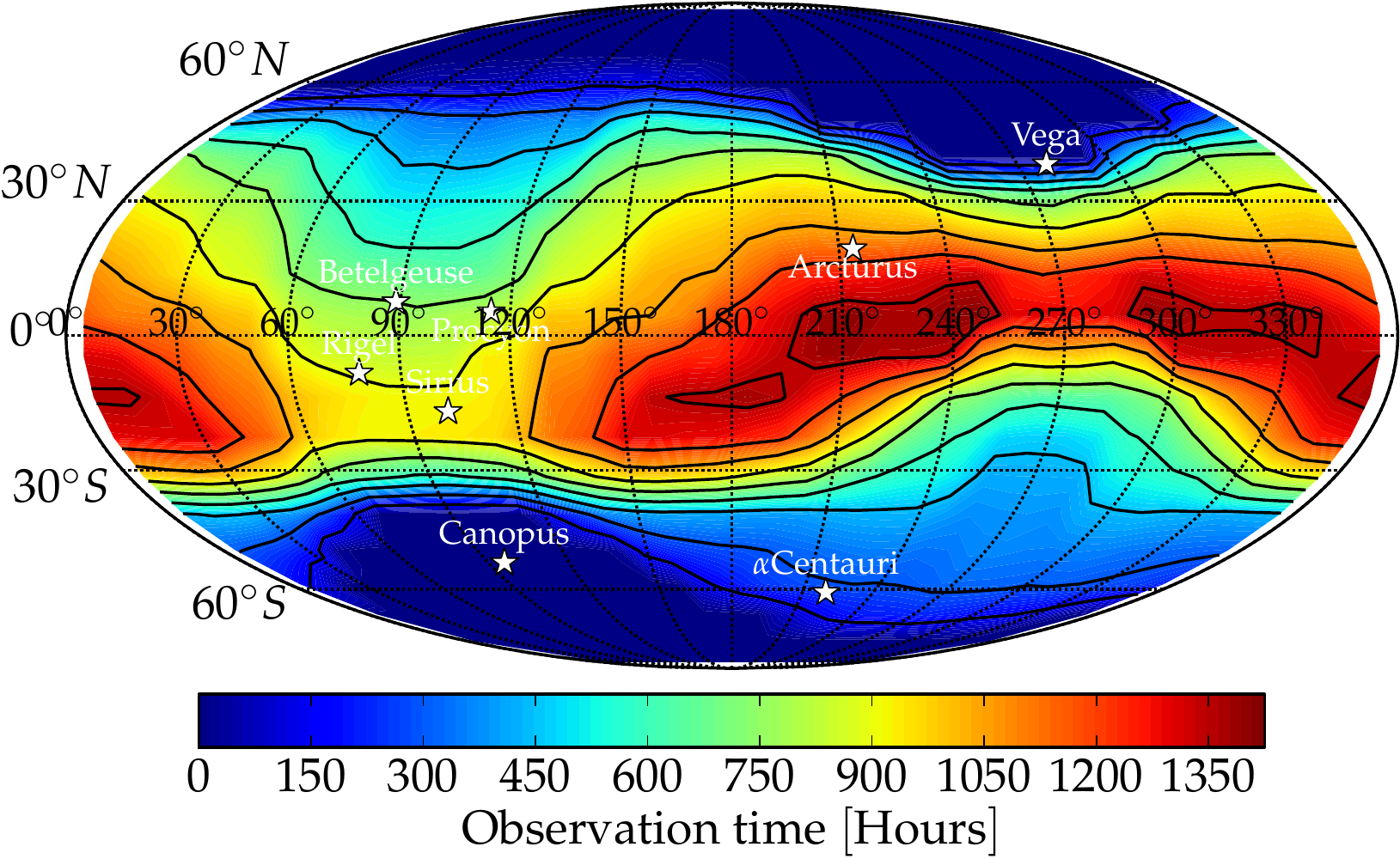} 

\caption{Cumulative observation of 50+6 minutes with the SAA. -- \emph{Left.} 620 km. -- \emph{Right} 700 km.}
\vspace{2em}
\includegraphics[width=0.49\linewidth]{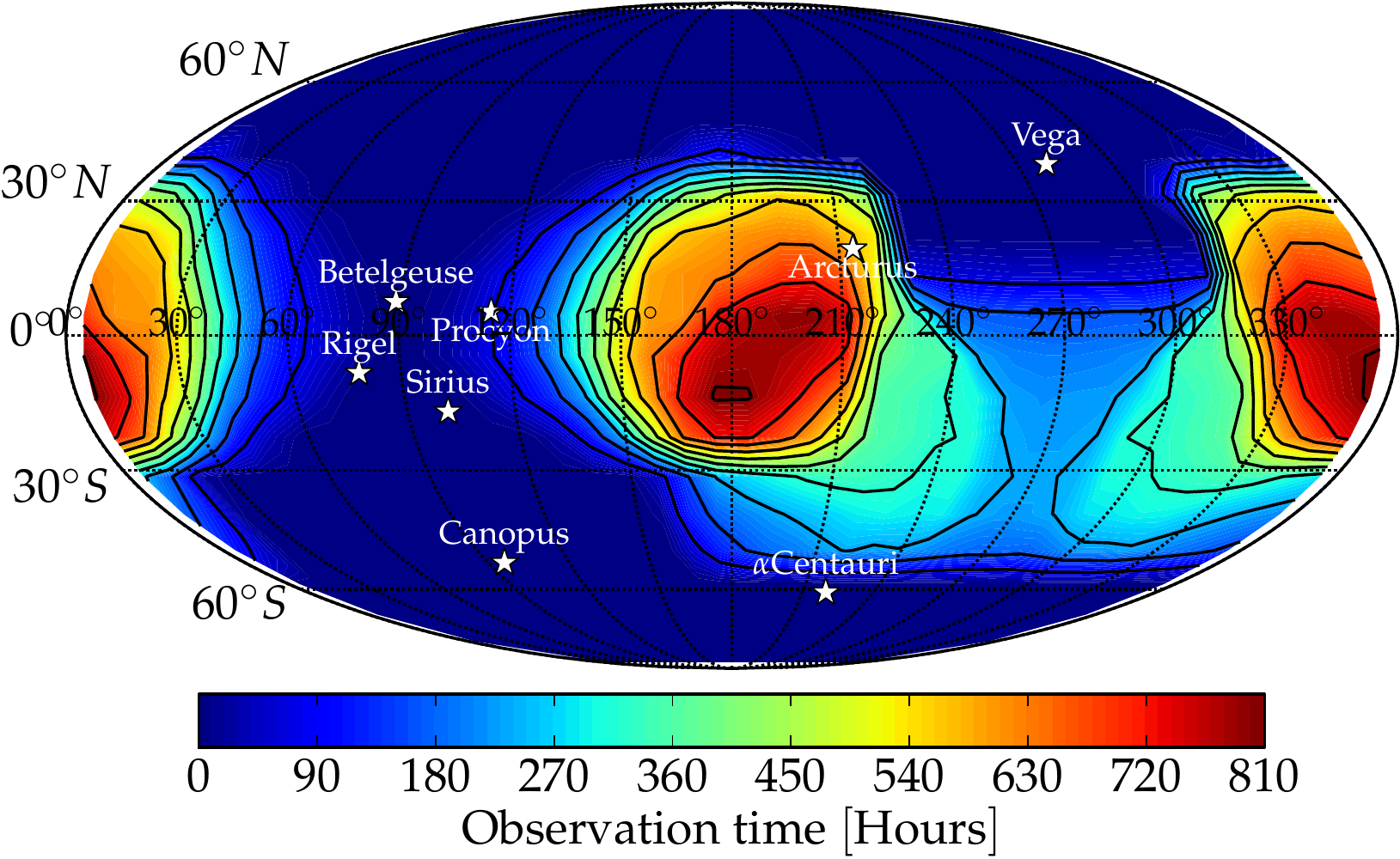} 
\includegraphics[width=0.49\linewidth]{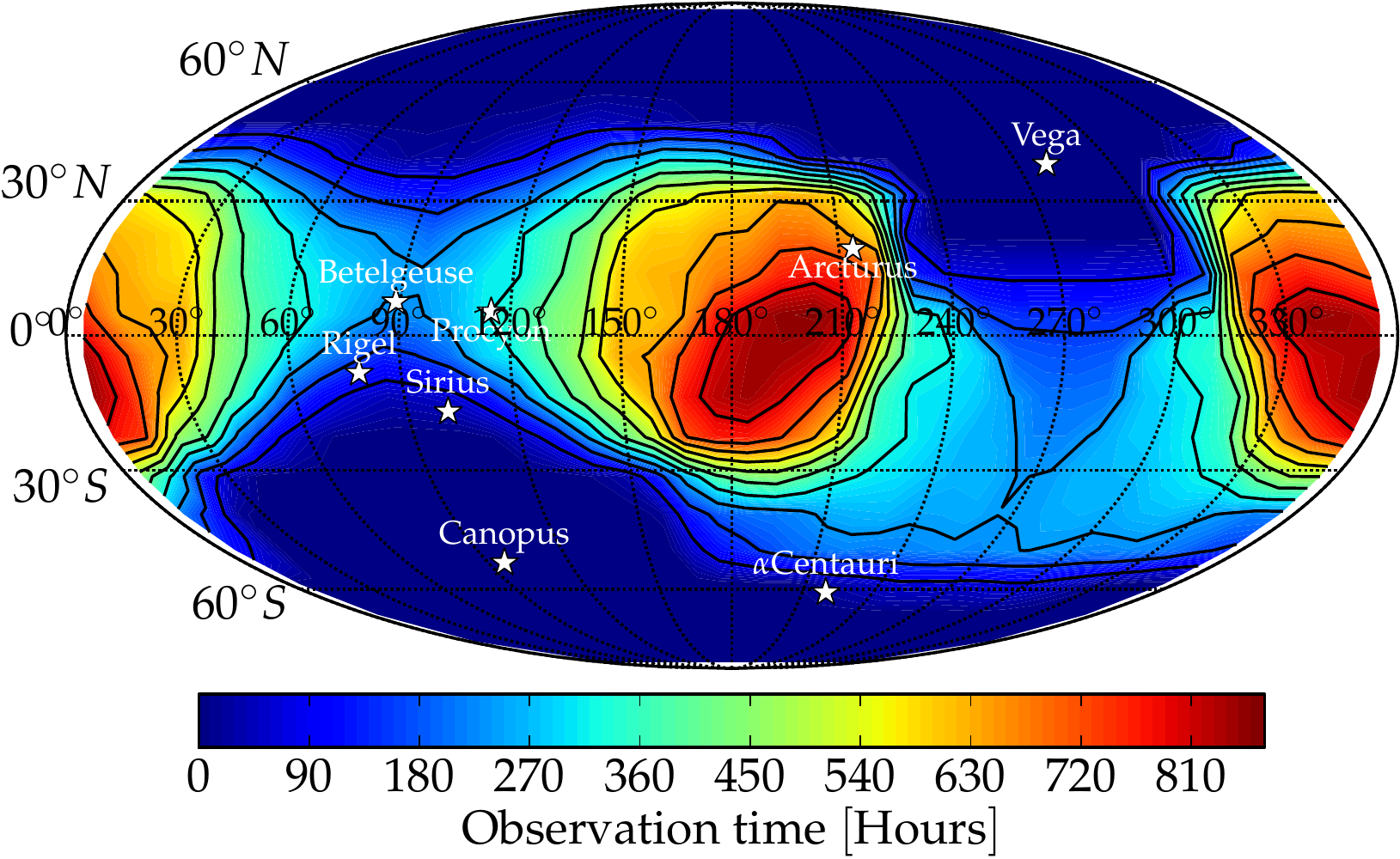} 

\caption{Cumulative observation of 50+6 minutes with the SAA for faint stars (12.5 mag\label{fig:result-faint-alt}). -- \emph{Left.} 620 km. -- \emph{Right} 700 km.}
\vspace{2em}

\includegraphics[width=0.49\linewidth]{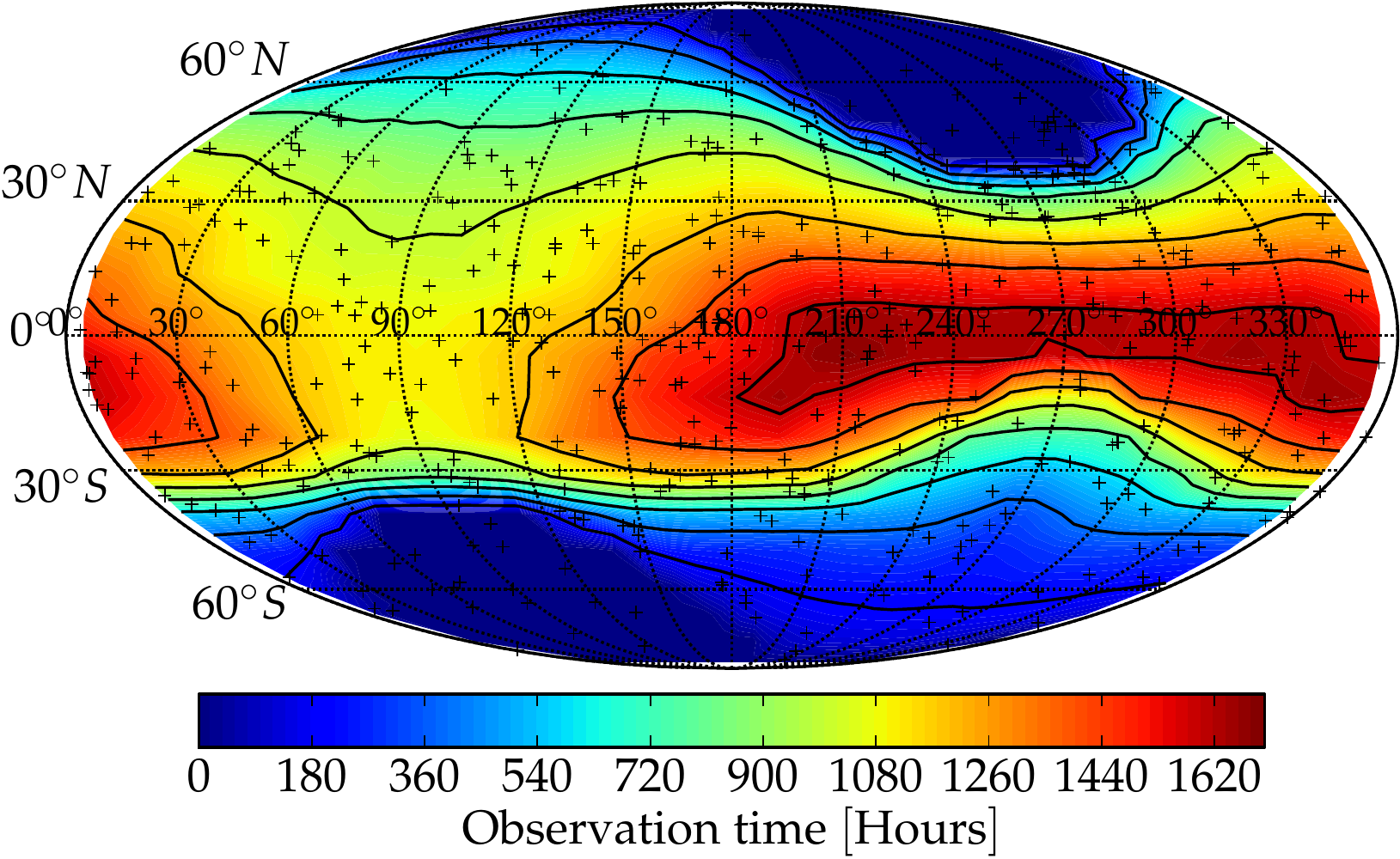} 
\caption{Cumulative observation of 50+6 minutes with the SAA at an altitude of 800 km. Stray light exclusion angle reduced to $25^\circ$ with a limiting magnitude of $V=6$. The crosses represent the currently known exoplanets. \label{fig:results-800-others}}
\end{center}
\end{figure*}

The effects of the observing time in a yearly cumulative map have been explained to some details, but there are other effects than the SAA and the amount of time spent observing. The stray light depends upon several parameters (altitude, stray light angle, PST,\dots) which can also be observed without surprises in the cumulative maps. It can clearly be seen that the altitude has an impact upon the observation time and affects the the shape. In Fig. \ref{fig:result-faint-alt}, the effect of the altitude on short observations of faint stars is depicted. The largest change (apart from the maximal observation time that increases from 810 hours to slightly more than 900) occurs in the winter: it is nearly impossible to observe in a region close to Betelgeuse. At 800 km, this region has become quite visible.
This is due to the angle subtended by the LOS to the illuminated cell becoming smaller. To contrast with this, regions between the two butterfly wings do not increase their visibility. They remain relatively shortly observable. 
The reason for this are that (1) the SAA always interrupts the observations and (2) $V=12.5$ is quite faint and therefore, it is difficult to observe such faint stars in negative declinations.
\paragraph{}

The effects of reducing the stray light exclusion angles have the expected outcome: bright stars can be observed longer while faint stars are not affected much. Bright stars observations can therefore be done during the solstices when faint stars are not well visible. Changing the PST to the INAF one has little effect on the global yearly cumulative observation time (Fig. \ref{fig:results-800-others}). 
Indeed, in section \S\ref{sec:dis-PSTs}, it was pointed out that the changes due to the change of PST from the older RUAG to the newer INAF would increase the stray light by between one and a half and two times which translates into changes of the order of half a magnitude. Therefore only subtle changes in the shape of the regions are seen. In case of the ``worst case'' PST, the change is dramatic as the equivalent limiting magnitudes is reduced by 2.5 mag and therefore, the visibility of faint stars at 800 km with this PST is worse than the RUAG/INAF ones at 620 km !

\subsection{Other Considerations}
One of the output of this work is the optimum altitude from the three Sun-synchronous orbit (SSO) at 620, 700 and 800 km. Purely from a viewing zone standpoint, the higher the better in every case. As the visible zone is defined by the constraints set in \S\ref{sec:numerics-visibility-constraints}, getting higher or/and relaxing the stray light exclusion angle will only increase the zone. Increasing the zone by reducing the exclusion angle will not improve the situation for the faint stars as those kind of observations are mostly constrained by the amount of illuminated surface of the Earth seen by the satellite. Adding the radiation issue to the picture makes the choice of the altitude more difficult. 
Indeed, the instrument cannot withstand a very high radiation level. Getting higher (up to 800 km) will also increase the time spent in the SAA and therefore the ionising dose received which may very well cause the satellite to fail earlier than its lifespan. 

It has to be noted that there will be a processing of the images taken by CHEOPS. The stray light will be subtracted hence loosening the consideration about the limiting magnitude. This will allow the observations of more faint targets. This analysis mirrors then a worst case scenario: that the noise from stray light cannot be removed.  
\paragraph{}
There are other considerations as well: CHEOPS is foreseen to be launched with another passenger on the rocket to reduce the cost of the launch. To accommodate more possible launch associates, the altitude could be chosen to be in the whole range from 620 to 800 km. Another important change to the orbital parameters could be the change of the LTAN from 6 am to 6 pm which would invert the maps: the butterfly wings would be centred around the winter solstice, which would mean that the Northern hemisphere would be easier to see with the disadvantage of losing observation time in the South\footnote{Most of the current ground capabilities in exoplanet exploration are located in the Southern hemisphere.}.
\paragraph{}

According to the science requirement 2.1 \citep{SciReq}, ``50\% of the whole sky should be accessible for 50 days of observation per year and per target with observation duration longer than 50\% of the spacecraft orbit duration''. To analyse the realisation of the percentage of the observable zone in the sky, the number of cells which can be observed for longer than 50 days with an efficiency of 50\% (or 600 hours) is divided by the total number of cells in the sky (800). Results, given in Tab. \ref{tab:dis-yearly-percent}, are not encouraging as this requirement is only met if the SAA is not considered. Reducing the stray light exclusion angle to $25^\circ$ at 800 km is the only way this requirement could be met. Another way to look at this requirement is to look at planets classifications and their minimal observable time. Tab. \ref{tab:dis-percentage-sky} shows the visible percentage of the sky in such cases.
\begin{table}[h]
 \begin{center}
 \begin{tabular}{r|rrrr} \toprule
$h$ [km] & 800 & 800 & 700 & 620 \\ 
$\alpha_\oplus$ [\textdegree] & 35 & 25 & 35 & 35 \\\midrule
   SAA   & 43\%& 49\%   & 39\%& 31\% \\
   No SAA& 62\%& 69\%   & 59\%& 50\% \\
  \bottomrule
 \end{tabular}
 \caption{Percentage of the sky that can be observed according to the yearly cumulative maps (RUAG PST). The first line gives the altitude $h$ of the orbit while the second yield the minimum angle from the limb of the Earth to the LOS. \label{tab:dis-yearly-percent}}
 \end{center}
\end{table}

\newpage \clearpage
\newpage \clearpage
\newpage \clearpage
\section{Targets Availability \& Scheduling} \label{sec:dis-targets-available}
\subsection{South Atlantic Anomaly} \label{sec:dis-SAA}
The South Atlantic Anomaly (SAA) is a region above the South Atlantic ocean where the magnetic field of the Earth is weaker for a given altitude than at a similar altitude around the globe. This perturbation is due to the fact that the Earth's dipole is slightly off-axis by about 500 km from the rotation axis \citep{Walt1994}. The magnetic field being weaker, more cosmic rays are measured in the SAA as the spacecraft enters the inner Van Allen belt. 
The flux radiation that suffers a spacecraft entering this zone is such that observations are to be stopped. Some of the failures of spacecrafts in the past are due to on-board electronics damaged by this augmented radiation \citep{Baker2005}. It is relatively easy to shield the electronics components against electrons. Protons, however, are difficult to stop and therefore much more dangerous for whatever -- or whoever -- that crosses the SAA. 

In this project, the SAA plays an important role. It is indeed impossible to observe stars during its crossing and thus the need to model the boundaries of this exclusion zone. The defining criteria is the sensibility to radiation of the satellite. At this point in the mission, this is not well defined and may evolve in the future due to technical constraints. 
The typical tolerance value given by the European Space Agency (ESA) is a proton flux of less than 2 protons of 50 MeV per squared centimetres and per second. Based on this value, the maximum crossing time\footnote{The crossing time varies from 0 to 20 minutes due to the parameters of the orbit which is not ground track repeating.} of the SAA is 20 minutes at 800 km altitude. This threshold value for the flux was adopted in this study, but the effect of different threshold values was investigated in order to find out the sensibility of the crossing time which impacts the observation time and therefore the schedule.

In 2018, the Sun will be in a minimum in its activity cycle of 11 years of periods which is a positive fact for the mission as the flux of protons will be smaller. The exclusion region will be reduced in comparison with Solar maximum. The model chosen to represent the SAA is the so-called AP--8 (minimum) \citep{Sawyer1976} which is widely used throughout the spatial community. This model is static in time -- \emph{i.e.} it does not produces ``SAA weather'' maps, but it can compute the extent of the zone at solar minimum or solar maximum.

\begin{figure}[h]
 \begin{center}
 \includegraphics[width=1\linewidth]{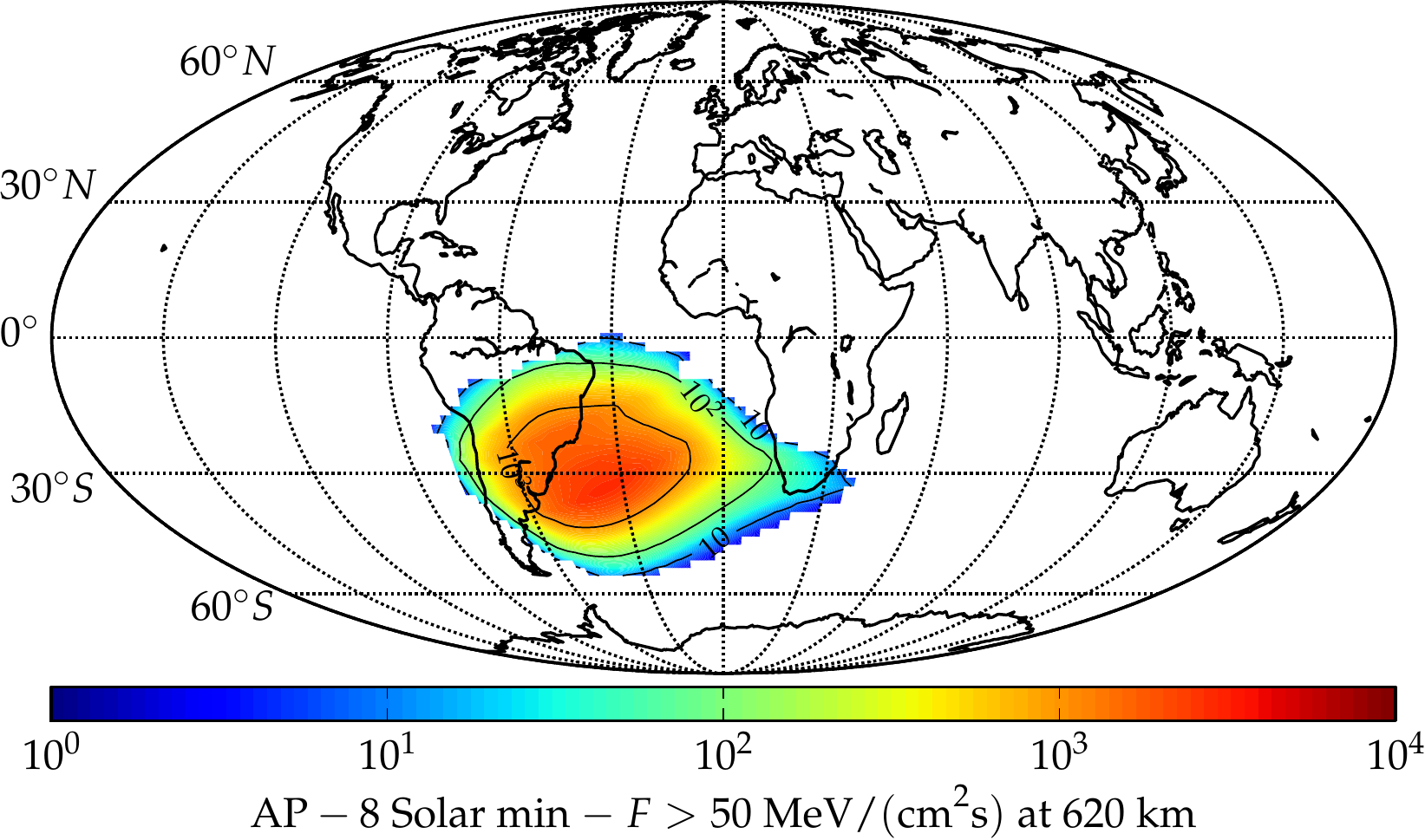}
  \includegraphics[width=1\linewidth]{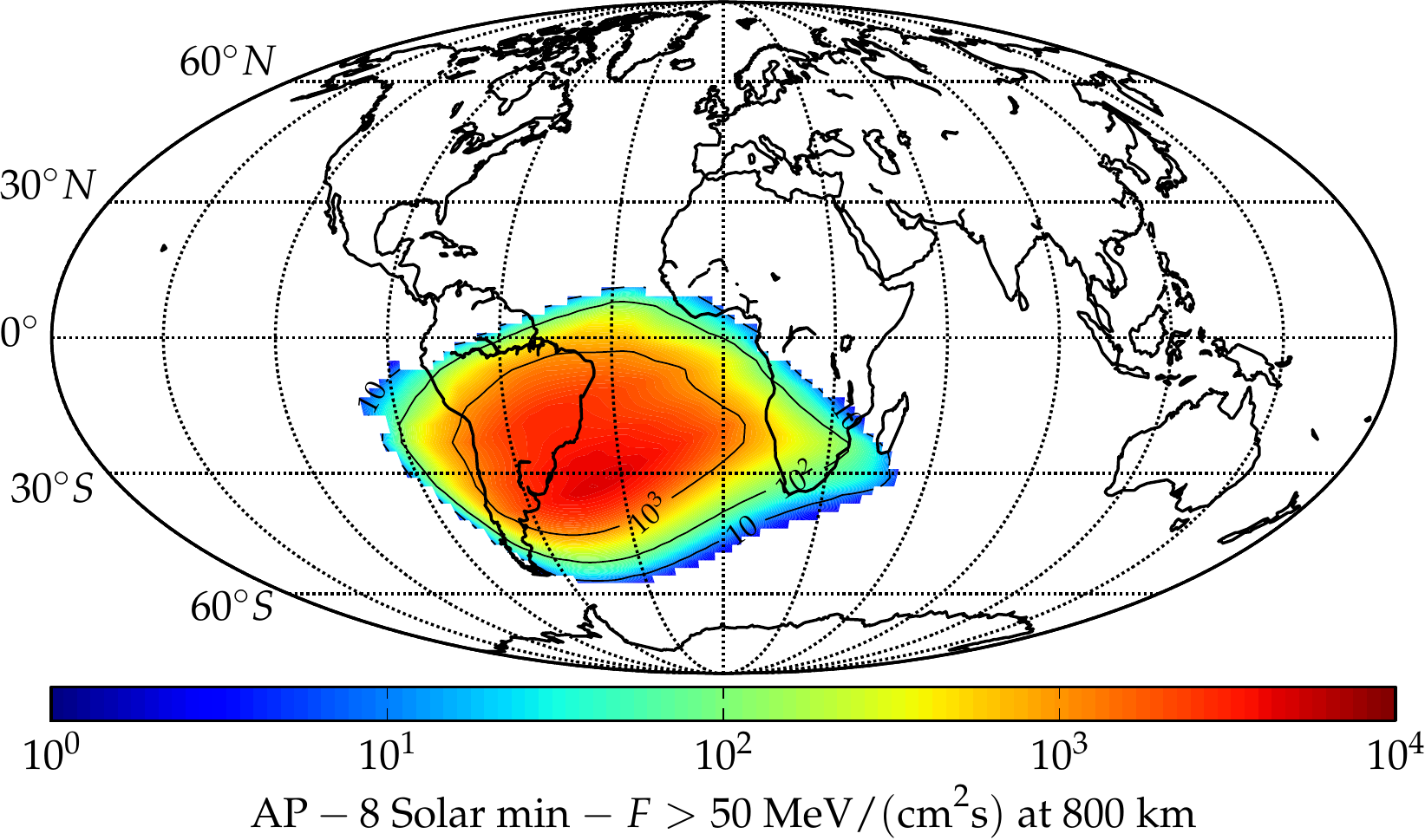}
  \caption{Extension of the South Atlantic Anomaly at an altitude of 620 (Top.) and 800 km (Bottom.) as predicted by the AP--8 Solar minimum model. The colour bar represents the amount of protons with energies of 50 MeV minimum.}
 \end{center}
\vspace{-2.em}
\end{figure}

\begin{figure}[h]
 \begin{center}
 \includegraphics[width=1\linewidth]{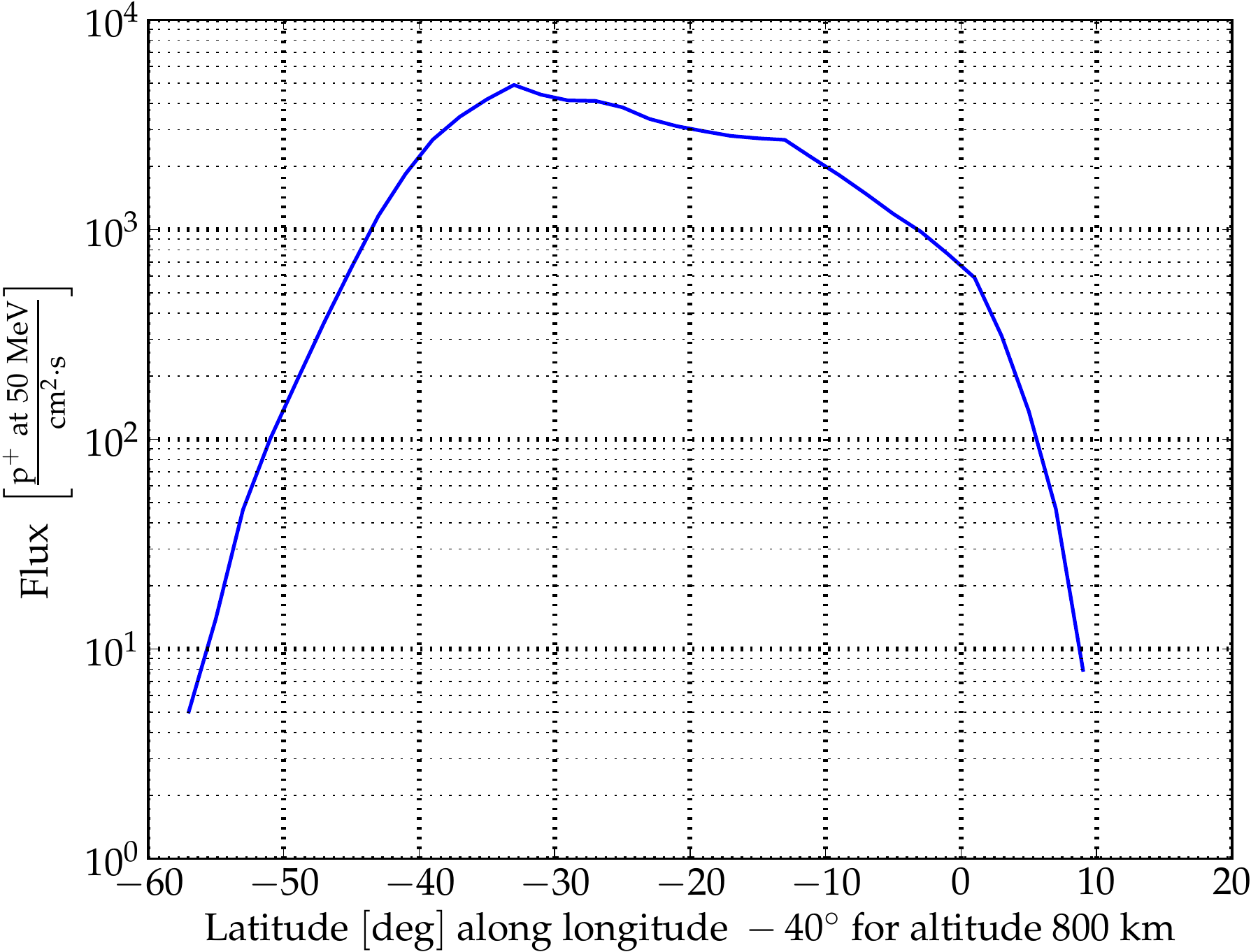}
  \caption{Typical radiation flux for a satellite flying along the longitude of -40\textdegree\ East at an altitude of 800 km. The crossing time in this case is about 20 minutes.}
  \label{fig:dis-saa-gradient}
 \end{center}
\vspace{-2.3em}
\end{figure}

It reveals that, as expected, the effect of the SAA is larger at 800 km: at 620 km altitude, the zone is centred around the South Atlantic ocean and possesses a tail that reaches the Horn of Africa. At 800 km altitude however, all the South American continent is perturbed and this region extends to Madagascar and the Atlantic coast of the African continent and past the Chilean coast on its Western end. The important criteria is how large the gradient of flux is. 
If the gradient of flux is small, it means that the sensitivity to the radiation will greatly impact the observation time whereas if it is very large, than the predicted observation time is a good approximation. Moreover, if the gradient is large enough, then the maximum radiation threshold -- 2 protons of 50 MeV for CHEOPS -- will not have a great effect on the interruptions. 
Figure \ref{fig:dis-saa-gradient} shows that the gradient is very steep for a satellite crossing the SAA along a meridian which is a sufficient approximation of the trajectory of CHEOPS in this case.

\subsection{Transit Detection Probabilities} \label{sec:dis-transit-detection}
\paragraph{Goals.}
The CHEOPS targets catalogue is not yet established. To observe those targets the list must contain at least the coordinates, the $V$ magnitude and predicted ephemerides for the transits. As defined in the science requirements of the mission \citep{SciReq}, there are several kind of objects that can be observed. Amongst them the Super-Earths and the Neptune-like planets. The detection capabilities are defined for a given orbital period $P$ and transit time $T$ (See \S\ref{sec:theo-transit} and in particular eq. \ref{eq:exo-transit-tf}). 
\begin{description}
 \item [SciReq. 1.1: Super-Earths] CHEOPS shall be able to detect an Earth-size planet transiting a G5 dwarf star (0.9 $R_\odot$) of the 9th magnitude in the V band, with a signal-to-noise ratio of 10. Since the depth of such a transit is 100 parts-per-million (ppm), this requires achieving a photometric precision of 20 ppm in 6 hours of integration time. This time corresponds to the transit duration of a planet with a revolution period of 50 days;
 \item [SciReq. 1.2: Neptunes] CHEOPS shall be able to detect a Neptune-size planet transiting a K-type dwarf (0.7 R $R_\odot$) star of the 12.5th magnitude in the V band (goal: V=13) with a signal-to-noise ratio of 85 ppm. Such a transit has a depth of 2500 ppm and last for nearly 3 hours for planets with a revolution period of 13 days. Hence, a photometric precision of 85 ppm is to be obtained in 3 hours of integration time.
\end{description}
This categorisation does not imply that Super-Earths have larger semi-major axis or brighter host star than Neptunes. It rather defines the typical classes of targets for this missions. The objects that will be observed will come from the currently known exoplanets as well as from the Next Generation Transit Survey \citep{Wheatley2013}.

The great advantage of a follow-up mission is that the knowledge of when the transit will occur is available (to a certain extent). For objects detected using the transit technique the ephemerides are known. For radial velocities detection, ephemerides are not given and the fact that the planet is actually eclipsing its planet is not certain (See \S\ref{sec:theo-detections}). 
However, using the RV curves, it can be deduced that the planet is closer or further away than its averaged distance from the shift in colour of the host star. As a transit happen when the planet is the closest to us, a first approximation of the date of the next transit can be obtained. 
In order not to miss this probable event, the star must be monitored for a small fraction of the period instead of the whole orbital period of the planet. Using the above data, one can predict what is the probability of detecting a transiting planet in a given direction of the sky \citep{Henry2011}.

\paragraph{Methodology.} To compute this probability, the sky is divided into grid points. Assuming that the planet transit the star, the probability of detecting the transit is computed. First, the visibility of the points are computed including constraints on the SAA and the stray light flux. Then, this data is translated into periods of observations of the duration of the minimum transiting time. Some interruption periods are allowed to account for the SAA and occultation of the targets due to objects in the LOS or the geometry. 
Targets must be visible for at least 50 minutes (plus the acquisition time of 6 minutes) to be considered into the maps of probabilities. To estimate the probability, firstly the number of times that an observation can be made is computed. 
This number is then divided by the orbital period of the planet. It can thus happen that the observation probability is more than 100\% as the target star would be observable for more than one orbital period of its planet of interest. 
A point that would have a probability of detecting a transit of 300\% means that if the planet is transiting its host star, it is possible to see this transit 3 times during the year 2018.

\begin{figure*}[!h]
\begin{center}
\includegraphics[width=0.49\linewidth]{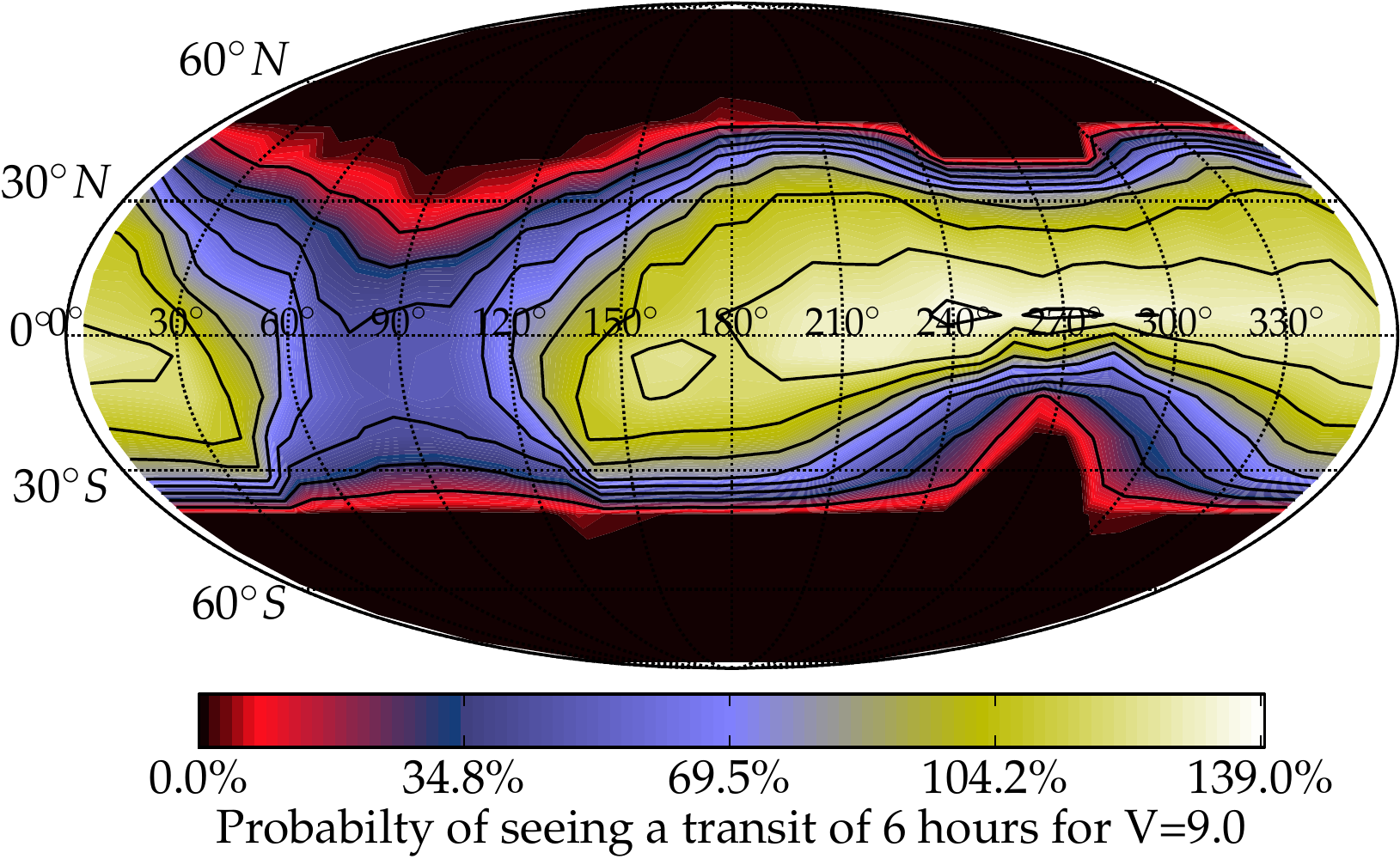} 
\includegraphics[width=0.49\linewidth]{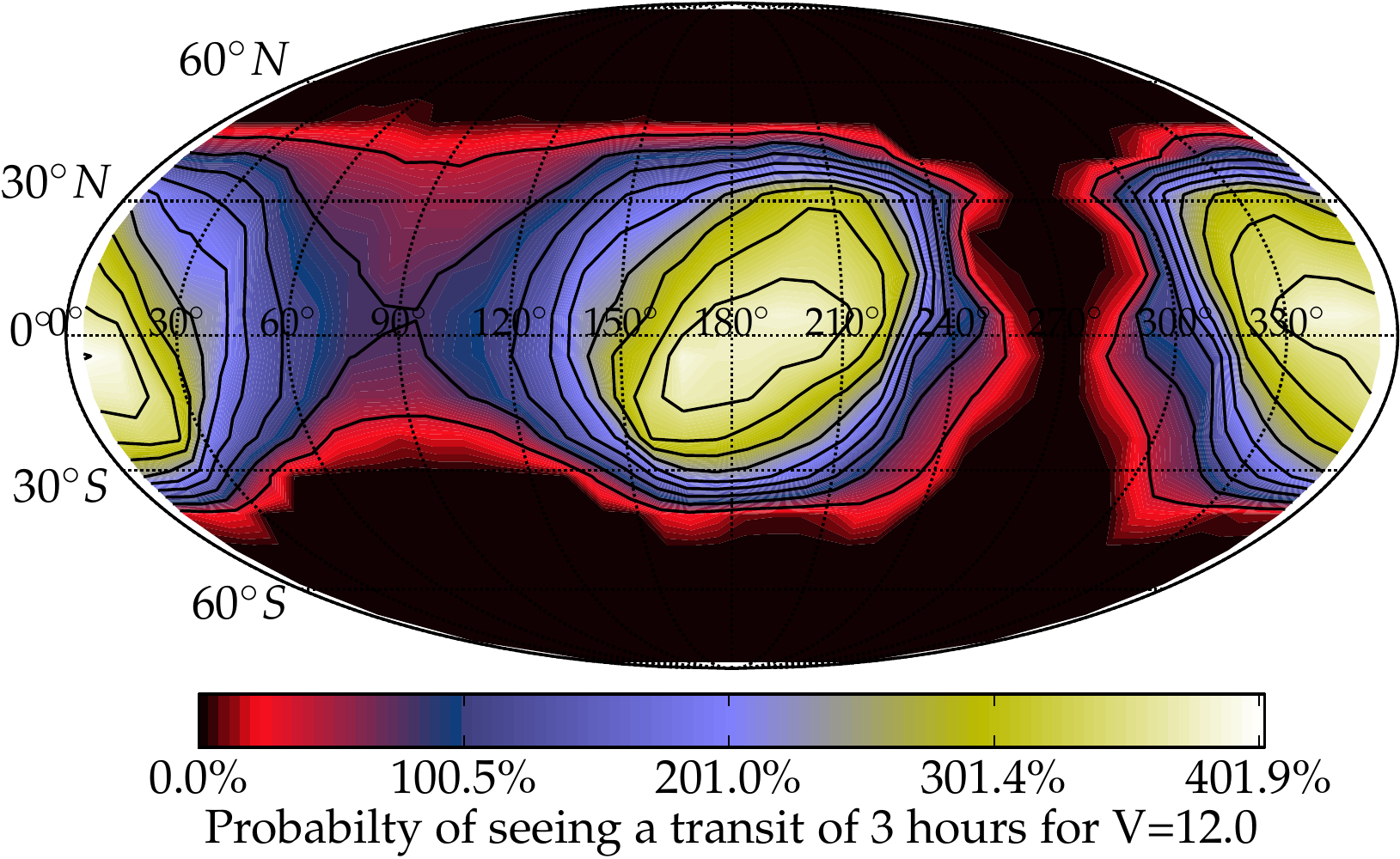} 

\caption{Probability maps to detect a transit. Altitude of 800 km. Interruptions maximal set to 50 min of the orbit, acquisition time set to 6 minutes. -- \emph{Left.} Map for a Super-Earth planet. -- \emph{Right} Map for Neptunes.\label{fig:proba-1}}
\vspace{2em}
\includegraphics[width=0.49\linewidth]{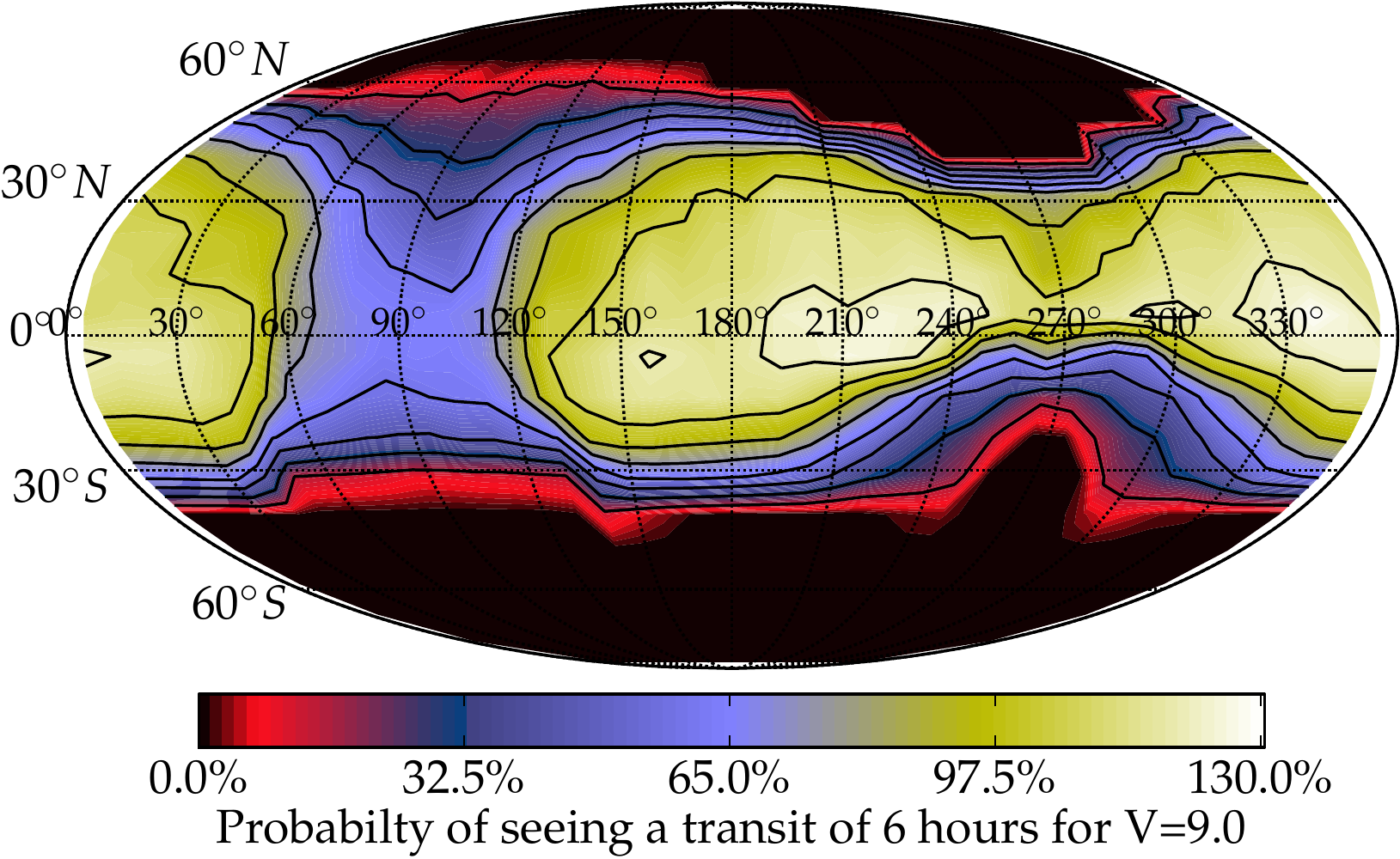} 
\includegraphics[width=0.49\linewidth]{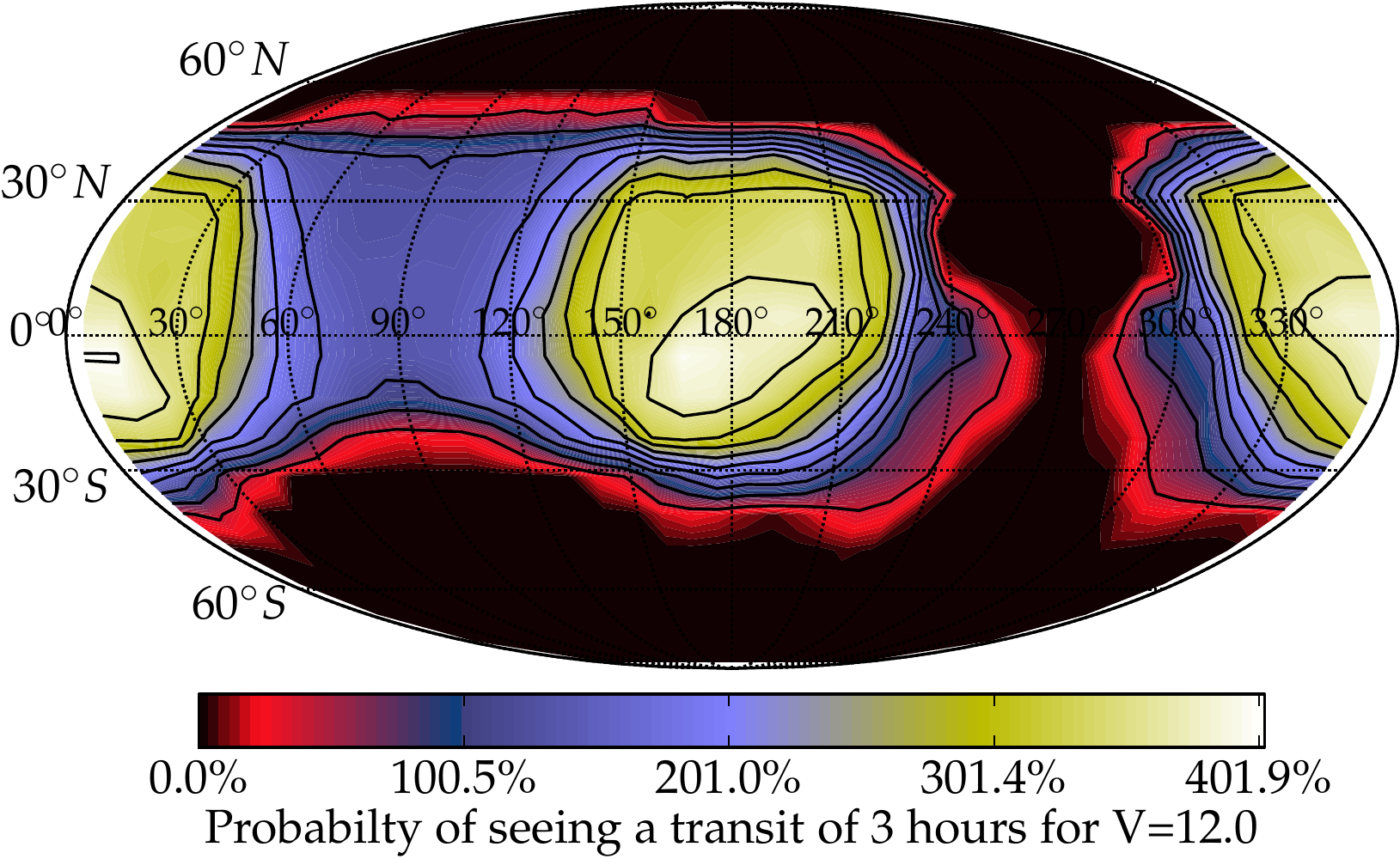} \label{fig:proba-2}

\caption{Probability maps to detect a transit. Altitude of 700 km. Interruptions maximal set to 50 min of the orbit, acquisition time set to 6 minutes. -- \emph{Left.} Map for a Super-Earth planet. -- \emph{Right} Map for Neptunes.}
\vspace{2em}
\includegraphics[width=0.49\linewidth]{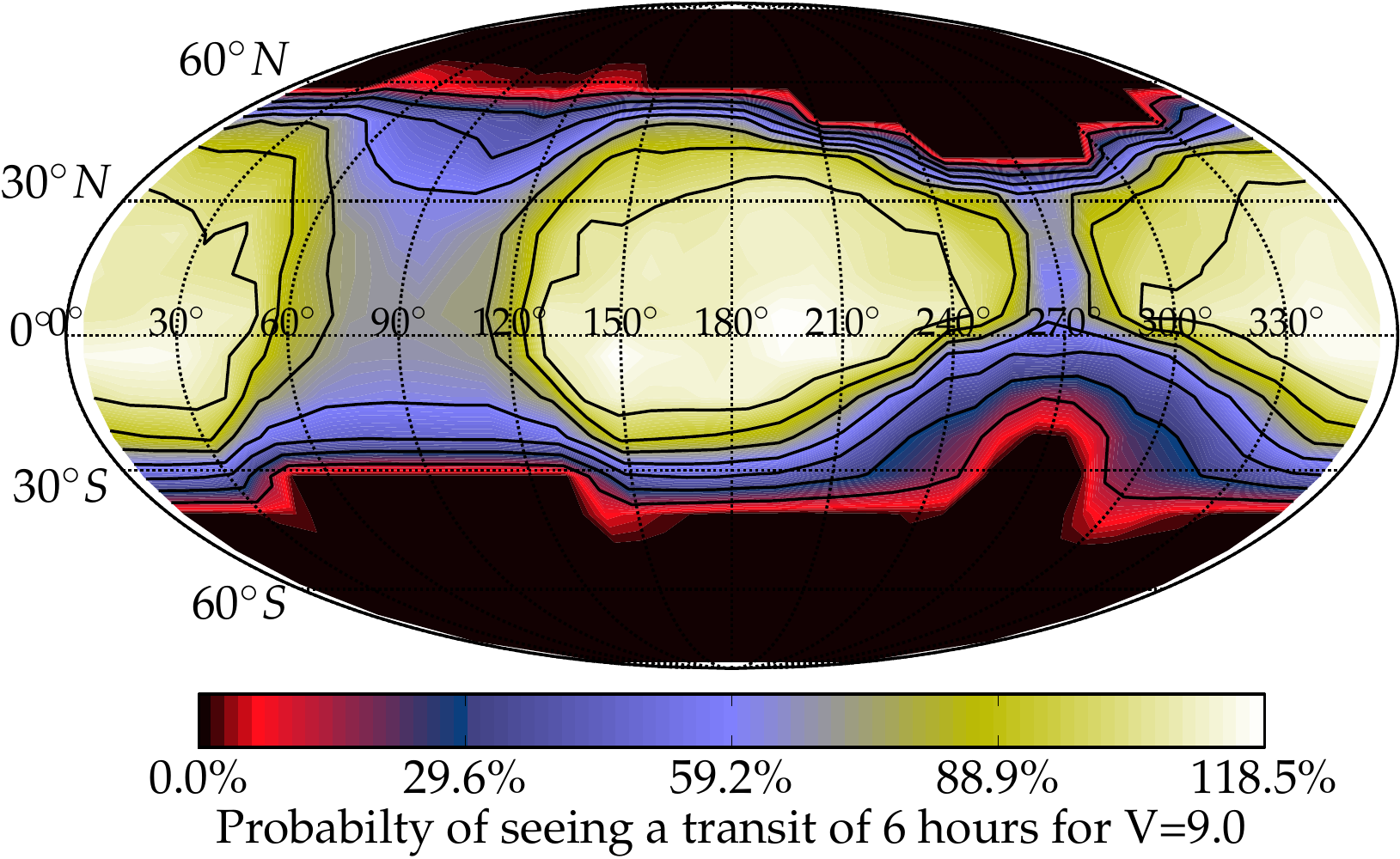} 
\includegraphics[width=0.49\linewidth]{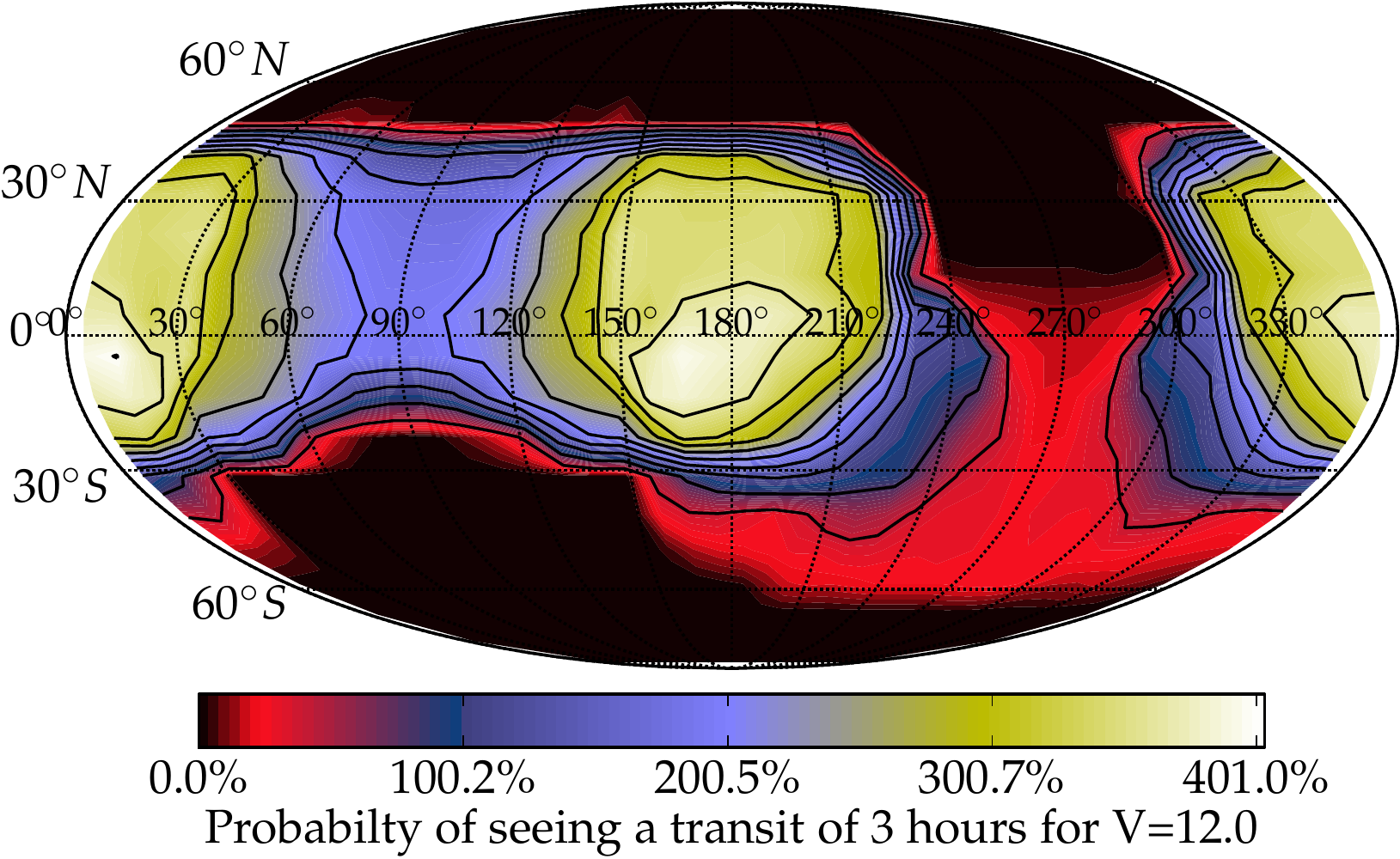} \label{fig:proba-3}

\caption{Probability maps to detect a transit. Altitude of 620 km. Interruptions maximal set to 50 min of the orbit, acquisition time set to 6 minutes. -- \emph{Left.} Map for a Super-Earth planet. -- \emph{Right} Map for Neptunes.}
\end{center}
\end{figure*}

\begin{figure*}[!h]
\begin{center}
\includegraphics[width=0.49\linewidth]{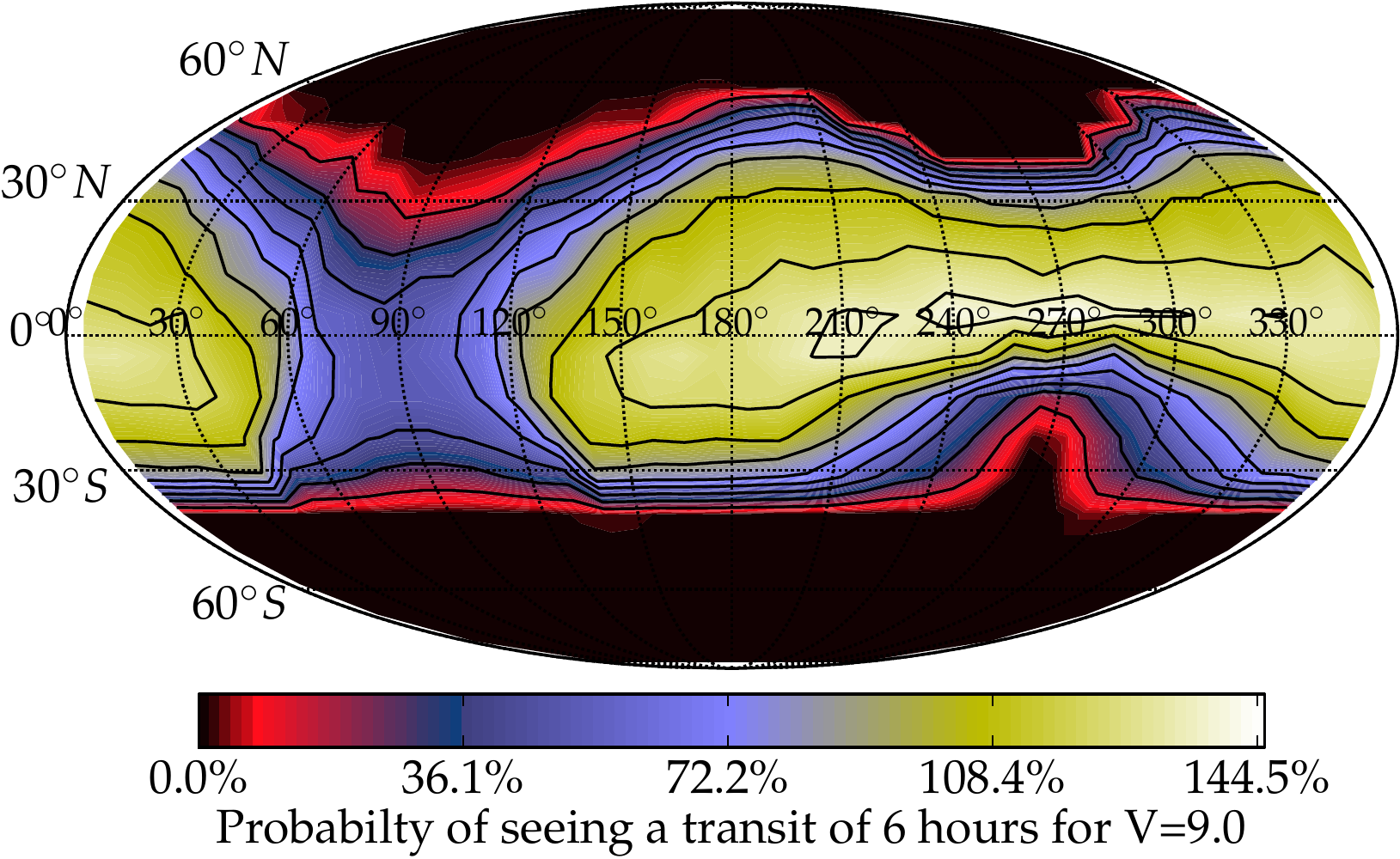} 
\includegraphics[width=0.49\linewidth]{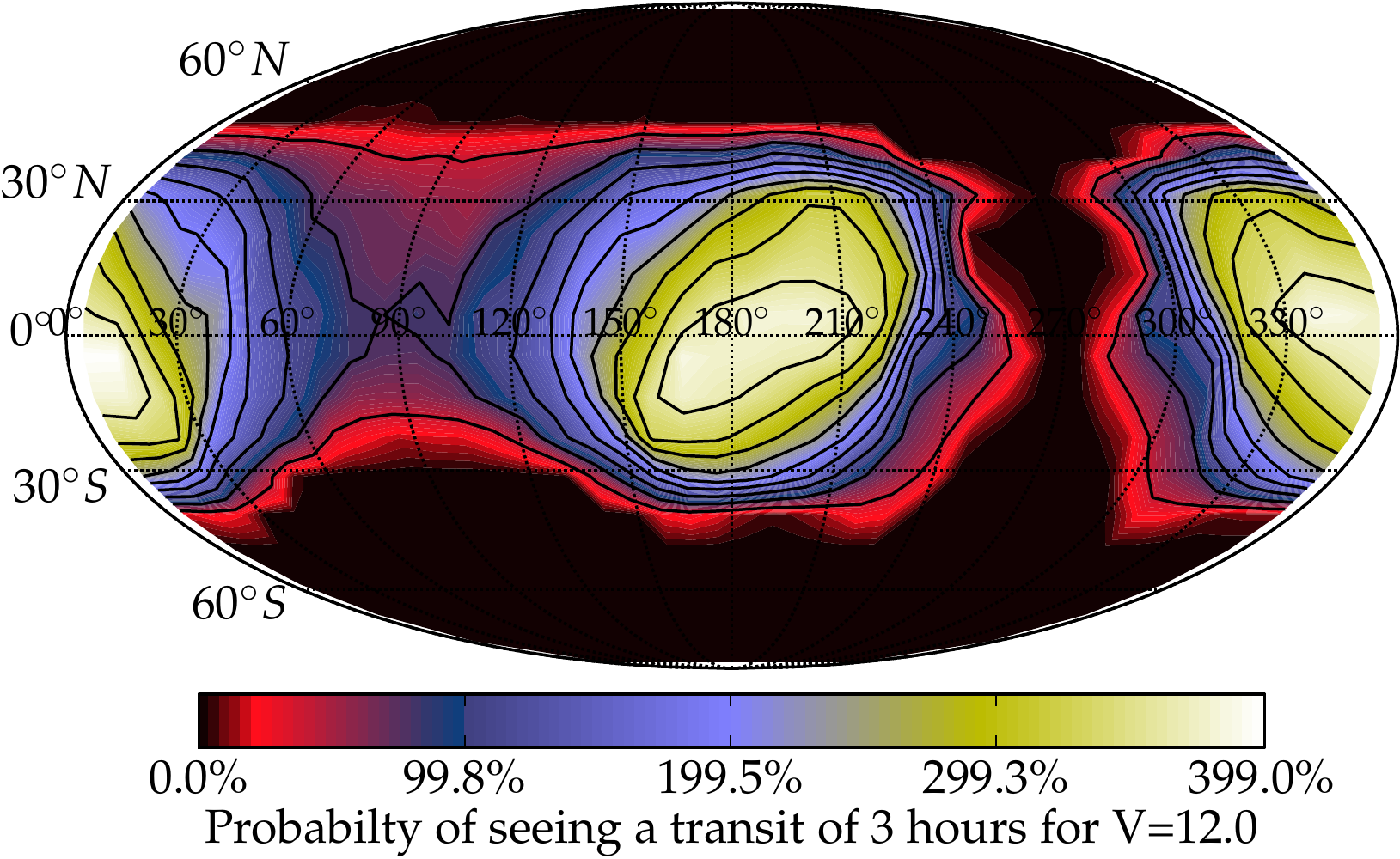} \label{fig:proba-4}

\caption{Probability maps to detect a transit. Altitude of 800 km. Interruptions maximal set to 50 min of the orbit, acquisition time set to 6 minutes and a stray light exclusion angle reduced to 25$^\circ$. -- \emph{Left.} Map for a Super-Earth planet. -- \emph{Right} Map for Neptunes.}
\vspace{2em}
\includegraphics[width=0.49\linewidth]{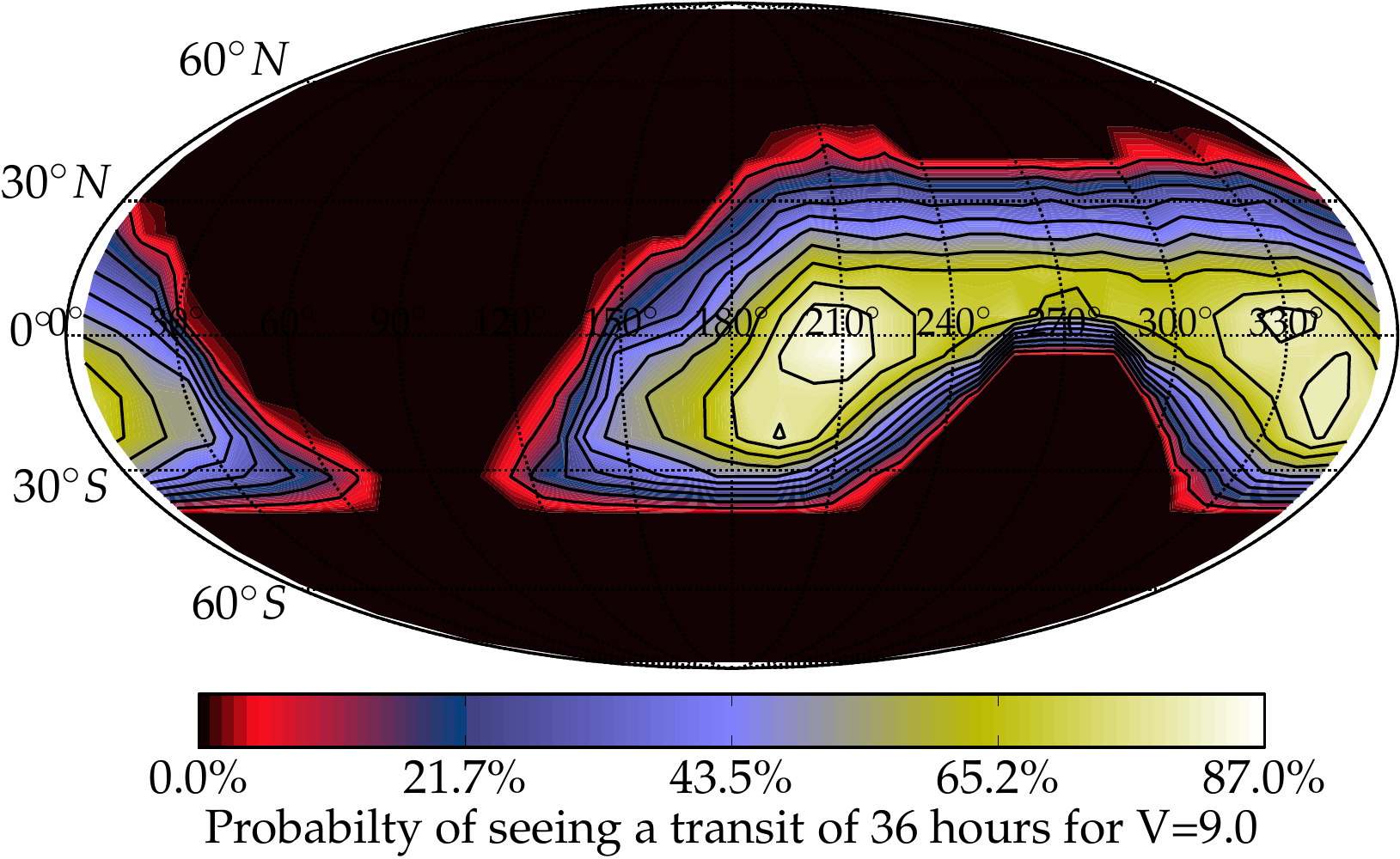} 
\includegraphics[width=0.49\linewidth]{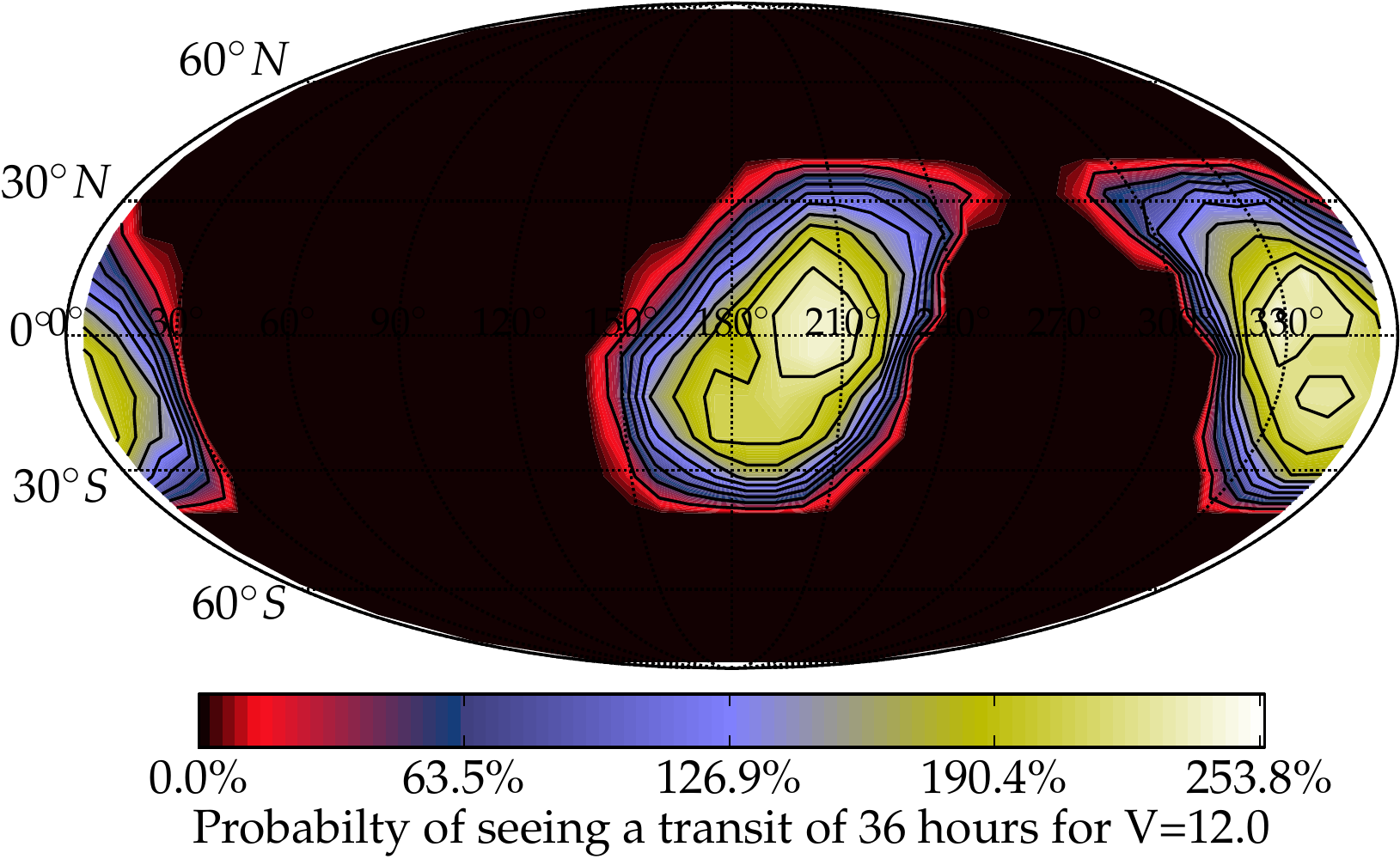} \label{fig:proba-5}

\caption{Probability maps to detect a transit for very large uncertainty on the transit time. Altitude of 800 km. Interruptions maximal set to 50 min of the orbit, acquisition time set to 6 minutes. -- \emph{Left.} Map for a Super-Earth planet. -- \emph{Right} Map for Neptunes.}
\vspace{2em}
\includegraphics[width=0.49\linewidth]{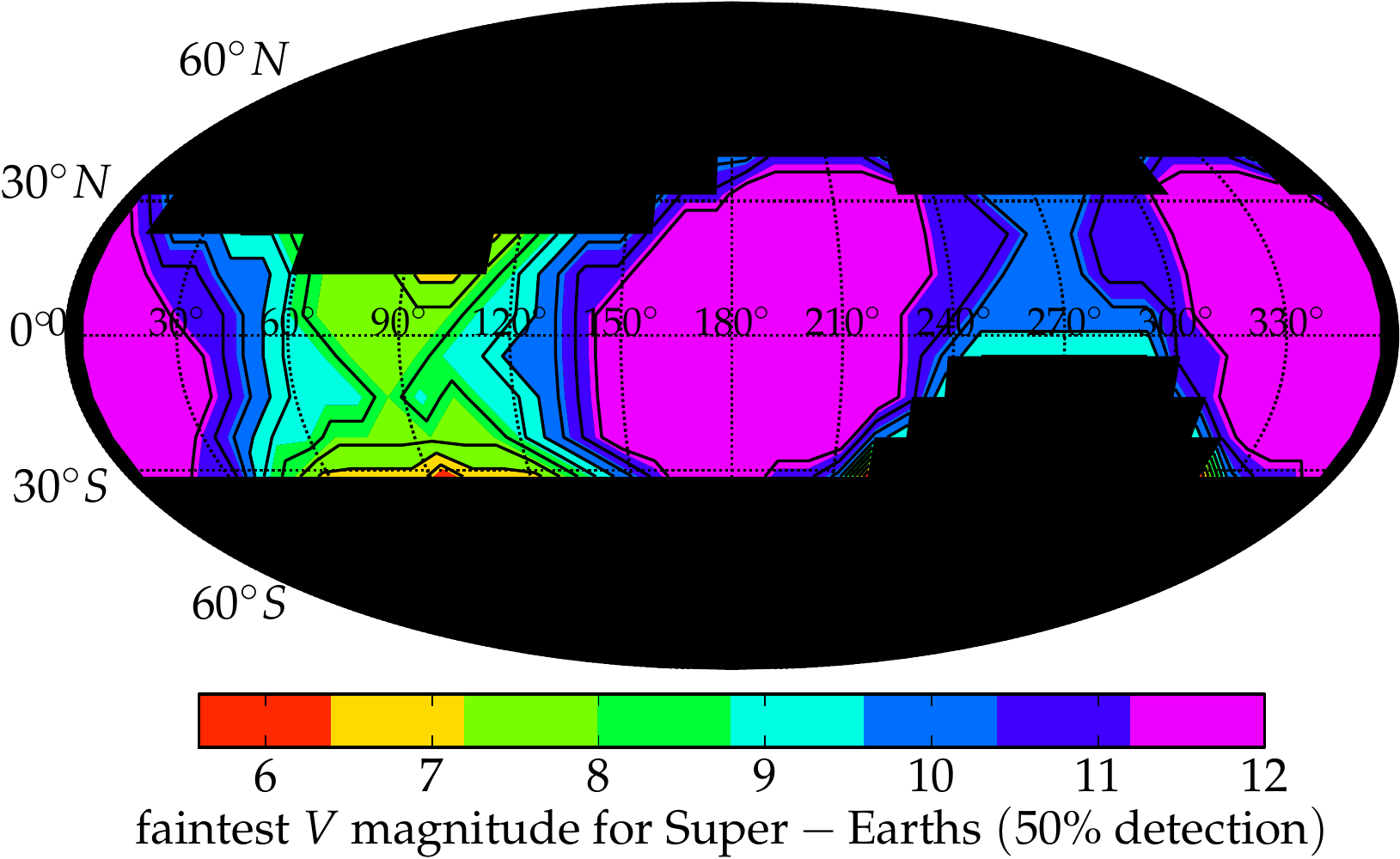}  \label{fig:proba-6}
\includegraphics[width=0.49\linewidth]{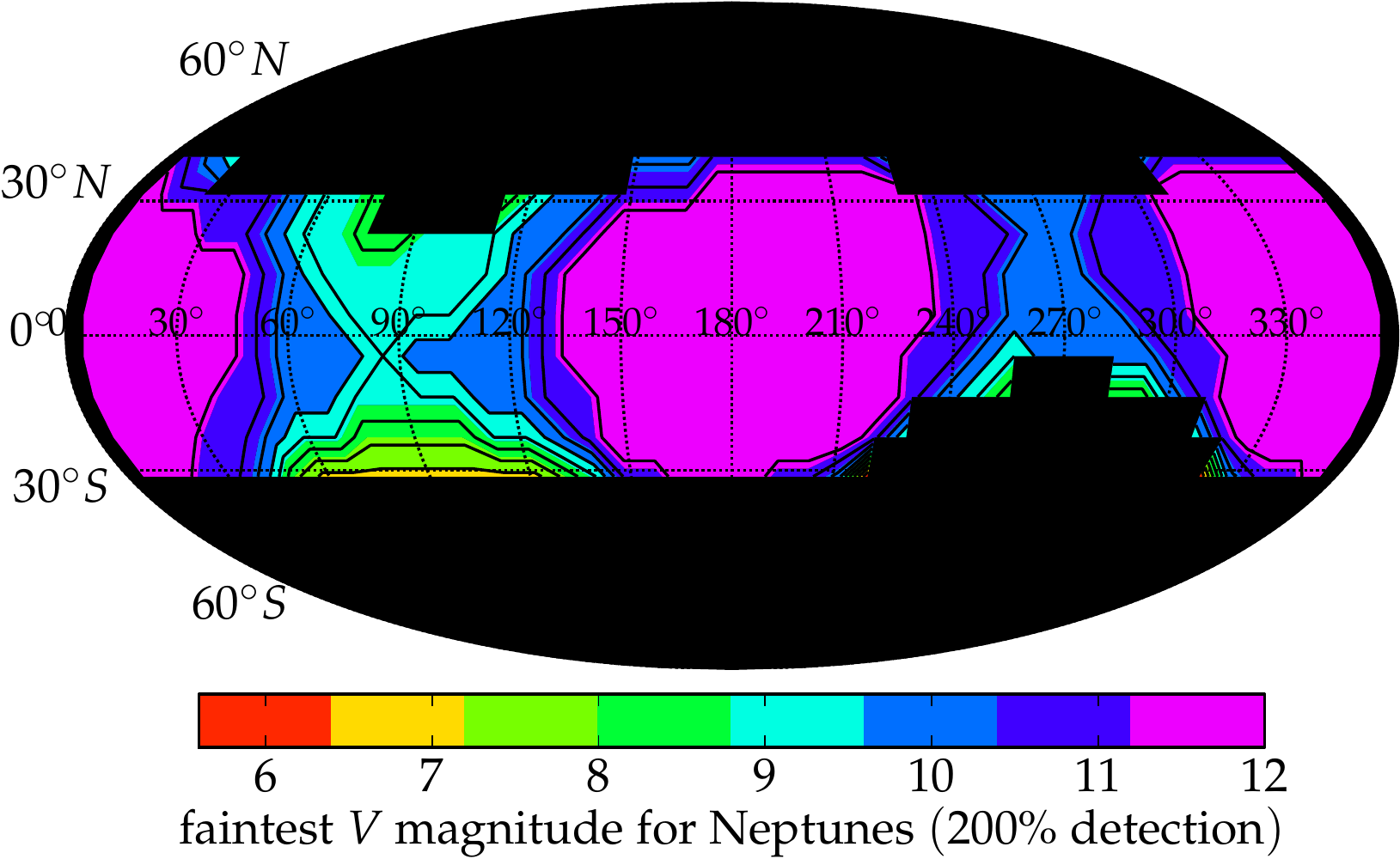} 
\hfill

\caption{Altitude of 800 km. Interruptions maximal set to 50 min of the orbit, acquisition time set to 6 minutes. Faintest magnitudes that can be observed\dots -- \emph{Left.} when looking at Super-Earths with at least a probability of 50\% of observing a transit. -- \emph{Right.} when looking at Neptunes with at least two observable transits.}
\end{center}
\end{figure*}

\paragraph{Discussion.} Maps of transit detection probabilities are presented in Fig. \ref{fig:proba-1}--\ref{fig:proba-6}. For course, the first striking feature in those plots are the now famous ``butterfly wings'' that appears again. Another important thing is that it will be easier to detect close Neptunes than far Super-Earths simply due to their shorter orbital period. The closer the planet is to its star, the shorter the orbital period and thus the easier it can be observed as transits occur more often. 
It has to noted that those two classes of planets do not imply that Super-Earths with an orbital period of the order of the day will not be observed. Again, this is study to explore the limits of CHEOPS performance.
Observing Neptunes around faint target can done very efficiently during the alignments of the terminator with the plane of the orbit, but also close to the December solstice. 
It would be completely impossible to detect them around the June solstice due to the SAA, but this is not the only reason. Otherwise, it would be the same for Super-Earths. The other reason is the magnitude of the host star. The ``forbidden region'' for faint star is very clearly centred around $\alpha = 270^\circ$ degrees. 
The visible region of the satellite is exactly centred on this region during the summer solstice. The maximum observation time does not vary much for either targets as it occurs when the plane of the terminator and the orbit merge.
\paragraph{}
Super-Earths that have a relatively long orbital period translate into low detection probabilities ($\lesssim150\%$). They can be observed at any time in the year, with a peak of observation time around the alignments of the terminator to the plane of the orbit, but also -- and contrary to the faint star targets -- close to the June solstice. The shape of the map is similar to the one of the Neptunes, with however two major differences: (1) the visibility in summer as already discussed and (2) the shape around the December solstice ($\alpha = 90^\circ$). The outer limit is imposed by the observation time of 6 hours for the Super-Earths.
This cannot be reached for high declinations ($\gtrsim30^\circ$ -- Fig. \ref{fig:proba-1}--\ref{fig:proba-3}). The reason is the visibility of those extreme points that can change from one orbit to the next due to the presence of the SAA that has a variability over typically 8 orbits. Combining the SAA with the appearance and disappearance of the point from the visible 
region yield that high declination points are harder to observe and that negative declinations are easier due to the less compelling magnitude limitation. 
\paragraph{}
Even with very large uncertainties on the transit duration (here 36 hours is used) (Fig. \ref{fig:proba-5}), long period of stare at targets are possible in or around the ``butterfly wings'' depending upon the kind of RV targets that could be observed. Searches for other planets using the TTV technique (\S\ref{sec:theo-TTV}) could be also performed in those regions as the amount of data required for the TTV is large \citep{Lithwick2012}.

\begin{table}[h]
 \begin{center}
 \begin{tabular}{r|rrrr} \toprule
$h$ [km] & 800 & 800 & 700 & 620 \\ 
$\alpha_\oplus$ [\textdegree] & 35 & 25 & 35 & 35 \\\midrule
   Super-Earths & 46\% 	 & 50\%   & 52\%   & 50\% \\
   Neptunes    & 46\%   & 46\%   & 46\%   & 51\% \\
  \bottomrule
 \end{tabular}
 \caption{Percentage of the sky that can be observed for a given type of target as defined above according to the probability maps for at least 50\%. The first line gives the altitude $h$ of the orbit while the second yield the minimum angle from the limb of the Earth to the LOS. The South Atlantic Anomaly is included in every simulation.\label{tab:dis-percentage-sky}}
 \end{center}
\end{table}

Table \ref{tab:dis-percentage-sky} present the percentage of the sky which can be observed by CHEOPS during the year 2018 for different categories of planets. It can be seen that the 50\% of the sky visible to CHEOPS as required by the SciReq 2.1 (see \S\ref{sec:dis-yearly}) is barely reached. It must be pointed that 700 km stands out with the maximal percentage for Super-Earth. 
This is justified by the trade-off between the time spent in the SAA against the viewing zone that increases with increasing altitude. At 620 km, the results are counter-intuitive as both type of target yield viewing zone larger than 50\%. The SAA plays therefore a very important role as the two other effects -- namely stray light flux and observable region that scales with the altitude -- are counter-balanced.

Given a probability of observing a transit (say 200\% or 2 transits for Neptunes and 50\% for Super-Earths), zones with the same limiting magnitudes can be plotted. Shown in Fig. \ref{fig:proba-6}, they reveal that for both type of targets, they are pretty similar. It is clear that Neptunes are easier to find (the probability limit is set to 4 times the one for the Super-Earths). Those plots also add to the fact that observing a Neptune-like planet in Northern winter around a faint star will be borderline impossible.
\paragraph{}
Interruptions in the observations will degrade the SNR, but are unavoidable and amount up to 50\% as it is a parameter of the simulation. Pristine observations can still be made, but those targets must be located in very specific regions, namely in the ``butterfly wings''. Despite this drawback, those maps remain a lower bound on the probability as the magnitudes used reflect the upper limit of their group and that orbital periods can be shorter than the one used. From those maps, it can therefore be seen that (1) CHEOPS will indeed be able to observe the target it intends to study and that (2) most of those targets can also be seen by ground-based observatories located in the Southern hemisphere (La Silla, host of the HARPS is at located $\sim29^\circ$~S). 

\subsection{Number of Stars in the Field} Several stars in the field can be an important source of noise. To estimate how many stars are in the field of view of the instrument for any given star, the Tycho-2 catalogue was used \citep{Hog2000} which contains 2'539'913 stars.
If there is another star close to the target, then the tails of the PSFs will overlap which is bad for the photometric signal. The photometry of the target star will indeed have another source that contributes to the signal: the near star. Even if the light curve of a star is given in relative values, the fact that there is a systematic contribution from the other star, decreases the depth of the transit and hence reduces the probability to detect the object. 
\paragraph{}

\begin{figure}
 \begin{center}
  \includegraphics[width=1\linewidth]{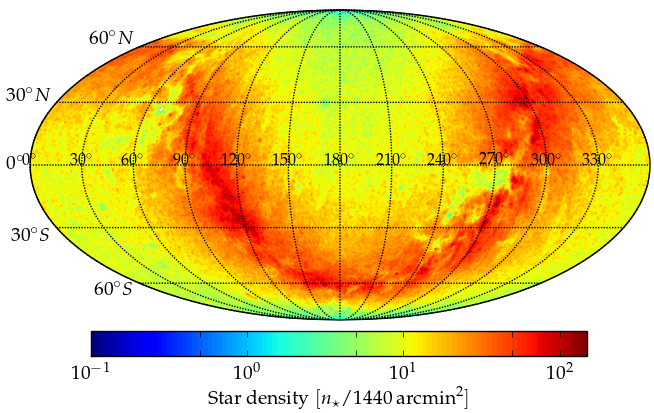}
  \caption{Star density in the sky in a window of equivalent size than the field of view if the CHEOPS detector.\label{fig:dis-fov-density}.}
 \end{center}
\vspace{-1.5em}
\end{figure}

There are two different fields of view in CHEOPS: (1) the whole detector (FOV) and (2) the region of interest. The former has a surface of 0.4 degree squared ($\equiv 1440 $ arcmin$^2\equiv1$ Mpx) and later 57.6 arcmin$^2\equiv 200\times200$ px. Between the two fields, there is a factor of 25.
An estimation is made by computing the density of the stars in a given surface of the sky. This was performed for two surfaces: the field of view of the whole telescope and the size of the region of interest. Here as well, the Tycho 2 catalogue was used as the reference for the position and the magnitude of the stars. The probability of having at least one $m_V<10$ star in the FOV when looking at a given position in the sky is almost 1 around the ecliptic plane. 
The most crowded field that can be observed in the region of interest is basically to look close to a binary star. As the brightness of the star increases, their density decreases. There are only around 77'000 stars brighter than magnitude 8.5 in Tycho 2 for more than 2'500'000 entries. A plot of the density of stars in the sky in the field of view of the instrument can be seen in Fig. \ref{fig:dis-fov-density}. The most densely populated regions in the sky are of course close to the plane of the Milky Way. However, there exist also poorly populated regions that could be useful for calibration purposes (see \S\ref{sec:disc-empty-regions}).

When cross-referencing this to the yearly observability maps, it can be seen that the plane of the Milky Way does not coincide with the plane of the orbit. This mean that the stellar density in the field varies across the year. For example, the butterfly wing in the observation map around May 21 ($\sim 240^\circ$) will be a region of sparse field. This is no longer the case at the next coincidence of the planes of the terminator and the orbit (centred on July 21, $\sim 300^\circ$). During the first months of the year 2018, CHEOPS is constrained to observe an area close to the plane of the Milky Way.

The target list is unknown at this time, however the distribution of exoplanets discovered by RV is fairly homogeneous\footnote{Based on data from \httplink{http://www.exoplanets.org}, retrieved April 30th 2013.}. Due to the nature of the RV measurements, the stars must be relatively bright (relatively to transit detection) to obtain good enough data. Thus crowded fields are not likely to be often an issue even if it cannot be excluded.

\subsection{The Use of Empty Regions in the Sky for Calibration Purposes} \label{sec:disc-empty-regions}
In an effort to reduce the noise on the signal in the long term, the ageing of the detector must be taken into account. Modelling this is not trivial and would yield a statistical answer and not give the state of one particular pixel. 
The dark signal in the frame can evolve with time and then so do the flat-fielding corrections. Pre-launch calibrations on the instrument will of course be performed on the ground. For a good estimation of the dark current and host spots, the detector must expose frame in complete darkness which means with a shutter closed or in a pitch dark room. For the flat-fielding, the detector must be illuminated by a constant flux. The hot spots or hot pixels are particularly sensitive pixel that saturate very quickly.

Once in orbit, those calibration measurements may prove difficult to obtain. Indeed, CHEOPS in its current version is neither equipped with a lamp that would be able to illuminate the detector nor with a shutter to see the dark current and the evolution of the hot spots. The question now is can the temporal evolution of the dark current be monitored? The effect of the vibrations due to the handling on the ground and most prominently the launch can be observed by taking a few frame before the opening of the cover which protects the telescope from contaminations. This technique can only be used once as the cover can only be opened once in order to ensure that it does open once CHEOPS in orbit. There are several possible solutions to calibrate without any shutter. 
For the Hubble Space Telescope, different techniques have been tried \citep{MacKenty2010} such as obtaining flats from on-board calibration lamp, crowded regions or observations of the Earth in the night or of the moonlit Earth. All of those techniques did not reveal any very satisfying results.

\begin{figure}[h]
 \begin{center}
  \includegraphics[width=1\linewidth]{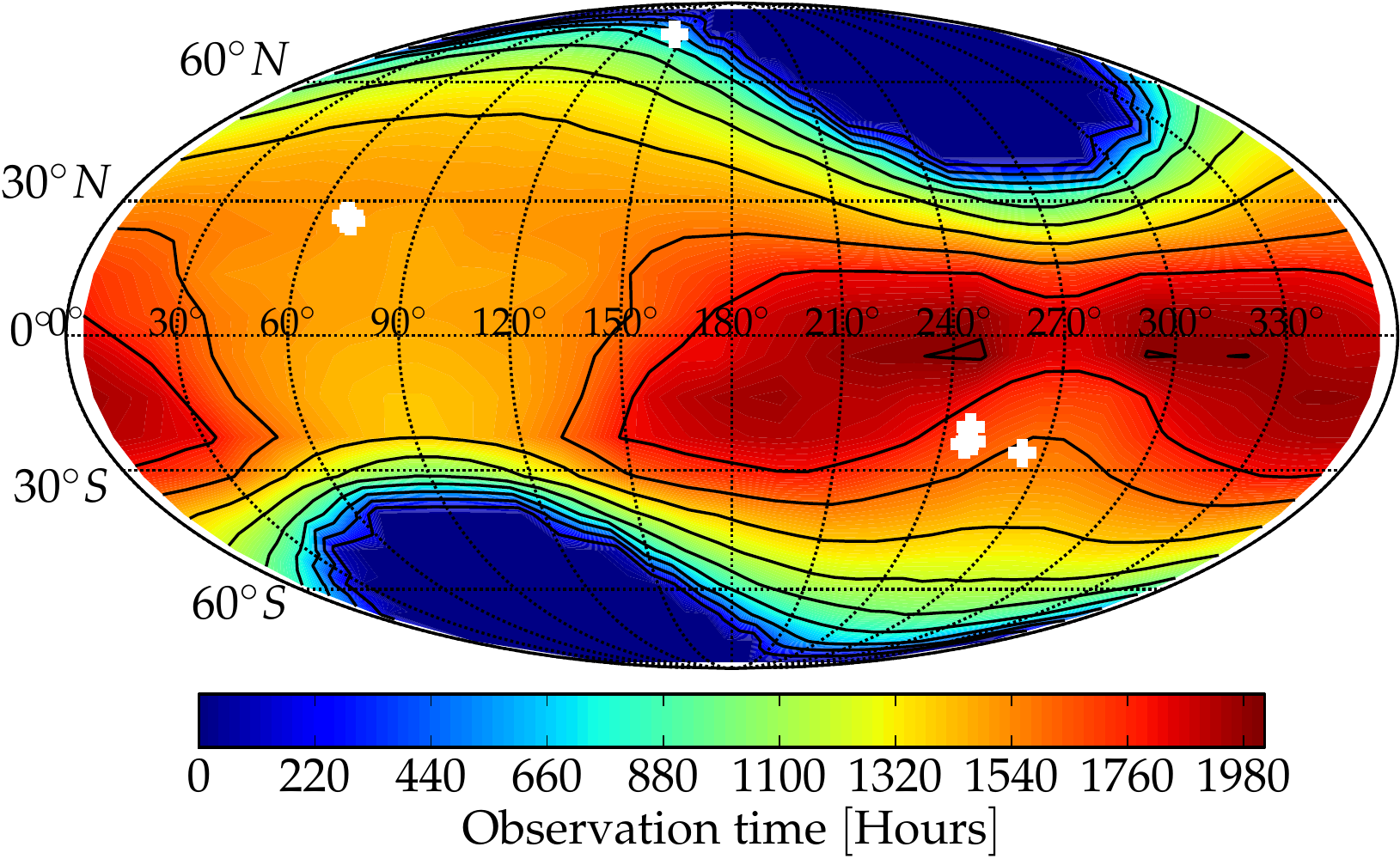}
  
  \caption{Distribution of possibly empty regions (white crosses) that could be used to calibrate the detector. The map represent a cumulative observation time for a minimum of 4 minutes of observations plus 6 of acquisition time at an altitude of 800 km.\label{fig:dis-empty-regions-distrib}}
 \end{center}
\end{figure}

The moonlit or the dark Earth could be tried on CHEOPS or even, if it is observable, an Earth-lit Moon that may be more uniform -- there are obviously no thunderstorms or cities on the Moon. Here another question was investigated: given the fairly small field of view of CHEOPS, doe there exist regions in the sky that would be faint enough? Of course, it all depends upon what \emph{faint enough} means. 
The detector chosen for CHEOPS is the ``CCD47--20 Back Illuminated High Performance AIMO'' from the European company e2v. Using the data sheet of the CCD \citep{aimo2006}, the characterisation of the typical dark signal yields that the brightest object in the field must be of the order of magnitude $V=23$ \footnote{see \S\ref{app:DarkCurrentNoise} for more details on this computation.}. 
The zodical light described by \cite{Levasseur1980} yield a equivalent magnitude of 23.3 to 22.3 per arcsec$^2$. A pixel of CHEOPS detector is equivalent to one arcsec and hence it would be faint enough to characterise the dark signal well. The question is: do sufficiently empty regions exist in the sky? 
The coordinates of several candidate regions are shown in appendix \ref{app:empty} with maps of the local sky. Dark current variations with time could therefore be monitored by looking at those regions. If proven that those are actually empty enough, then a calibration could be performed in those regions.
To answer this question the Tycho-2 catalogue was -- again -- used to find less denser regions in the sky. It has to be noticed that this catalogue is not complete for objects fainter than a magnitude of about 12 \citep{Hog2000}. To probe fainter regions, the catalogue galaxy Sloane Digital Sky Survey (SDSS) where possible \citep{Paris2012}. Some other regions remain to be explored carefully to look for bright objects. Figure \ref{fig:dis-empty-regions-distrib} present the distribution of those regions in the sky.

There are three main regions: two within $\pm30^\circ$ degrees of the ecliptic and one about $70^\circ$ North. As the observable zone moves from left to right in Fig. \ref{fig:dis-empty-regions-distrib} (increasing right ascension) with time, the region close to $\alpha=60^\circ$ is visible from end of November to about mid January. The second near $150^\circ$ is more difficult to see due to its high declination, but can be observed in early spring while the last group of empty zones in the sky could be observed between the spring terminator--orbital plane alignment and the summer solstice.

The small field of view of CHEOPS -- $1440 $ arcmin$^2$ -- can therefore look at regions in the sky in which the faintest objects have V magnitudes of the order of 22-23 according both to the Tycho-2 and the SDSS catalogues. Further explorations of those regions should be performed to ensure that no bright object were missed. 
The in-flight calibration of the dark current and of the hot spots could be done in the best case about three times a year which could be enough to characterise the degradation of the dark signal with time as well as the map of the hot pixels. Flat fielding on the other hand may prove difficult with this method: although the illumination is fairy constant and homogeneous, it might be too faint. 
HST observations of the zodical light (see cited paper before) showed that the flux of photons was too low to get a sufficient flat out of the data. But the detector of the two space telescopes are difficult to compare.

\newpage \clearpage

\newpage \clearpage

\cleardoublepage
\chapter{Conclusions \& Outlooks} \label{ch:conclusion}
\lettrine[lraise=0., nindent=0em, slope=-.5em]{C}{HEOPS} is the first mission of its kind: the ``Small-Class'' missions of ESA. With tight budgets in a lot of different areas -- cost, mass, volume, data transmission, \dots -- it is still able to reach scientific goals of great importance in the field of exoplanets (\S\ref{sec:CHEOPS}). CHEOPS is a follow-up mission; it does not aim at necessarily discovering new planets, but at better constraining the parameters of those already discovered. 
The characterisation of different types of exoplanets -- Super-Earths, Neptunes and hot-Jupiters -- will allow a more in-depth understanding of the evolution and formation model of planetary systems (\S\ref{sec:theo-formation-evolution} and following). CHEOPS will primarily measure radius of small planets and probe atmospheres which require pristine observations. The technique used by the satellite is the transit method (\S\ref{sec:theo-detections}) which measures the dimming of light that occurs when a planet eclipses 
its host star. One 
of the requirements is to reach a noise of maximum 20 ppm over 6 hours of integration for Super-Earths around stars brighter than $V=9$. This can only be reached by observing from space and by analysing and mitigating the different sources of noise. 
\paragraph{}

The objectives of this work were to study and understand the behaviour of the stray light flux for different quantities in terms of orbits, exclusion angles or optics. Using a software developed in a previous phase of the project to generate the limit of the observation region seen by CHEOPS, the stray light flux was computed. 
It was showed that this first software had flaws in the way the stray light exclusion angle was computed which could generate erroneous observation maps (\S\ref{sec:numercis-MATLAB-details}). \emph{i.e.} that the flux could be higher than the requirement of 1 photon per second per pixel. 
In order to compute the stray light, an existing code was extensively rewritten (\S\ref{numercis:stray_light.f}) to remove errors in the computations and speed up the process. To handle this code, a pipeline was developed (\S\ref{sec:numerics-pipeline}) which manages the inputs and outputs and transfer them to a suite of analysis tools (\S\ref{sec:numerics-analysis}).
\paragraph{}

The analysis focuses on the year 2018, the year following the launch. The peculiar orbit of the satellite -- Sun-synchronous orbit (SSO) -- imply that the viewable region moves with time and that the relative angle between the planes of the night--day terminator and the orbit evolves (\S\ref{sec:dis-sl}). 
The effect of this is that the stray light flux has a strong seasonal variation. The general shape is a smooth W-shaped that is centred around the June solstice\footnote{As discussed in \S\ref{sec:dis-sl}, this depends on the Local Time of the Ascending Node. The baseline is to set it to 6 am (\S\ref{sec:CHEOPS}).}. 

Unsurprisingly, the altitude of the orbit impacts the amount the stray radiation received. The higher, the less flux measured. It reduces by about 20\% when going from 700 to 800 km altitude. An increasing altitude allows also for a larger viewing zone. However, there is a complication as there exists a zone of very high radiation called the South Atlantic Anomaly which greatly perturbs the operations on-board satellites that fly in the region (\S\ref{sec:dis-SAA}). 
Through the analysis of sky maps of the cumulative observation time (\S\ref{sec:dis-yearly}) and the probability of detecting a transit (\S\ref{sec:dis-transit-detection}), it can clearly be seen that there are two favoured regions that combine long possible observation times and low stray light contamination. Those regions correspond to the periods in the year when the two planes are close to be aligned to one another. 
The effect of changes in the optical assembly (\S\ref{sec:dis-PSTs}) materialised by a change in the point source transmittance (PST -- \S\ref{sec:theory-sl}) must remain within a factor of 2 or 3 to the preliminary RUAG design to enable the pristine observations required by the CHEOPS mission. The diminution of the stray light exclusion angle does not increase the visibility of faint stars ($9\lesssim V\lesssim12.5$), but does increase the observation capabilities for bright stars ($6\lesssim V\lesssim 9$). A thus post-processing of the stray light flux is necessary to eliminate regions that does not comply with the requirements.

This work also pointed out a solution to the calibration of the dark current and of the hot pixels of the instrument. As the field of view of the instrument is very narrow, there exist regions in the sky that are ``empty'' to the sensitivity of the CHEOPS detector (\S\ref{sec:disc-empty-regions}). Those regions can be looked at regularly throughout the year and by their observation, the ageing of the detector can be measured and therefore decorrelated from the data. 
\paragraph{}
A few recommendations can be made from the results obtained during this master thesis:
\begin{enumerate}\itemsep0em 
 \item \emph{Simulate the stray light} in order to assess the limiting magnitude due to the stray light and correct the visible region for the targets;
 \item \emph{A stray light exclusion angle of $25^\circ$ or $28^\circ$ can be used} in the generation of the visible zone. An exclusion angle of 35\textdegree\ does not ensure that the stray light will be lower than the requirements. It may very well be much higher and in that case, potential targets would be lost;
 \item \emph{A high altitude may not be the solution}: even if increasing the altitude does increase the visible zone, it does not necessarily improve the quality of the observations. This fact is due simply to the SAA whose crossing imposes to stop any scientific activity on-board and damages the CCD. When assessing the observable zone in terms of transit detection probability, the orbit SSO620 or SSO700 yield the best results;
 \item \emph{There are two favoured observation times that must be used for pristine observation}. In April and mid-September, the visible zone is the less interrupted by the stray light exclusion angles and the stray light flux minimal. Those regions in the sky are limited to short periods of times around those dates, but are prone to long observation of faint stars;
 \item \emph{Short period planets should be observed in Northern winter}. Regardless of the magnitude of the host star, the period close the December solstice is difficult to observe as the satellite enters deep in the illuminated Earth. Short period transiting planets would be easy to observe as their eclipse occurs often and can be observed several times.
\end{enumerate}

Future work should continue the efforts started here to prepare the scheduling of the observations to increase the science efficiency of the satellite. Moreover, a similar detailed stray light analysis should be carried out once a more finalised PST is computed. This ``ideal'' PST does not take into account the different contaminations or misalignments that arise during manufacturing and handling of the optical assembly. 
A worst-case study should also be able to characterise the losses in terms of observations or inversely, from a tolerable amount of losses, the maximum difference to the ideal PST can be quantified. The effect of the \emph{Moon} stray light could also prove interesting. Furthermore and maybe most importantly, the stray light calculator and viewing zone simulator should be merged to have an efficient code that could predict the observations of CHEOPS.
\paragraph{}

The goals of this Master thesis have fully been reached. The study of the stray light gave rise to several high level considerations for the mission as well as allowed a first step towards the optimised scheduling of the observations. This work has proven that the science foreseen in the CHEOPS proposal can be achieved in terms of sky visibility and observation time. It also quantified the stray radiation throughout the year and lead to considerations about the altitude of the orbit, the definition of the exclusion angles and the performance of the optical assembly. 
\paragraph{}
There are several other space-borne missions that are planned to fly in the near future with the objective of surveying a large amount of star to detect transits (\emph{e.g.} Transiting Exoplanet Surveying Satellite -- TESS, \cite{Ricker2010}) or other characterisation missions (Exoplanet Characterisation Observatory -- EChO, \cite{Tessenyi2010}). The former -- the small NASA TESS --, which will discover planets in the whole sky, will be launched about 6 months before CHEOPS and therefore, the synergies between those two missions are evident. While TESS is foreseen to discover up to several thousand planets, CHEOPS could possibly continue the characterisation of potentially interesting targets. The later -- ECho -- will notably perform spectroscopy of the atmosphere on interesting targets from the mid 2020s. CHEOPS could establish a list of potentially interesting targets and with planet with atmosphere.
\paragraph{}
CHEOPS is an ambitious mission that will provide information of great importance for the exoplanetary science. In this context, this work paved the way towards an efficient use of the observation time and towards a high quality of data.

 \vfill%

\newpage
\onecolumn
\vfill


\bibliographystyle{apj}
\nocite{*}
\bibliography{master.bib}

\begin{appendices}
 \chapter{List of Abbreviations}
 \begin{table*}[!h]
 \begin{center}
  \begin{tabular}{lp{13.5cm}} \toprule
    CCD & Charged Coupled Device \\
    CHEOPS & CHaracterizing Exoplanets Satellite \\
    EPFL & Ecole Polytechnique F\'ed\'erale de Lausanne \\
    ETHZ & Eidgen\"ossische Technische Hochschule Z\"urich \\
    EChO & Exoplanet Characterisation Observatory \\
    ESA & European Space Agency \\
    ESO & European Southern Observatory \\
    FOV & Field Of View \\
    HARPS & High Accuracy Radial velocity Planetary Search \\
    HST & Hubble Space Telescope \\
    IAU & International Astronomical Union \\
    ICRF & International Celestial Reference Frame \\
    INAF & Istituto Nazionale di Astrofisica / Italian national institute for astrophysics\\
    LEO & Low Earth Orbit \\
    LOS & Line Of Sight \\
    LTAN & Local Time of the Ascending Node\\
    MOA & Microlensing Observations in Astrophysics \\
    NASA & National Aeronautics and Space Administration \\
    OGLE & Optical Gravitational Lensing Experiment \\
    ppm & Parts per million \\
    PSF & Point Spread Function \\
    PST & Point Source Transmittance function \\
    RAAN & Right Ascension of the Ascending Node \\
    RMS & Root mean square \\
    RV & Radial Velocity detection technique \\
    SAA & South Atlantic Anomaly \\
    SDSS & Sloane Digital Sky Survey \\
    SL & Stray Light \\
    SNR & Signal-to-Noise Ratio \\
    SciReq & Science Requirements \citep{SciReq} \\
    SSO & Sun-Synchronous Orbit \\
    TAPS & Theoretical Astrophysical and Planetary Science Group at UniBe\\
    TESS & Transiting Exoplanet Survey Satellite\\
    UniBe & University of Bern \\
    UniGe & University of Geneva \\
   \bottomrule
  \end{tabular}
  \end{center}
\end{table*}

\chapter{Additional Computations}
\section{Analysis of Stray Light Exclusion Angle from the Terminator} \label{app:terminator-angles}
\begin{figure}[h]
 \begin{center}
  \includegraphics[width=0.59\linewidth]{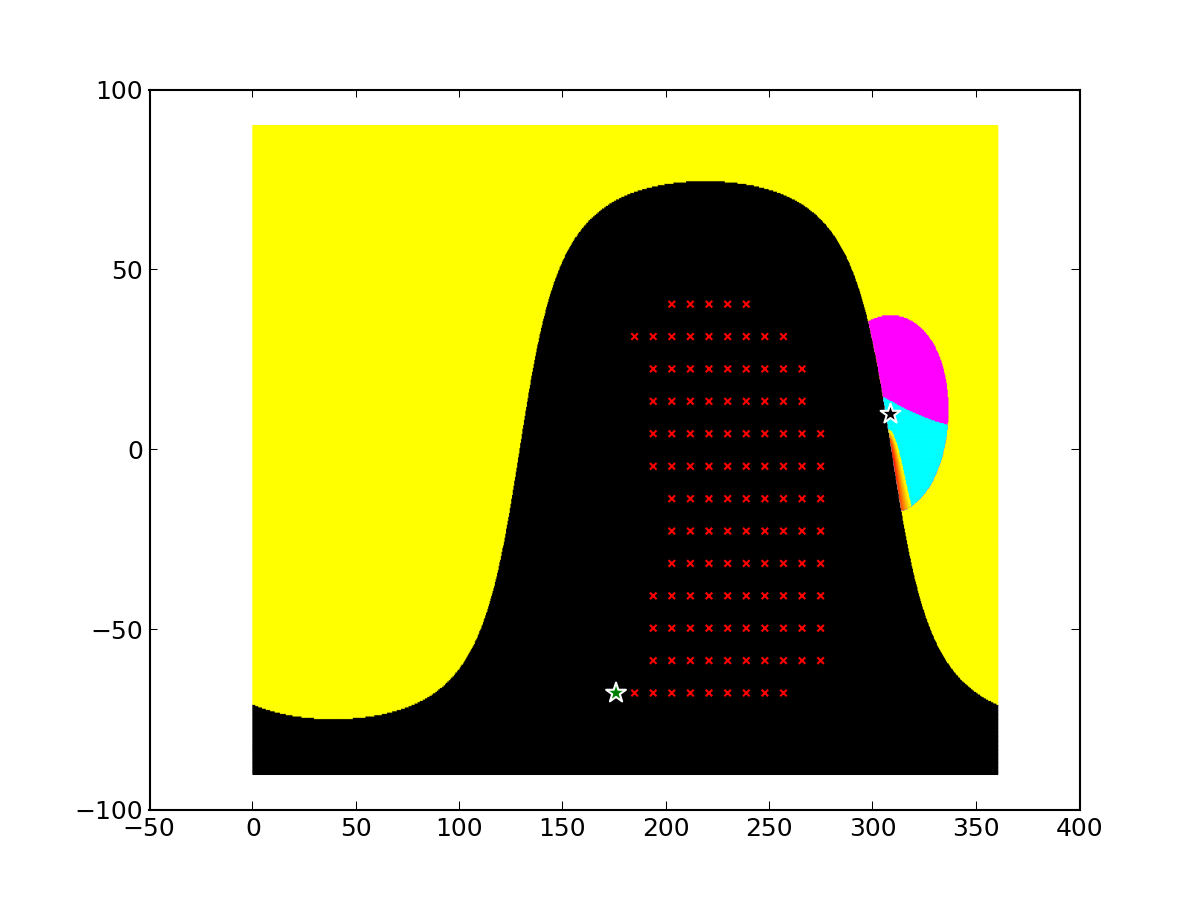}
  \includegraphics[width=0.4\linewidth]{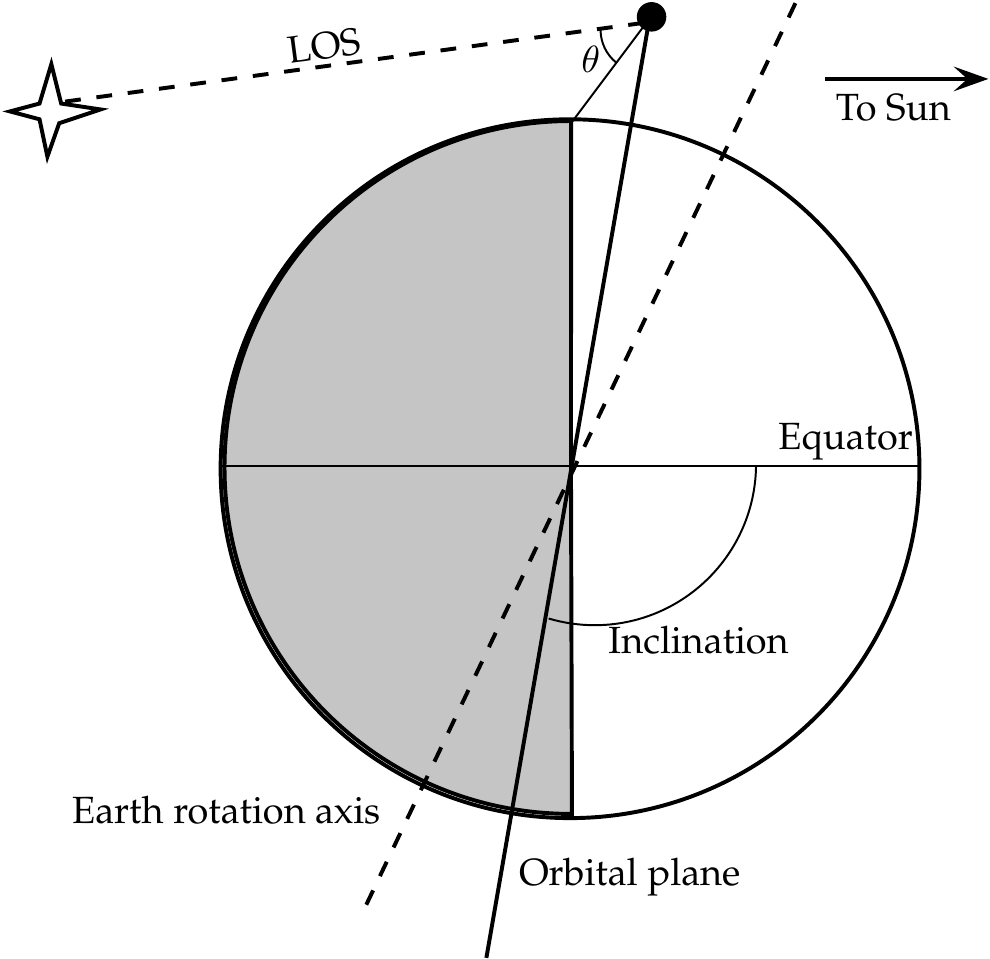}
  \caption{\emph{Left.} Example of a typical faulty target (green star) on the projected map of the Earth. The illuminated part is in yellow, the night in black. Red crosses are all cells that are visible in the sky. The pink region is the illuminated part of the Earth seen by the satellite while the blue one is illuminated, seen and contributing to the stray light. Wrong angles are shown in red to yellow near the terminator. The x-axis represents the right ascension in degrees and y-axis is the declination. -- \emph{Right.} Sketch of the position of the satellite (black dot). $\theta$ is the stray light exclusion computed only to the terminator.\label{fig:app-map_error}}
  \vspace{-1.5em}
 \end{center}
\end{figure}

As explained in section \S\ref{sec:numerics-visibility-constraints}, the stray light exclusion constraint is relaxed when observing over the night. The stray light exclusion angle is measured from the terminator along the line of sight. The condition is that this angle $\theta$ is larger (or equal) than the threshold angle $\theta^\ast$. This threshold angle is the same as the stray light exclusion angle $\theta^\ast=\alpha_\oplus=35^\circ$. In the following, it will be shown that this computation is necessary, but not sufficient and thus dangerous. 
Indeed, in certain configurations\footnote{Those configurations are basically when the plane of the orbit is close to the plane of the terminator.}, photons reflected close to the terminator can enter the telescope with angle less than the threshold. An example of this bad configuration is shown in Fig. \ref{fig:app-map_error}.

The consequence of having angles lower than the threshold is the presence of spikes in the stray light flux. This considerably slows the computational process down the process as these spikes vary from one orbit to another.
\begin{figure}[h]
 \begin{center}
  \includegraphics[width=0.55\linewidth]{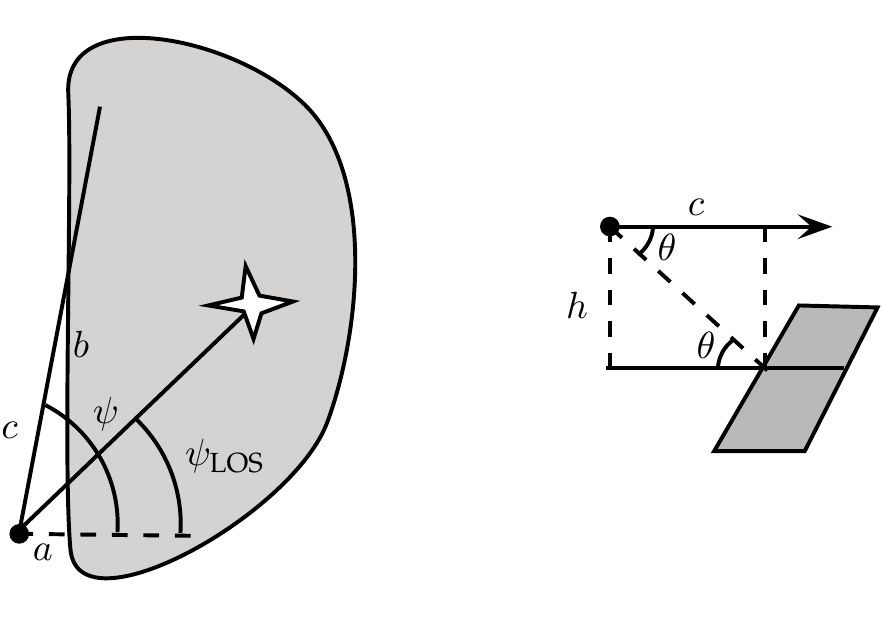}
  \caption{The grey zone is the night side of the Earth and the black dot the satellite (which is over the day side). $\psi_\text{LOS}$ is the angle measured from the normal of the local terminator line to the line of sight -- \emph{Left.} View from the top. The distance $b$ is the distance from the intersection between the line $a$ and the terminator and the intersection between the line $c$ and the terminator. -- \emph{Right.} Side view.\label{fig:app-angle-terminator}}
  \vspace{-1.5em}
 \end{center}
\end{figure}

The situation is sketched in figure \ref{fig:app-angle-terminator}. The horizontal distance of the satellite to the terminator $a$ is fixed for a given time and so is $h$, the altitude of the satellite. It has to be noted that what follows intends to show in a simplified way that angles can drop below 35\textdegree. The angle $\psi$ is measured from the normal of the terminator to the line of sight or to the line joining the telescope and an illuminated surface. Therefore:
\begin{equation}
 a = \text{cst},\quad b=a\tan\psi\implies c=a\sqrt{1+\tan^2\psi}
\end{equation}
When looking at this in the plane of the vertical, $h$ is constant as it is the altitude of the satellite, $c$ varies in function of $\theta$:
\begin{equation}
 \tan\theta=\frac{h}{c}
\end{equation}
Bringing those two equations together by eliminating $c$ yields:
\begin{equation}
 \frac{h}{a\sqrt{1+\tan^2\psi}}=\tan\theta \implies \arctan\left( \frac{h}{a\sqrt{1+\tan^2\psi}}\right)=\theta\geq\theta^\ast
\end{equation}
This is calculated in the observability map code for the line of sight only. However, if $\psi$ is sufficiently decreased, then $\theta$ may drop below $\theta^\ast$. Indeed:
\begin{equation}
 \text{if}\ \psi\nearrow \implies a\sqrt{1+\tan^2\psi}\nearrow\implies\frac{h}{c}\searrow\implies \arctan() = \theta\searrow
\end{equation}
and therefore, if the target is placed such that $\psi_\text{LOS}$ is large enough, there exist parts of the illuminated Earth that can be seen with an angle less than $\theta^\ast$. Note that this is valid only for angles $\psi \sim \psi_\text{LOS}$.

\section{Dark Current Noise} \label{app:DarkCurrentNoise}
This section is dedicated to the characterisation of the dark current. One idea is to point at an empty region of the sky and take frames just like it would be done during calibration, but without a shutter. Here, the emphasis is on what does empty mean. What is the limiting magnitude of the remaining objects in the field of view ?
At the time of writing, the selected detector is the CCD AIMO from the company e2v. In the data sheet \citep[][private communication]{aimo2006}, the dark signal $Q_d$ is given by the empirical law valid between 230 and 300 K:
\begin{equation}
 \frac{Q_d}{Q_{d0}}=1.14\cdot10^6 T^3\exp\left( -\frac{9080}{T} \right)
\end{equation}
where $Q_{d0} = 250$ e$^-$/px/s is the dark signal at 293 K. The $1\sigma$ variation is given to be $\pm60$ e$^-$/px/s. Below 230 K, additional contribution to the dark current with weaker temperature dependence can arise. The temperature the detector on-board CHEOPS will be 233 K which therefore yield:
\begin{equation}
 Q_d^* = (4.9\pm1.0)\cdot10^{-2}\ \frac{\text{e}^-}{\text{px\ s}}
\end{equation}
In order to characterise correctly the dark current across the detector, the lowest estimation of the quantum efficiency of the CCD is taken into account ($Q_E=0.75$ e$^-$/ph) and therefore the equivalent flux of photon would be:
\begin{equation}
 N_d^*=(5.72\pm1.37)\cdot 10^{-2} \frac{\text{ph}}{\text{px\ s}}
\end{equation}
This flux $N_d^*$ would be equivalent of observing a star which would trigger a response from the CCD of the same value. Therefore, a limiting magnitude can be computed for the field of view. To convert the flux to magnitude in the right units, the following formula is used (see \S\ref{numerics:functions} for justification):
\begin{equation}
 m_V=-\frac{5}{2}\log_{10}\left[\frac{N_d^*}{J_VF_v\frac{\delta \lambda}{\lambda}\left( \frac{R_\text{tel}}{R_\text{PSF}} \right)^2}\right]
\end{equation}
which yields a magnitude of $m_V(N_d^*)=19.6\pm1.6$. The brightest objects in the sky should not perturb the measurement, so the limiting magnitude $V_l$ is set to one fainter than the best case for the magnitude which is:
\begin{equation}
 V_l = 19.6+1.6+1\approx 23
\end{equation}
This limiting magnitude is extremely faint and therefore very difficult to observe without other sources of noise -- such as the stray light -- can arise.


\chapter{Examples of Visibility Charts}\label{app:visibility-charts}
The above analysis was performed using only a few generic assumptions such as the transit time, orbital period and limiting magnitude. CHEOPS catalogue will contain the coordinate of the object as well as the predicted time of transit. In order to make the first steps towards this very important tool in the CHEOPS mission, a script was developed that, in the very same idea as above, computes when a particular object of a catalogue is be visible. 
An example of such a chart is given in Fig. \ref{app:visibility-charts} which depicts visible stars for an altitude of 800 km and the first week of January 2018 for a few known exoplanets that could be potential targets. It can clearly be seen that the SAA does not influence the observation in the same fashion for all stars. In this chart, only targets of negative declinations are shown. This script automatically computes the observable periods of the target 
using a minimum observation time and a given magnitude. Of course, for development and understanding purposes, the limiting magnitude or SAA criteria can be switched off. In the future, ingress and egress times will enable to build a list of possible observations from which the schedule will be deduced. It can be seen that although the SAA perturb a lot the observations, there are still targets that can be observed almost continuously: they are points that are always visible, except for the SAA interruptions.
\begin{figure}[h]
 \begin{center}
   \includegraphics[width=1\linewidth]{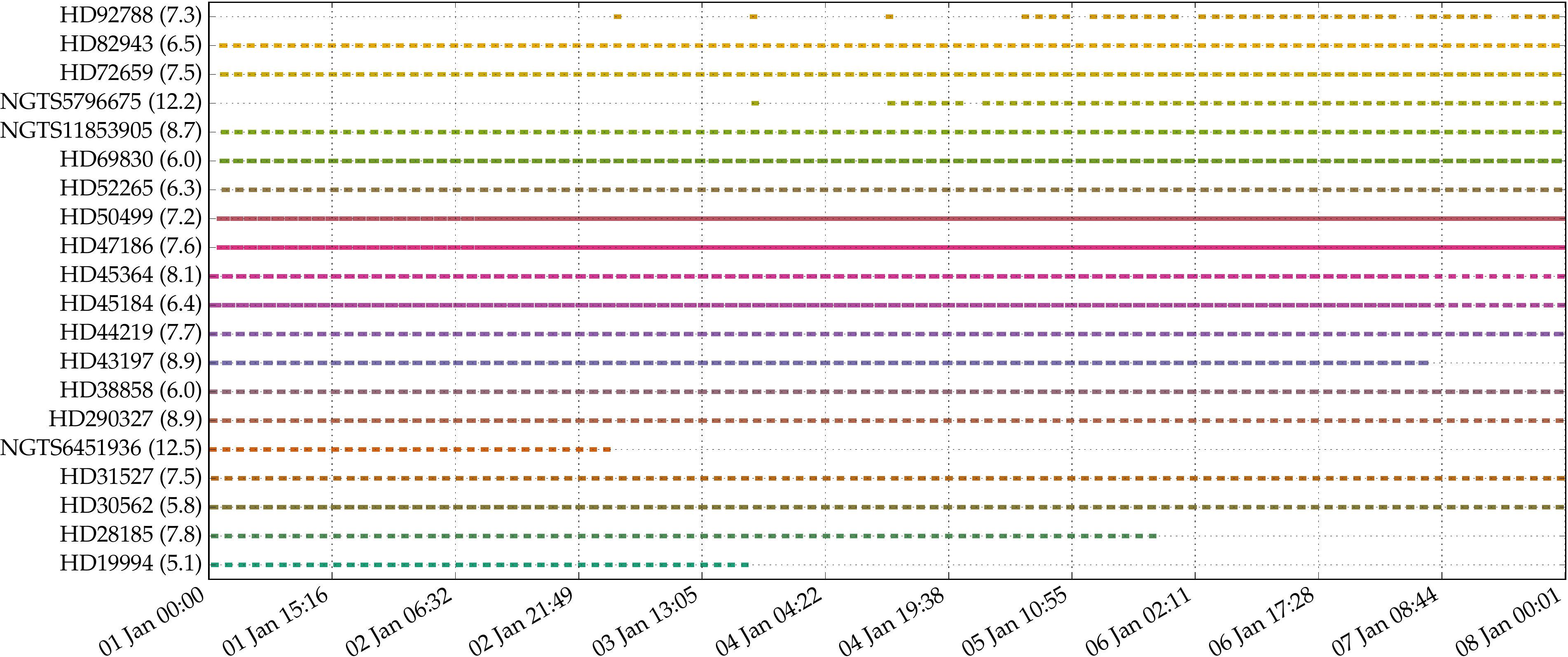}
   
   \vspace{2em}
  \includegraphics[width=1\linewidth]{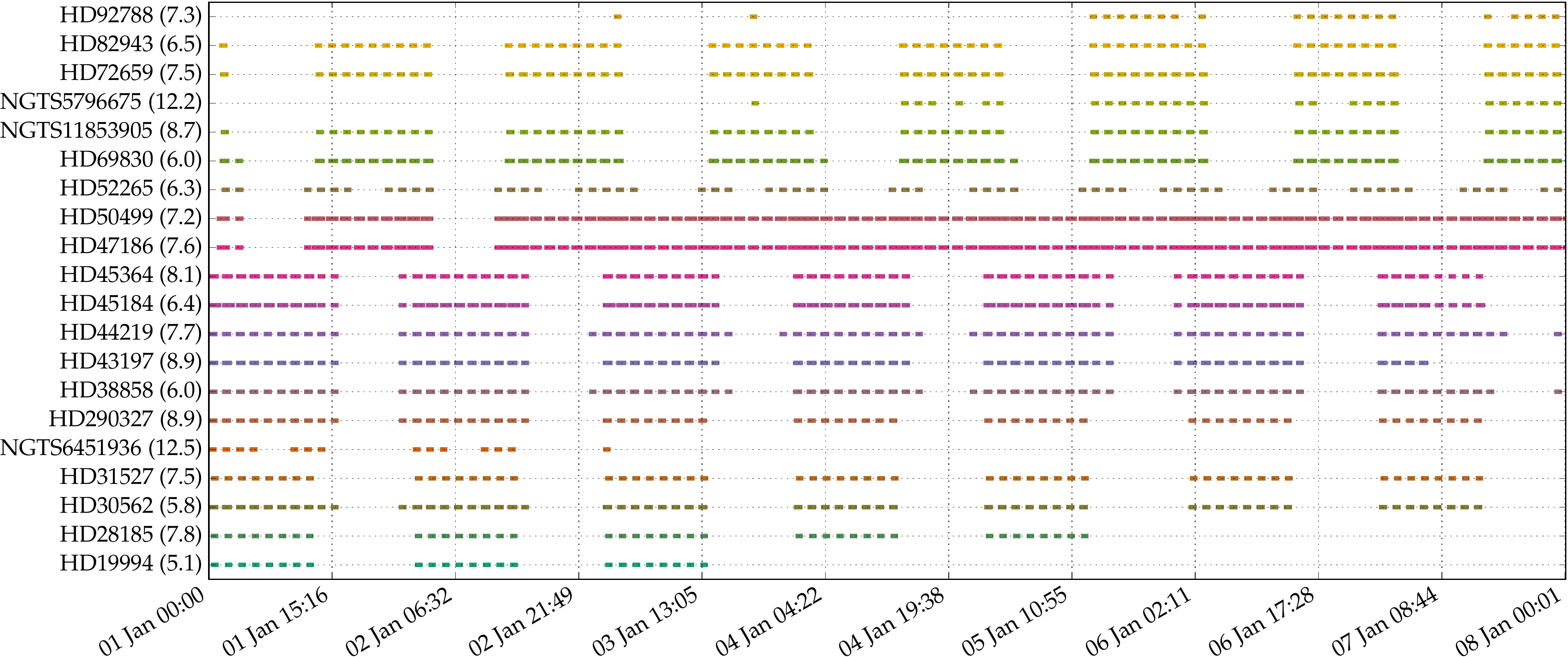}
  \caption{Visibility Charts for the first week of 2018 for an orbit at 800 km. There are 20 objects, all located in negative declinations area. The object's name and the $V$ magnitude in parenthesis are given. -- \emph{Top.} Without the constraint of the SAA. -- \emph{Bottom.} With the SAA.\label{fig:visibility-charts}}
  \end{center}
\end{figure}

\chapter{Empty Regions in the Sky} \label{app:empty}
In the following, regions that would match the criteria of ``empty'' (as defined in \S\ref{sec:disc-empty-regions}) are presented. The red square in the plot represent the whole field of view of CHEOPS (1440 arcmin$^2$) while the inner red circle depicts the region of interest. All regions presented on this page were checked again the SDSS. The next page presents regions that were not checked against a catalogue of faint objects. The stars in the maps are plotted according to their data in the Tycho-2 catalogue. The $V$ magnitude is shown for stars brighter than the 10$^\text{th}$ magnitude.

\begin{figure}[h]
 \begin{center}
  \includegraphics[width=0.49\linewidth]{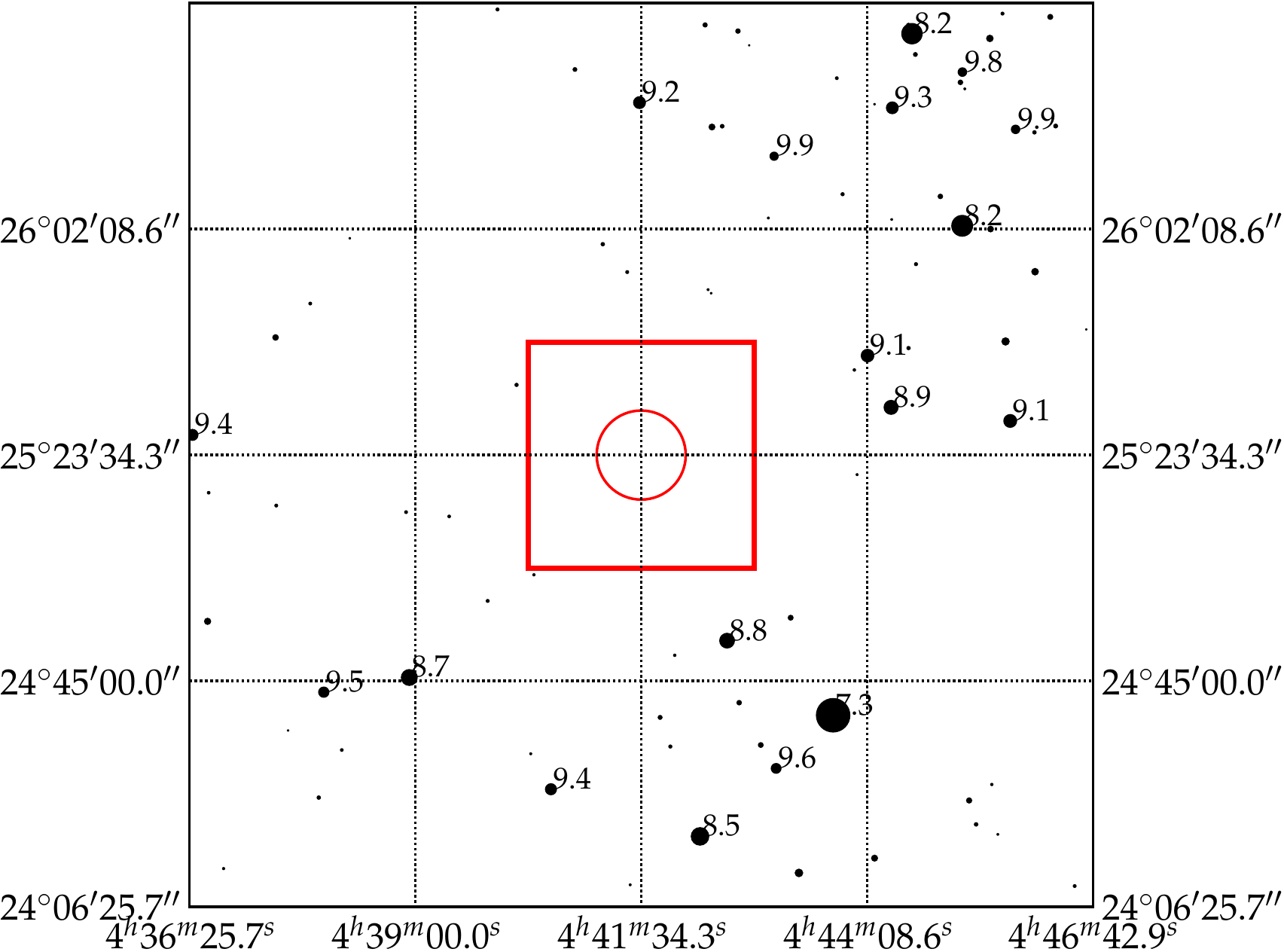}
  \includegraphics[width=0.49\linewidth]{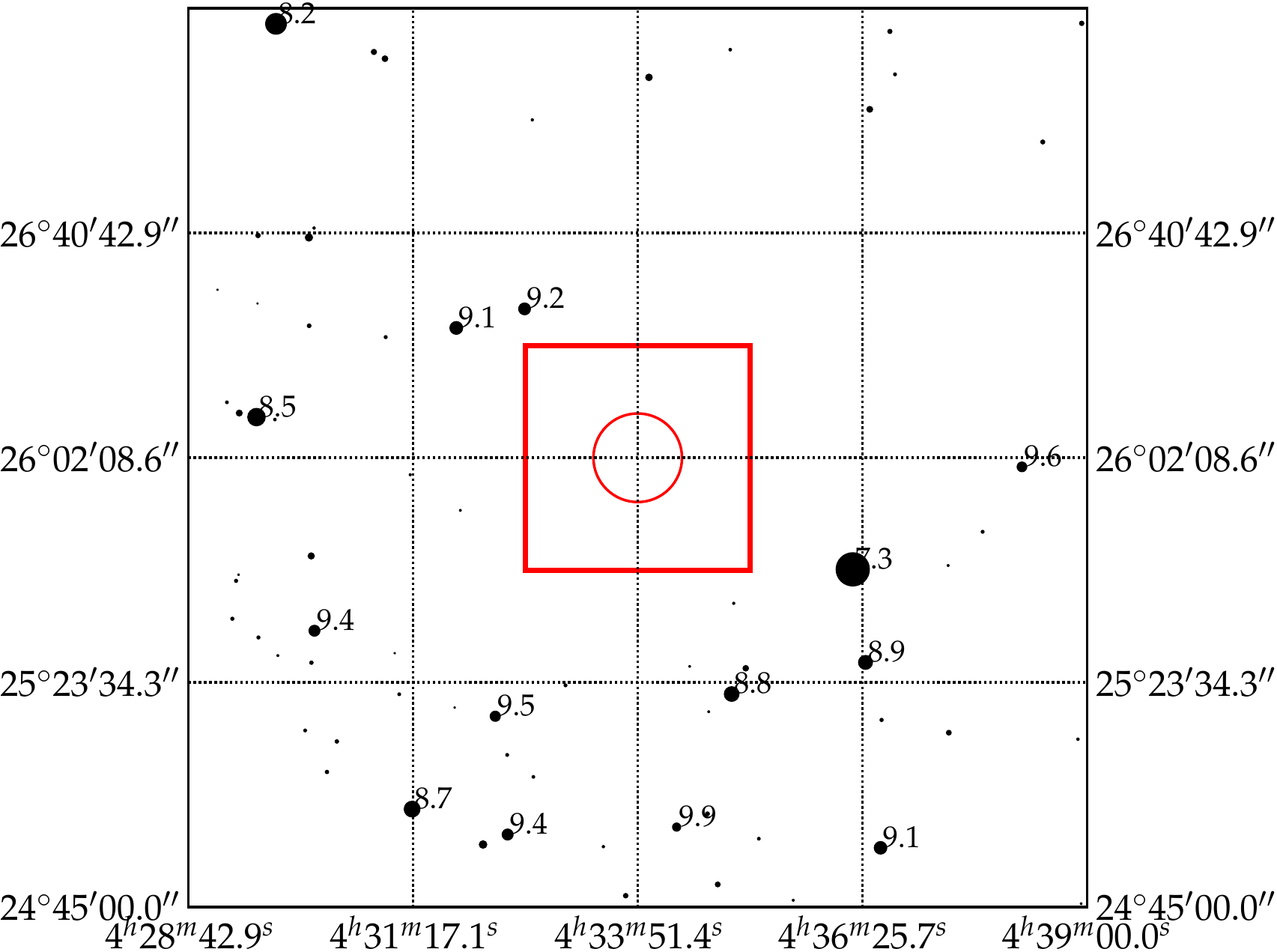}
  \includegraphics[width=0.49\linewidth]{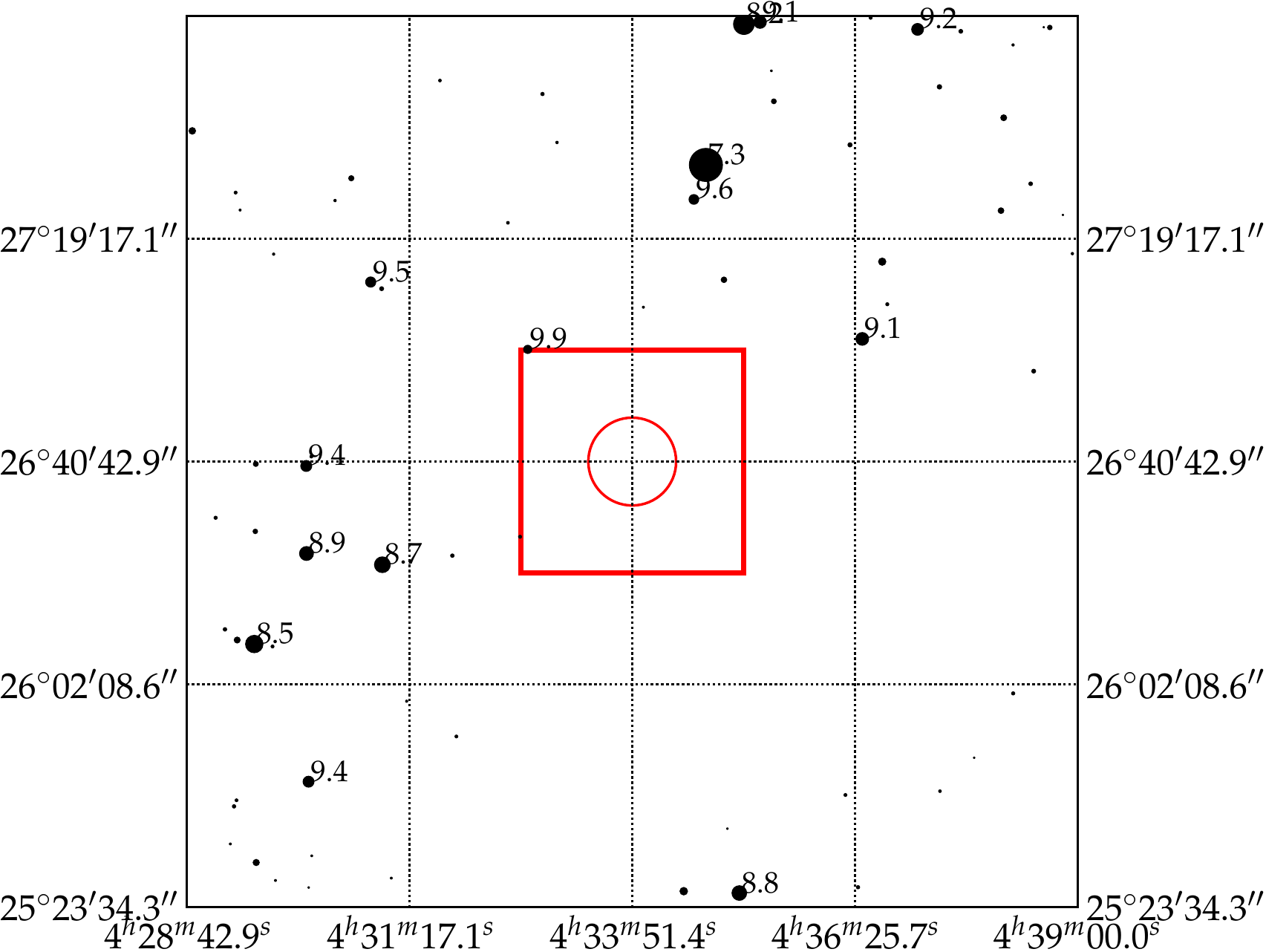}
  \includegraphics[width=0.49\linewidth]{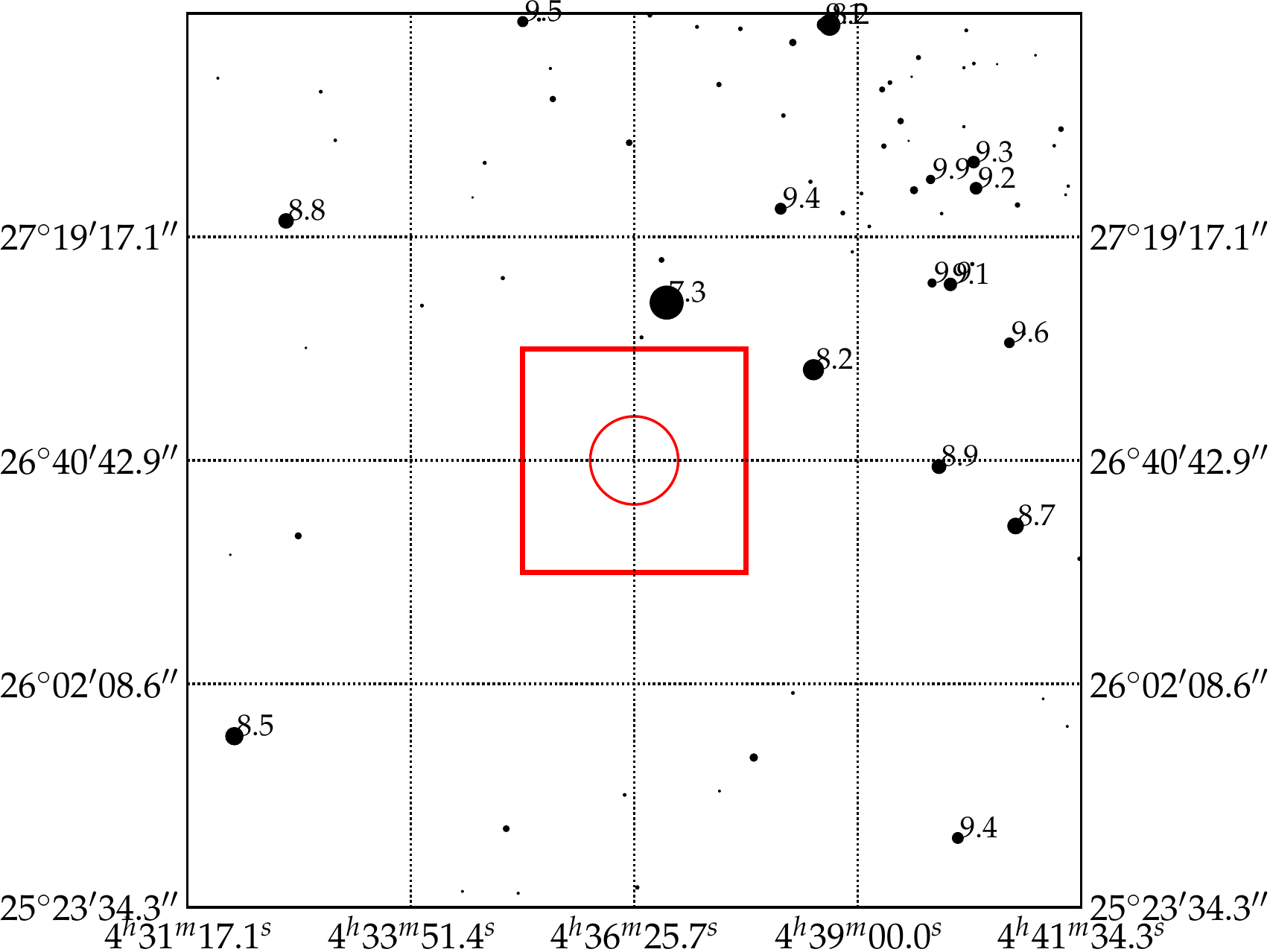}
  \caption{Sky maps for empty regions checked against the SDSS catalogue.\label{fig:empty-1}}
  \end{center}
\end{figure}
\newpage
\begin{figure}[h]
 \begin{center}
  \includegraphics[width=0.49\linewidth]{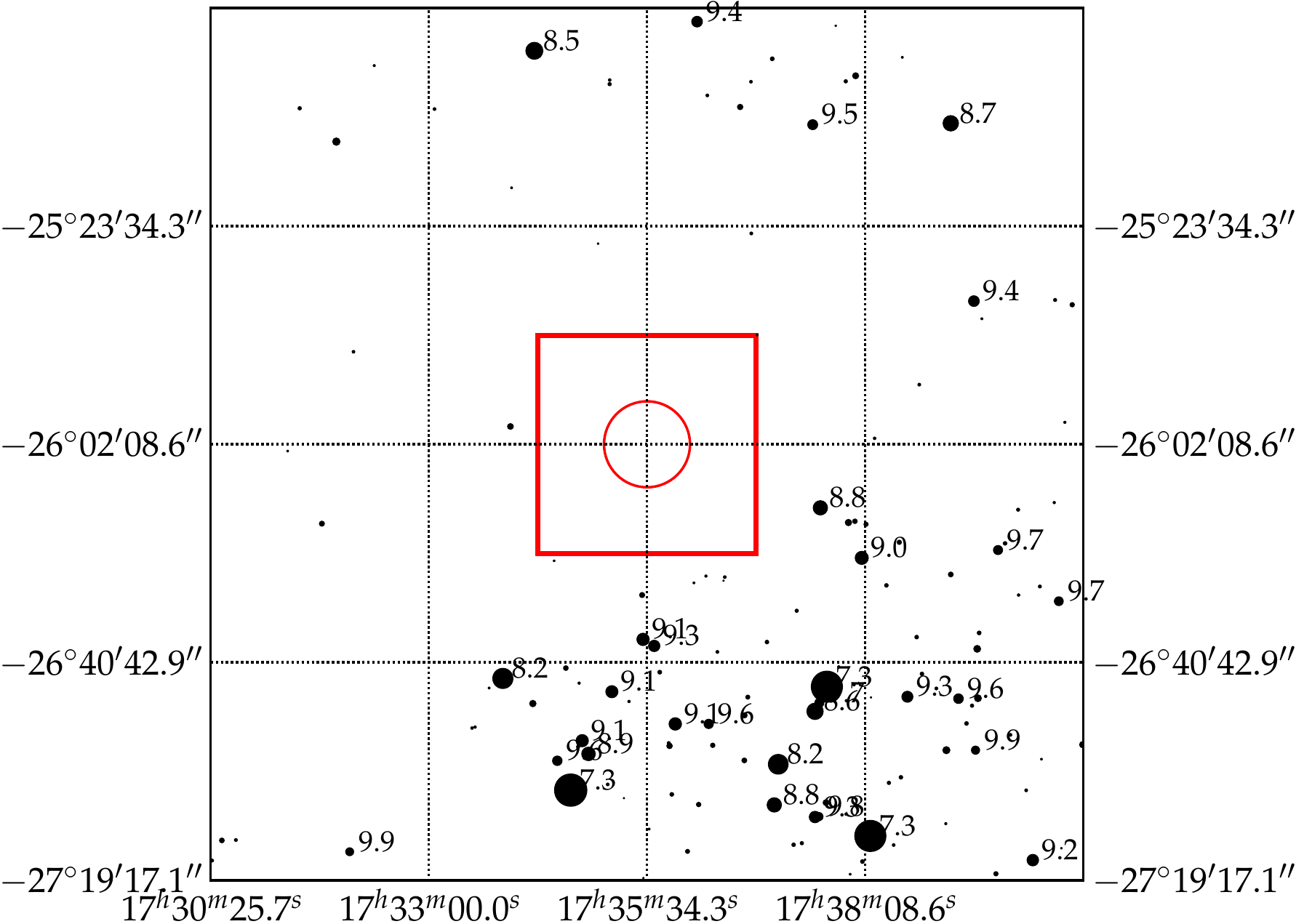}
  \includegraphics[width=0.49\linewidth]{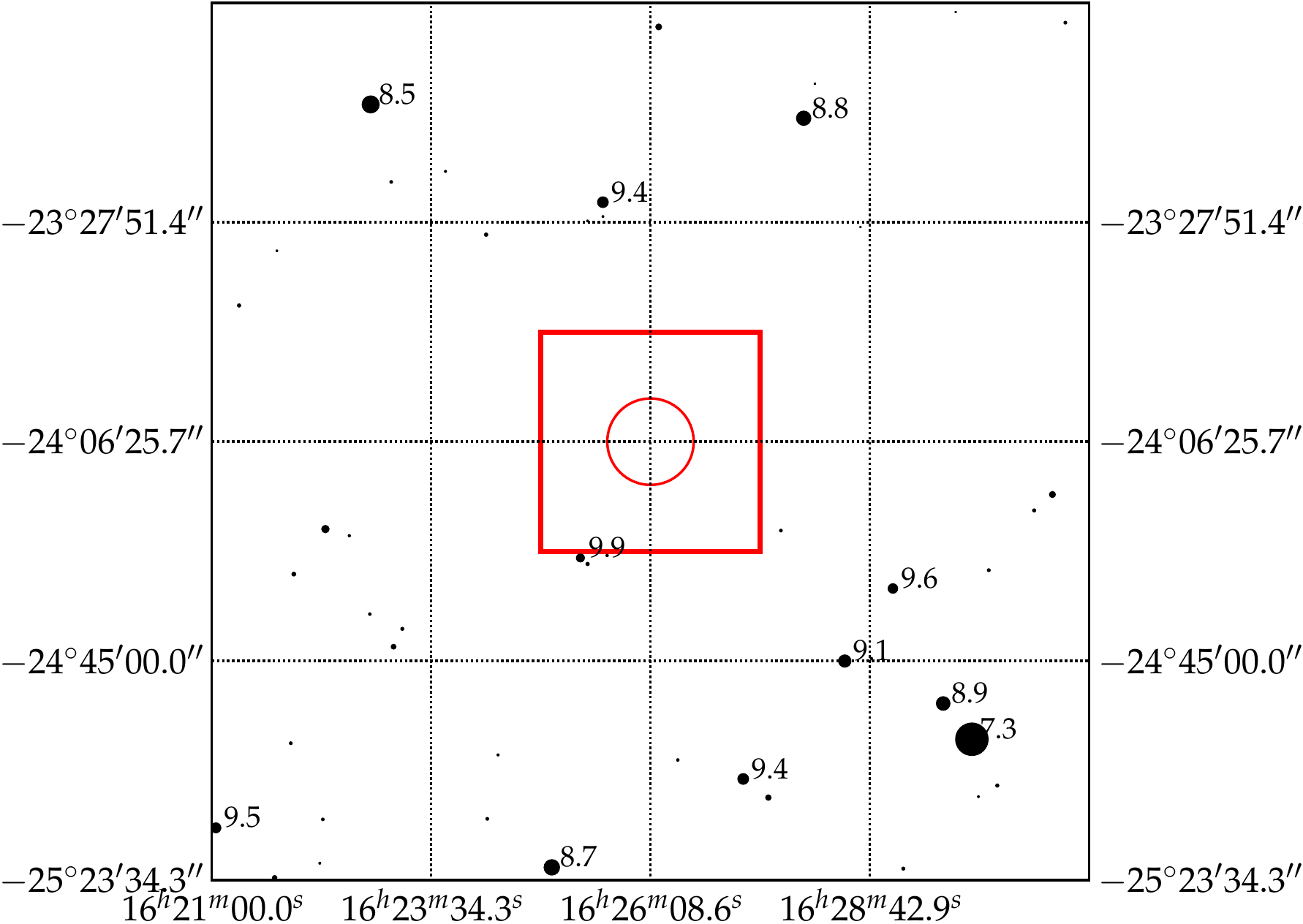}
  \includegraphics[width=0.49\linewidth]{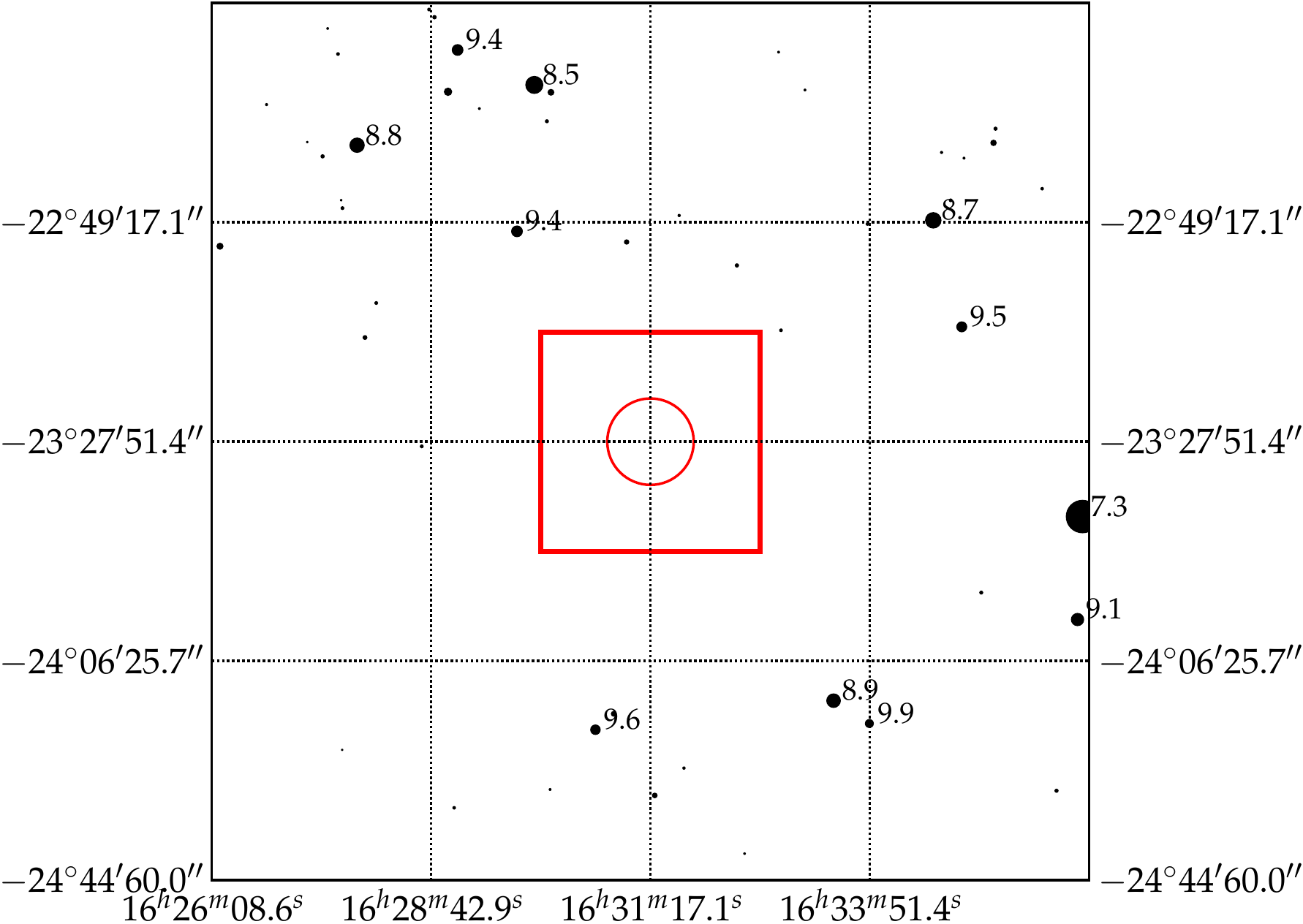}
  \includegraphics[width=0.49\linewidth]{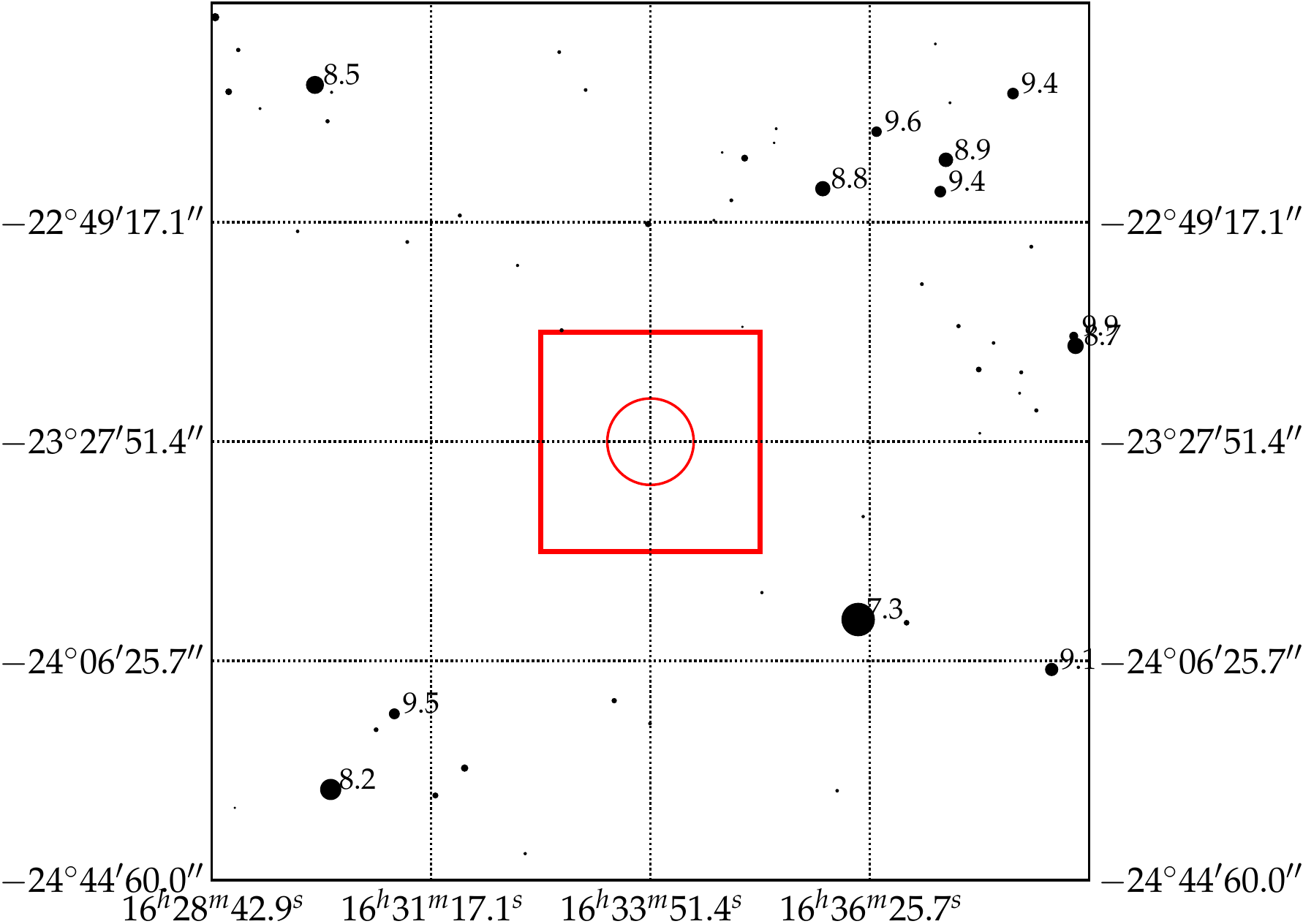}
  \includegraphics[width=0.49\linewidth]{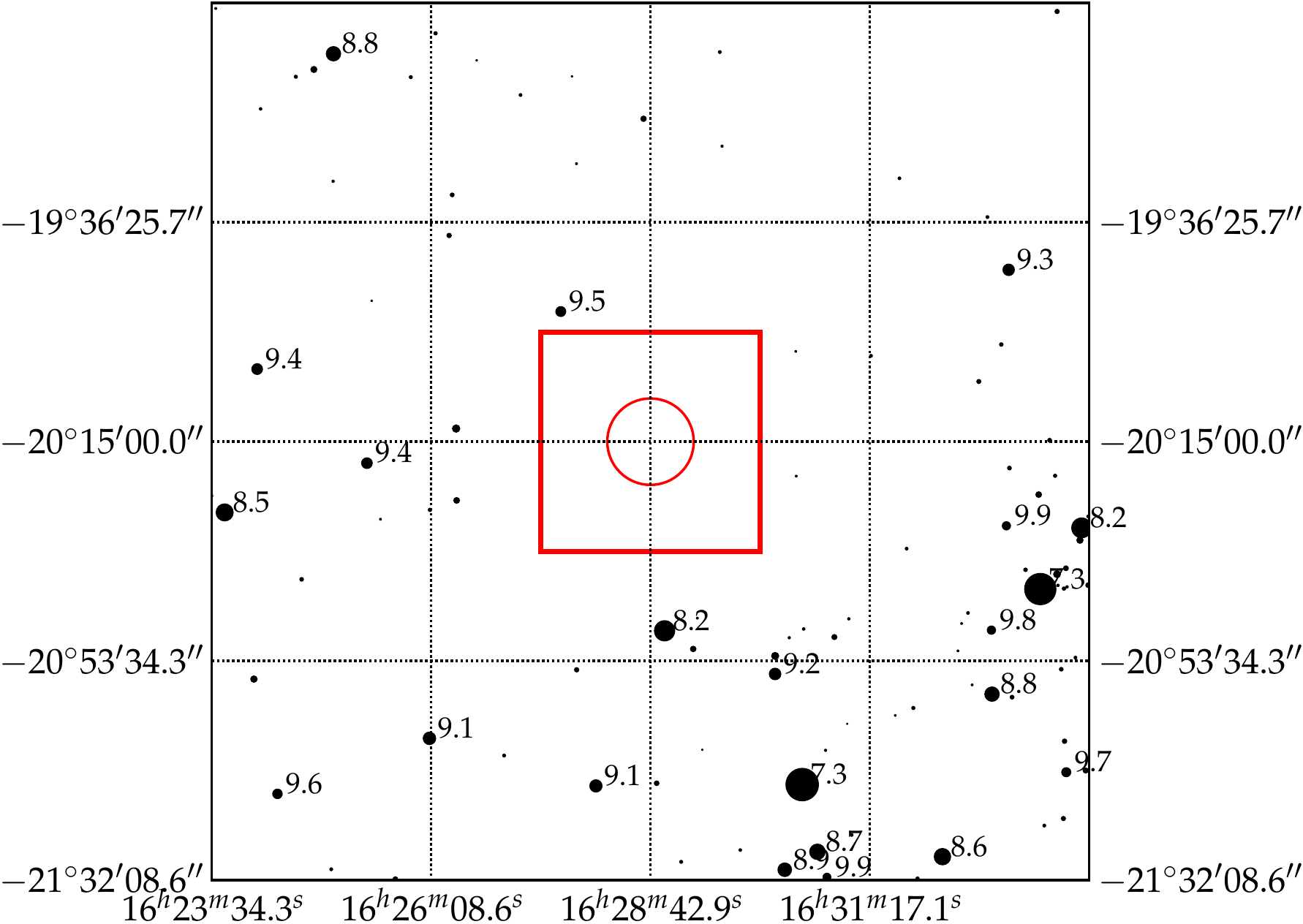}
  \includegraphics[width=0.49\linewidth]{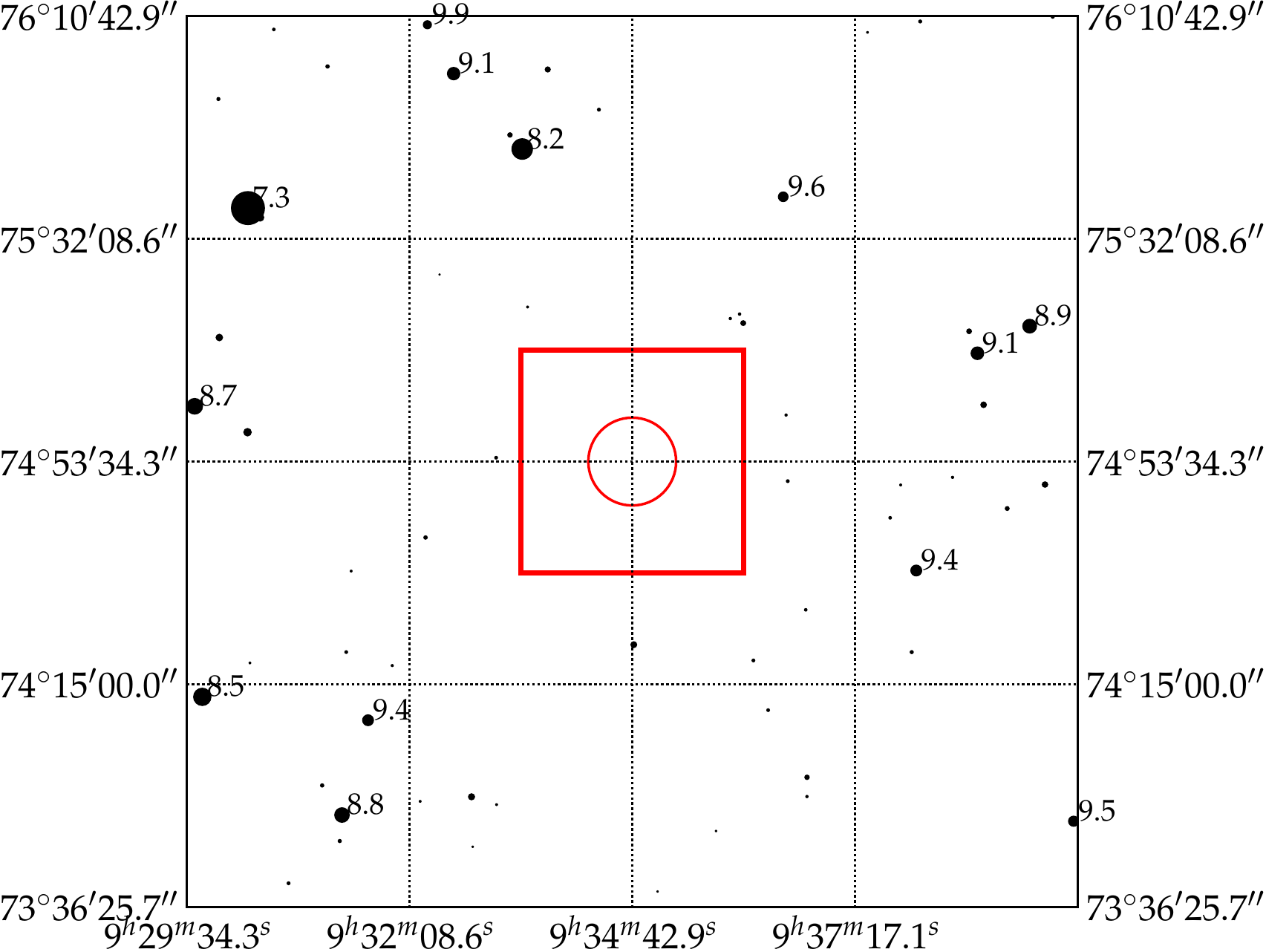}
  \caption{Sky maps for empty regions that still need further confirmation.\label{fig:empty-2}}
  \end{center}
\end{figure}

 \chapter{Presented Poster}
 The following poster was presented during the science meeting \textnumero 1 of the CHEOPS mission which took place mid May 2013 in Bern. The purpose of this meeting was to present the CHEOPS mission and its science to the (scientific) community. The section dedicated to the science with CHEOPS in this work is based on talks given during this meeting.
%
 The poster was prepared in collaboration with Adrien Deline from UniGe whose Master thesis also focussed on CHEOPS. His objective was to assess which targets could be observed without taking into account the detailed stray light analysis performed here. He was followed by David Ehrenreich. Luzius Kronig wrote the code that produces observability maps.
 
 \begin{figure}
 \begin{center}
  \includegraphics[height=1\linewidth,angle=90]{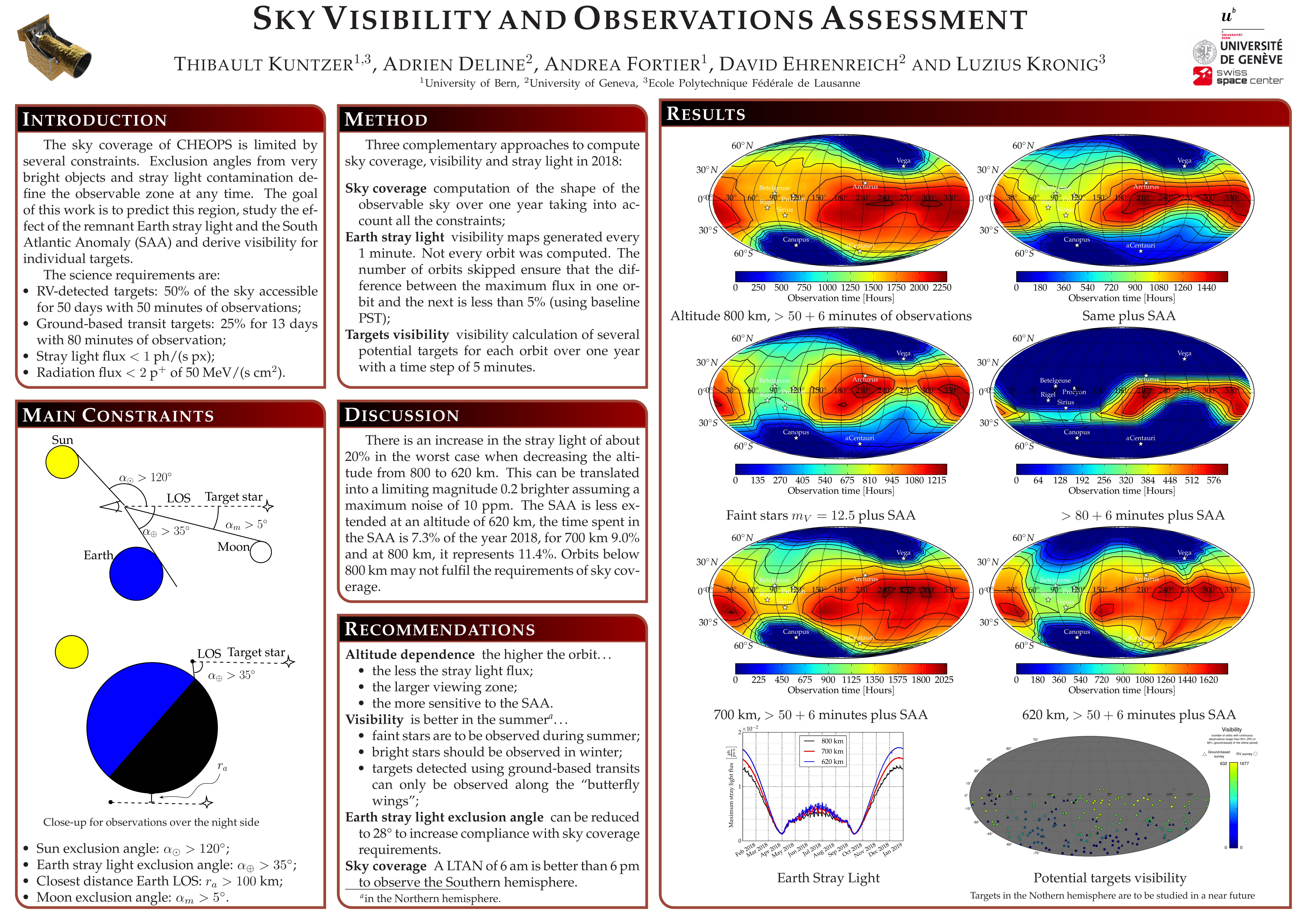}
 \end{center}
\end{figure}

%
%

\end{appendices}
\end{document}